\RequirePackage{fix-cm}
\documentclass[smallextended]{svjour3}       
\smartqed  
\usepackage{graphicx}
\usepackage{breqn}
\usepackage{graphicx}
\usepackage{epstopdf}
\usepackage{pdfpages}
\usepackage{tikz}
\begin{document}

\title{A multidimensional combustion model for oblique, wrinkled premixed flames}

\author{Michael Pfitzner, Junsu Shin, Markus Klein}

\institute{M. Pfitzner \at
	Bundeswehr University Munich, Werner-Heisenberg-Weg 39, Neubiberg, Germany \\
	Tel.: +49-49-60042103\\
	Fax: +49-49-60042116\\
	\email{michael.pfitzner@unibw.de} 
}

\date{Received: date / Accepted: date}

\maketitle

\begin{abstract}
A new  premixed turbulent combustion model is proposed. It is based on one-dimensional (1D) filtering of density times progress variable and of the reaction source term of laminar premixed flame profiles using a filter kernel which reflects the variation of the slicing area of planar flame fronts as they move through multidimensional filter volumes. It is shown that these multidimensional effects qualitatively change  the relation between the filtered reaction source term and the Favre-filtered reaction progress variable compared to 1D filtering, particularly at large filter widths. Analytical results for the filtered quantities are achieved by approximating density times progress variable and reaction source term by suitable Ansatz functions. Filtered data from Direct Numerical Simulations (DNS) of statistically planar turbulent premixed flames at different free stream turbulence levels and heat release parameters is used to develop and validate the model.  Two wrinkling factor models as function of filter width and subgrid turbulence level are proposed. For small filter widths up to two times the laminar flame thickness, minor effects from subgrid flame folding are observed. For larger filters, the filtered reaction source term rises linearly with filter width at a rate which increases with subgrid turbulence level. The modelled reaction source term as function of Favre averaged progress variable and filter width shows excellent agreement with filtered DNS data for all investigated free stream turbulence levels, filter widths and heat release parameters. 

\keywords{Turbulent premixed combustion \and Flamelet pdf \and LES combustion model \and Filtered reaction source term \and Wrinkling factor}
\end{abstract}

\section{List of symbols}
\begin{tabular}[h!]{|p{10mm}p{38.mm}p{1pt}p{10mm}p{39.mm}|} \hline
\textbf{Arabic} & && & 	\\
$c$ & reaction progress variable && $c_p$ & specific heat at constant pressure \\
$p$ & pdf &&    $t$ & time\\
$u$  & velocity &&   $y$ &  mass fraction \\
$x$ & spatial coordinate &&   $A,B$ & BML parameters \\
$B_1$ &	Arrhenius prefactor &&   $D$  &  diffusion coefficient, fractal dimension \\
$I$  & correction factor	&&  $H$    & Heaviside function \\  
$Le$  & Lewis number   &&    $N$ & pdf normalization factor \\
$s_L$ & laminar flame speed    &&   $T$ & temperature \\
$x_m$ & centre of filter interval &&   & \\
\textbf{Greek} & && & \\
$\alpha$ & temperature rise parameter && $\beta$  & 	Arrhenius activation temperature coefficient \\
$\delta$ &	delta function, flame thickness && $\tau$ &	density ratio \\
$\rho$ & density && $\phi$ & product mass fraction \\
$\lambda$ &	heat conductivity && $\omega$ & reaction source term \\
$\xi$    &  canonical spatial coordinate && $\Delta$ &	filter size \\
$\Gamma$ &	scale ratio &&  &	 \\
$\Sigma_f$ &	flame surface density && $\Xi$ &	wrinkling factor \\
$\Sigma$ &	isosurface area && $\Omega$ & filter volume \\
\textbf{Subscripts}& && &  \\
$b$ & burnt && $f$ & flame  \\
$n$ & index analytic profile && $u$ & unburnt \\
$th$ & thermal && 1D & one-dimensional\\
$A$ & Arrhenius && $0,1$ & Arrhenius rate temperature exponents \\
$\xi$ & canonical coordinates &&  &  \\
\textbf{Superscripts} & && & \\ 
$+$ & upper edge of filter && $-$ & lower edge of filter \\
$*$ & $c$ value on isosurface &&  &  \\
\hline
\end{tabular}

\section{Introduction}

Turbulent premixed flames of fuel-oxidizer combinations with large activation energies are present in many technical combustion applications. Prominent examples are flames using hydrogen or hydrocarbons as fuel and pure oxygen or air as oxidizer. Large activation energies and low diffusivities of gaseous components result in very thin reaction layers of premixed flames even at atmospheric pressure. With increasing pressure, density and thus chemical reaction rates raise while the diffusivities and heat conductivity drop, reducing the reaction layer thickness further. 

These thin reaction layers usually cannot be resolved numerically in LES where the typical computational cell is more than an order of magnitude larger than necessary to numerically resolve the laminar flame structure embedded in the turbulent flow field. Thus combustion models are required also for LES of premixed combustion processes in most technical applications and academic configurations.

Reaction layers are folded and stretched by the turbulent flow field, but there is evidence that even at quite large Karlovitz numbers, their inner structure remains largely unaffected \cite{nilsson2018structures}. This is also supported by experimental observations by Driscoll et al. \cite{driscoll2008turbulent} and DNS results by Luce et al. \cite{luca2019statistics} indicating that even at quite high levels of turbulence intensity $u'/s_L$, the profiles of species conditioned on progress variable behave as in a 1D laminar flame and that the filtered fuel consumption rate increases in proportion to the folded surface of the reaction layer. Additional effects such as flame stretch, flame curvature and thickening of the reaction layer through small scale turbulent eddies modify this proportionality only moderately. 

After specification of a (monotonous) reaction progress variable $c$, all other quantities (species, temperature, density) in the 1D flame can be tabulated as function of this single variable. It is convention to normalise the progress variable to $c=0,1$ in the fully unburnt/burnt regions, respectively, although this is not strictly being necessary.

Many turbulent premixed combustion models using a single progress variable $c$ have been developed in the past. The artificially thickened flame  (ATF) model \cite{colin:2000} makes the flame front resolvable on the LES grid by artificially increasing the diffusivity while reducing the reaction term such that the local laminar flame propagation speed remains unchanged. The effect of non resolved subgrid flame wrinkling on the reaction source term is taken into account by empirical efficiency functions. 

Some models assume the existence of infinitely thin flame fronts propagating locally at a turbulent flame speed $s_T$, for which empirical expressions are used. Examples are the G-equation level-set approach \cite{pitsch2002large} and models calculating a subgrid flame surface density $\Sigma$. In the latter type of models, the sum of the molecular diffusion term and the chemical reaction source term in the $c$ transport equation is replaced by $\rho_u s_L \Sigma_f$, where $\rho_u$ is the density of the unburnt medium, $\Sigma_f$ is the flame surface density and $s_L$ is the flame propagation speed. $\Sigma_f$ is either determined by a transport equation \cite{poi05} or approximated as $\Sigma_f= \Xi \mid\nabla \overline{c}\mid$, evoking algebraic (or again transport based) models for the wrinkling factor $\Xi$  \cite{ma2013posteriori} and often replacing $\mid\nabla \overline{c}\mid$ by $\mid\nabla\tilde{c}\mid$. Models of this type change the mathematical character of the  progress variable transport equation, preventing a recovery of the laminar flame front structure in the DNS limit.

Progress variable pdfs were evaluated from DNS data of a methane-air gas  turbine burner by Moureau et al. \cite{moureau2011large} and from data in a DNS database of turbulent n-heptane-air flames simulated with detailed chemistry by Lapointe and Blanquart \cite{lapointe2017priori}. Since the filtered source terms calculated with the flamelet pdf derived from 1D laminar flame profiles considerably underestimated the filtered DNS values, Moureau et al. \cite{moureau2011large} proposed filtering at a reduced filter width $\Delta'<\Delta$ where $\Delta'$ is calculated from the condition that the mean and variance of the 1D pdf agreed with the filtered DNS ones. $\Delta/\Delta'$ can be viewed as a flame wrinkling factor since both flame wrinkling and a filtering at a smaller $\Delta'<\Delta$ increase the pdf by a roughly constant factor. Similar scalings were observed by Pfitzner \cite{pfitzner2020pdf} when analysing an analytic model pdf of a 2D sinusoidally folded flame front. Lapointe et al. \cite{lapointe2017priori} studied the influence of differential diffusion and concluded that its effect can be neglected at larger filter sizes and for high Karlovitz number. 

Wrinkling factors were evaluated from the ratio of filtered DNS gradients and the resolved LES gradients by several groups. DNS databases using single-step Arrhenius chemistry were evaluated in Klein et al. \cite{klein2018flame},\cite{klein2019priori} while Proch et al.  \cite{proch2017flame2} used quasi-DNS data where the chemistry is evaluated from tabulated premixed flame tables but flame folding is fully resolved. A simple analytical 2D sinusoidal flame folding model analysed by Pfitzner \cite{pfitzner2020pdf} showed wrinkling factors which were nearly constant over a large range of $c$ values for large filter volumes. Similar results were obtained in Pfitzner and Klein \cite{pfitzner2021near}, where pdfs resulting from DNS data from statistically planar turbulent flames filtered at larger, RANS-like filter width were analysed. These authors also found that the effect of subgrid flame wrinkling on the subgrid pdf of $c$ could be mimicked by performing the 1D filtering operation at a smaller, effective filter width $\Delta'=\Delta/\Xi$ where $\Xi$ is a suitably chosen
wrinkling factor.

In the Filtered Laminar Flame (FLF) model \cite{fiorina2010filtered}, the chemical source term of a 1D laminar flame is filtered at a filter width $\Delta$ and then tabulated as function of the similarly filtered progress variable. Although the natural choice of the filter size would be the LES grid size $\Delta$, the authors choose filter sizes larger than the LES filter to avoid numerical oscillations of their LES during run time. In case of a flat laminar flame front moving parallel to one side of a cubical filter volume, the FLF model is exact. 3D effects like oblique propagation of flames through the filter volume, subgrid flame folding and modifications of the inner reaction zone through flame stretch and flame thickening require empirical modifications. 

The 1D laminar steady-state transport equation of progress variable $c$ cannot be solved analytically even in the case of one-step Arrhenius chemistry. Analytical laminar flame profiles and pdfs can be obtained by approximation of the Arrhenius source term $\omega(c)$ through appropriate surrogate functions. A simple source term, which is piecewise linear in $c$, was proposed by Echekki and Ferziger \cite{ferziger1993simplified}. A more accurate approximation to the Arrhenius source term with analytical solution was recently introduced by Pfitzner  \cite{pfitzner2020pdf}. The analytic expressions for the reaction progress variable, reaction source term and their filtered counterparts provide an analytic FLF model which closely approximates the solution using single-step Arrhenius chemistry. 

The FLF method provides reaction source term $\overline{\omega}_\Delta$ in relation to the filtered reaction progress variable $\overline{c}_\Delta$. Since in combustion large eddy simulations, a transport equation of the Favre-filtered reaction progress variable $\tilde{c}$ is solved, an additional model relation between $\overline{c}$ and $\tilde{c}$ is required to use the FLF method as LES combustion closure. 

Analyses of DNS's of statistically flat turbulent premixed flames with single-step Arrhenius chemistry by Pfitzner and Klein \cite{pfitzner2021near} showed that for large, RANS-like filter sizes, the subgrid pdf of progress variable is similar to a 1D flamelet pdf scaled with a constant factor over most of the $c$ range while being cut off earlier near the edges towards $c=0,1$. A similar pdf is created by filtering the 1D flame profile at a reduced $\Delta'=\Delta/\Xi$.

DNS analyses of Hansinger and Pfitzner \cite{hansinger2020statistical} indicated on the other hand that for medium filter sizes typical for LES, the subgrid pdf of progress variable of turbulent premixed flames can be quite complicated and there is a strong variation of subgrid flame surface density caused by oblique propagation of flame fronts through the (in their case cubical) filter volume while the effect of subgrid flame folding reduces with filter size.  \\ \\
The goals of the present paper are fourfold: \\ \\
a) To present a combustion model which correctly reproduces the effect of flame fronts moving obliquely through multidimensional filter volumes. \\ \\
b) To provide a methodology to represent laminar flame profiles of $c(x)$, $\rho(x) c(x)$ and $\omega(x)$ through functions which yield analytical results for the 1D filtered quantities $\overline{\omega}_\Delta$, $\overline{c}_\Delta$ and $\tilde{c}_\Delta$. \\ \\
c) To derive an LES combustion model for filtered reaction source term $\overline{\omega}_\Delta$ as function of $\tilde{c}_\Delta$ and $\Delta/\delta_{th}$, representing the effect of subgrid flame folding through a wrinkling factor $\Xi$. \\ \\
d) To derive accurate models for the subgrid wrinkling factor $\Xi$ in terms of variables available in LES. \\ \\
The paper is structured as follows:
After introducing the analytical flame profile and source term, we propose a fitting methodology for flame profile and source term based on series of analytically integrable functions. We present area distributions of planar surfaces slicing through 3D filter volumes and we show how to represent this effect in 1D filtering through use of a symmetrical filter kernel. After introduction of the DNS database of statistically planar turbulent premixed flames, we compare the filtered source terms from the 1D model with  conditionally averaged DNS data. We present two wrinkling factor models and validate the complete combustion model with DNS data before drawing some conclusions. In appendices we provide a step-by-step recipe to implement the new combustion model and some helpful mathematical formulae. 

\section{Laminar flame profiles}
The 1D transport equation for a normalized temperature progress variable $c$ reads
\begin{equation} 
	\rho\frac{\partial c}{\partial t}+\rho u \frac{\partial c}{\partial x}=\frac{\partial}{\partial x}\left(\frac{\lambda}{c_p} \frac{\partial c}{\partial x}\right)+{\omega(c)}              
	\label{eq:ctrans}
\end{equation}
In case that $c$ is a normalized mass fraction, the term $\frac{\lambda}{c_p}$ is replaced by $\rho D$ where $D$ is the corresponding diffusivity. In steady-state condition, the continuity equation requires $\rho u = const. = \rho_u s_L$. In that case one can introduce a rescaled canonical spatial coordinate  \cite{poi05}  $\xi = \int_0^x \rho_u s_L c_p / \lambda  dx$ yielding a simpler $c$ differential equation:
\begin{equation} 
	\frac{\partial c}{\partial \xi}=\frac{\partial^2 c}{\partial \xi^2}+\omega(c)              
	\label{eq:cstrans}
\end{equation}
For single-step Arrhenius chemistry and constant-pressure combustion, with $c$ being a normalized temperature progress variable, the source term can be written as \cite{poi05}
\begin{equation} 
	\omega(c)=\Lambda \left(1-\alpha(1-c)\right)^{\beta_1-1} (1-c)exp\left(-\frac{\beta(1-c)}{1-\alpha(1-c)}\right)              
	\label{eq:omstrans}
\end{equation}
$\alpha$ represents the normalized temperature raise $\alpha=(T_b-T_u)/T_b$ and $\beta=\alpha T_{a1}/T_b$ is a measure of the activation temperature $T_{a1}$. Normalized density is given by $\rho(c)/\rho_u = (1-\alpha)/\left( 1-\alpha(1-c)\right)$. The temperature exponent $\beta_1$ is usually taken as $\beta_1=0$ or $\beta_1=1$ and the eigenvalue $\Lambda$ has to be chosen such that $c = 0,1$ for $\xi\rightarrow \mp \infty$ is fulfilled.

In \cite{pfitzner2020pdf}, an approximation to the progress variable profile has been proposed, which admits analytical calculation of $\omega(c)$ from eq.(\ref{eq:cstrans}):
\begin{equation} 
	c_n(\xi)=[1+exp(-n\xi)]^{-1/n}              
	\label{eq:cnnew}
\end{equation}
with free parameter $n$ which can be adapted to mimic the effect of $\alpha,\beta$ in the Arrhenius profile. The profile $c_n(\xi)$ can be inverted analytically:
\begin{equation} 
	\xi_n(c)=\frac{1}{n} ln\left(\frac{c^n}{1-c^n}\right)              
	\label{eq:xnnew}
\end{equation}
and the thermal flame thickness is given by:
\begin{equation} 
	\delta_{f,n}=\frac{1}{\left( dc_n(\xi)/d\xi \right)_{max}}=\frac{(n+1)^{\frac{n+1}{n}}}{n}
	\label{eq:deltafm}
\end{equation}
The reaction source term is:
\begin{equation} 
	\omega_n(c)=(n+1)(1-c^n)c^{n+1}              
	\label{eq:omnew}
\end{equation}
The $c$ profile with adapted $n$ and the corresponding $\omega(c)$ are used here to evaluate the DNS data for consistency.

\section{Alternative numerical representation of $c$, $\rho c$ and $\omega$ profiles}
Unfortunately, the analytically defined $c_n(\xi)$ and $\omega_n(\xi)$ do not yield analytical results 
when integrated with some of the filter kernels discussed below. We therefore present alternative approximations of $c(\xi)$, $\rho(\xi) c(\xi)$ and $\omega(\xi)$ yielding more tractable results upon integration. Although the canonical coordinate $\xi$ is used mostly, the method works as well when using profiles defined in physical coordinates $x$.

\begin{figure} [ht]
	\begin{minipage}[b]{.42\linewidth} 
		\begin{tikzpicture}
			\node[] (Grafik) at (0,0) {\includegraphics[width=1\textwidth]{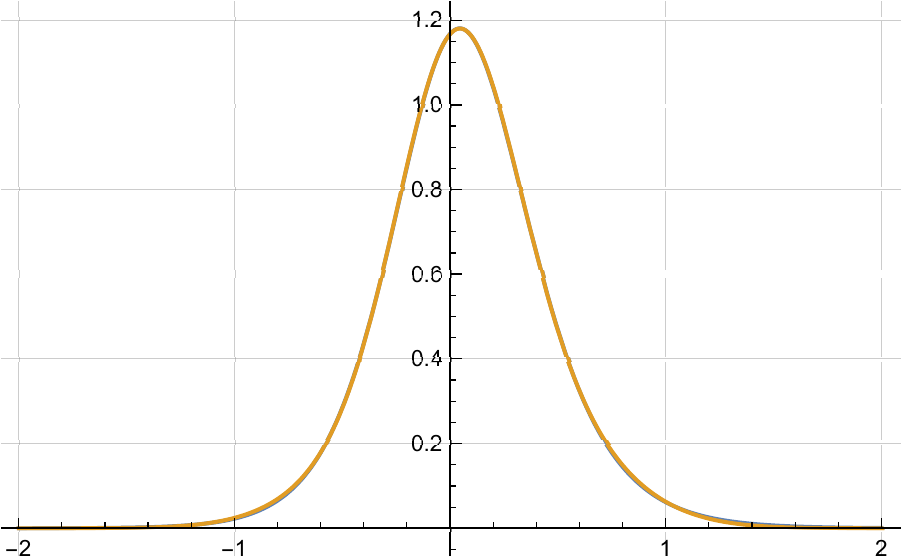}};
			\node[anchor=south,yshift=-10pt,xshift=10] at (Grafik.south) {$\xi$};
			\node[rotate=90,anchor=south,yshift=0,xshift=0pt] at (Grafik.west) {$\omega(\xi)$};
		\end{tikzpicture}
	\end{minipage}
	\hspace{.1\linewidth}
	\begin{minipage}[b]{.42\linewidth} 
		\begin{tikzpicture}
			\node[] (Grafik) at (0,0) { \includegraphics[width=\linewidth]{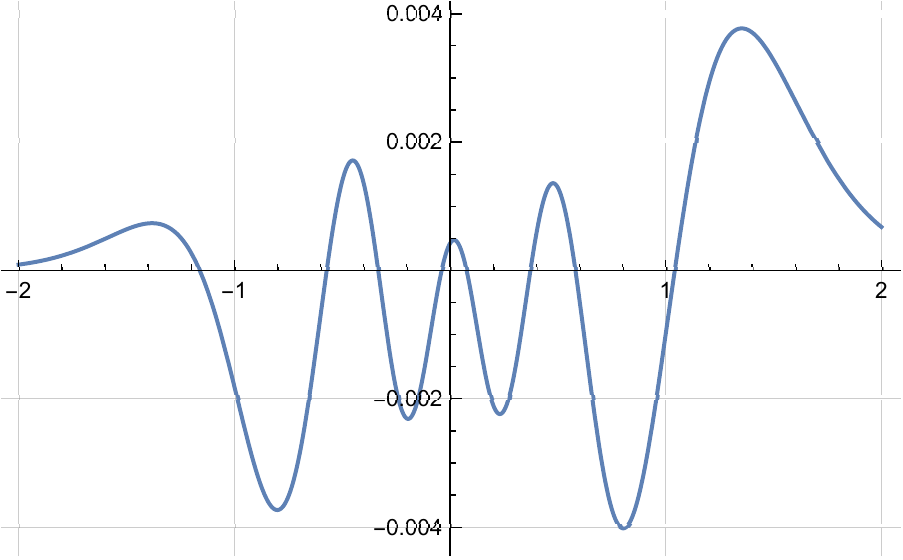}};
			\node[anchor=south,yshift=-10pt,xshift=10] at (Grafik.south) {$\xi$};
			\node[rotate=90,anchor=south,yshift=0,xshift=0pt] at (Grafik.west) {$\delta$};
		\end{tikzpicture} 
	\end{minipage}
	\caption{Left: $\omega_n(\xi)$ (blue) with approximation (orange); note that the blue line is completely covered by the orange line; right: difference between $\omega_n(\xi)$ and approximation with 3 Gaussians}
	\label{fig:omxiapprox}
\end{figure}
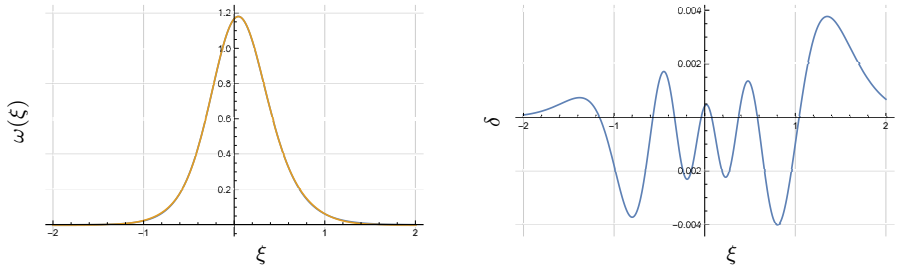

We start from the observation that the source term $\omega_n(\xi)=\omega_n(c_n(\xi))$ is very narrow in space and bell-shaped. The normalized Gaussian centred at $\xi=0$ is defined as:
\begin{equation}
	g_a(\xi)=\frac{\sqrt{a} e^{-a \xi^2}}{\sqrt{\pi }}
	\label{eq:omgauss}
\end{equation}
We find that an excellent fit to $\omega_n(\xi)$ can be achieved using only a few scaled and shifted Gaussians: 
\begin{equation}
	\omega_n(\xi) \approx \sum_i a_{i,1} g_a( a_{i,2} \xi-a_{i,3})
	\label{eq:approxomgauss}
\end{equation}
where the constants $a_{i,j}$ are fit coefficients. Fig.(\ref{fig:omxiapprox}) shows an overlay of $\omega_n(\xi)$ with $n=4.45$ and its approximation through 3 Gaussians together with their difference. 

$c(\xi)$ profiles fully consistent with this approximation to $\omega(\xi)$ can be calculated analytically. The solution of the differential equation (\ref{eq:cstrans}) with a single Gaussian source term $\omega(\xi)=g_a(\xi)$ is provided in appendix 2. $c(\xi)$ can be calculated as superposition of the solutions for the Gaussian sources representing $\omega(\xi)$, using the same scaling and shifting constants from eq.(\ref{fig:omxiapprox}). In addition, 
the resulting $c(\xi)$ can also be integrated analytically to yield  $\overline{c}$.

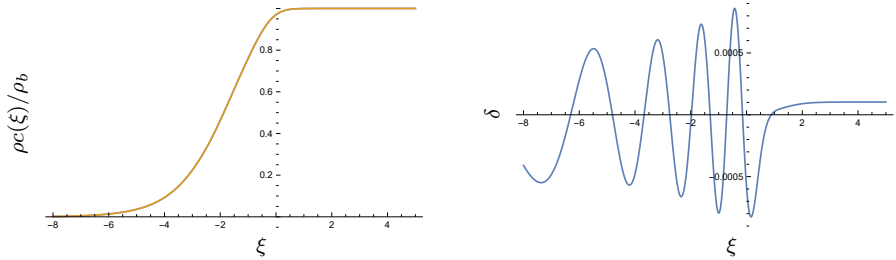
\begin{figure} [ht]
	\begin{minipage}[b]{.42\linewidth} 
		\begin{tikzpicture}
			\node[] (Grafik) at (0,0) {\includegraphics[width=1\textwidth]{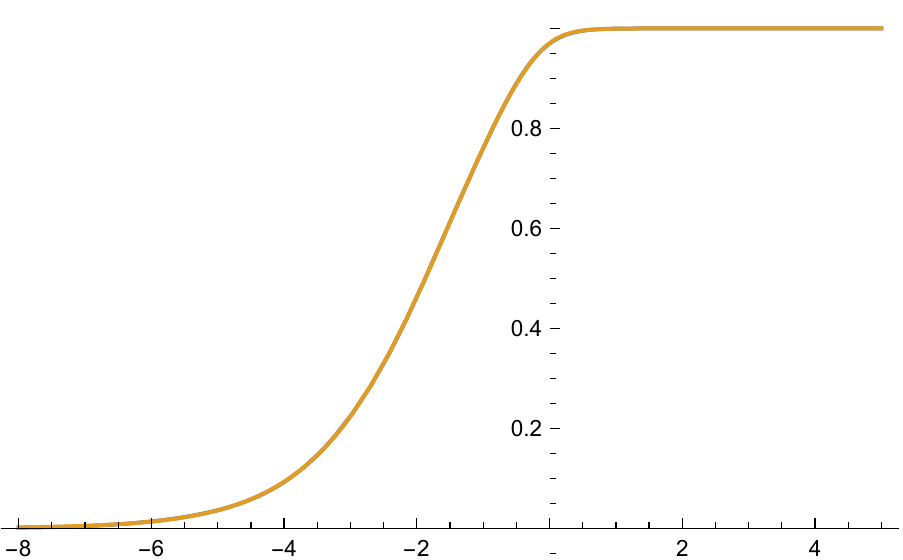}};
			\node[anchor=south,yshift=-10pt,xshift=10] at (Grafik.south) {$\xi$};
			\node[rotate=90,anchor=south,yshift=0,xshift=0pt] at (Grafik.west) {$\rho c(\xi)/\rho_b$};
		\end{tikzpicture}
	\end{minipage}
	\hspace{.1\linewidth}
	\begin{minipage}[b]{.42\linewidth} 
		\begin{tikzpicture}
			\node[] (Grafik) at (0,0) { \includegraphics[width=\linewidth]{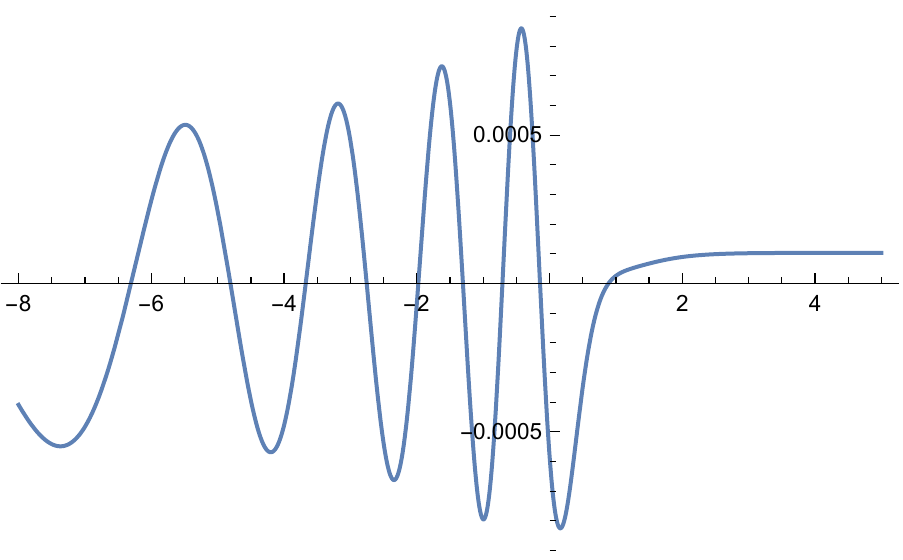}};
			\node[anchor=south,yshift=-10pt,xshift=10] at (Grafik.south) {$\xi$};
			\node[rotate=90,anchor=south,yshift=0,xshift=0pt] at (Grafik.west) {$\delta$};
		\end{tikzpicture} 
	\end{minipage}
	\caption{Left: $\rho(\xi) c_m(\xi)/\rho_b$ (blue) with approximation (orange); note that the blue line is completely covered by the orange line;  right: difference between  $\rho(\xi) c_m(\xi)/\rho_b$ and approximation with 4 error functions}
	\label{fig:rhocxiapprox}
\end{figure}

$\rho(\xi) c(\xi)$ can also be approximated through a series of suitable functions. We found excellent approximations using only a few shifted/scaled tanh, sigmoid or error functions. Such approximations are equivalent to a single-layer artificial neural network. Alternatively, $\rho(\xi) c(\xi)/\rho_b$ could be approximated (less accurately) by a single scaled and shifted $c_n(\xi)$, also adapting the $n$ parameter. In this case, the integral of $\rho c$ can be calculated by the method presented in \cite{pfitzner2020pdf}. Fig.(\ref{fig:rhocxiapprox}) shows the approximation of $\rho(\xi) c_n(\xi)$ through four error functions with the difference. Similar results are achieved when using tanh or sigmoid Ansatz functions.

\section{LES filtering}
In LES of flows with variable density, usually Favre filtered transported variables  $\tilde{z}$ are used ($\tilde{z}=\overline{\rho z}/\overline{\rho}$). The $\tilde{c}$  transport equation reads (using the gradient diffusion hypothesis)
\begin{equation} 
	\frac{\partial	\overline{\rho} \tilde{c}}{\partial t}+\frac{\partial\overline{\rho} \tilde{u_i} \tilde{c}}{\partial x_i}=\frac{\partial}{\partial x_i}\left(\frac{\mu+\mu_t}{Pr_t} \frac{\partial c}{\partial x_i}\right)+\overline{\omega}              
	\label{eq:ctransFavre}
\end{equation}
with turbulent viscosity $\mu_t$ and turbulent Prandtl number $Pr_t$ (to be replaced by a turbulent Schmidt number $Sc_t$ in case of a species based $c$ variable). The filtered source term $\overline{\omega}$ needs to be modelled in terms of variables which are known during the LES.

\begin{figure} [ht]
 	\centering
 	\includegraphics[width=0.6\textwidth]{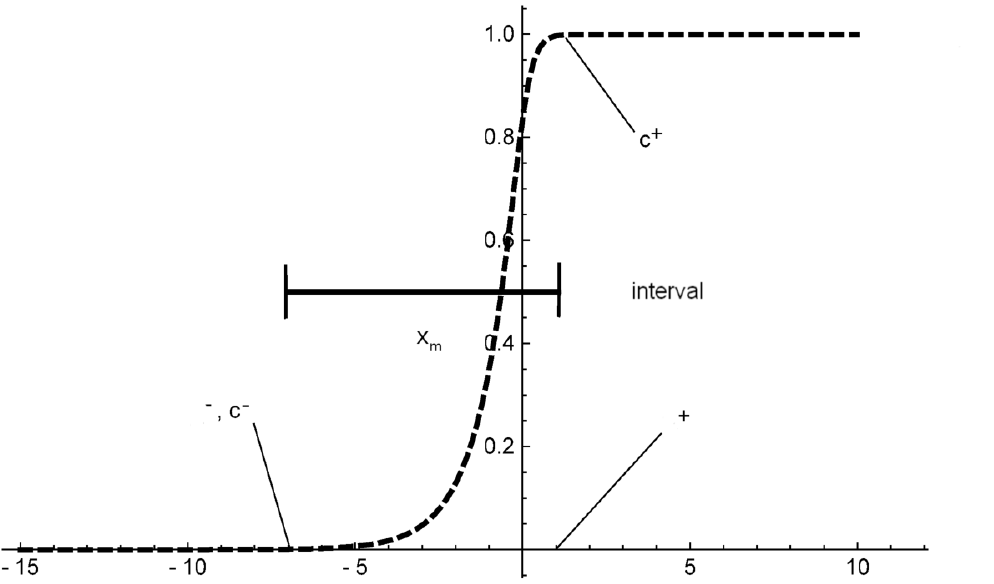}
 	\caption{LES cell width for integration of premixed reaction source term}
 	\label{fig:LESint}
 \end{figure}

The combustion model proposed here approximates $\overline{\omega}$ through quantities filtered from a 1D laminar flame profile $c(x)$. Fig.(\ref{fig:LESint}) shows such a profile together with a filter interval of width $\Delta \sim 2\cdot\delta_{th}$ positioned symmetrically around the filter midpoint location $x_m$.
We can denote the position of the left filter boundary as $x^-$, the position of the right filter boundary as $x^+$, yielding $x_m=(x^++x^-)/2$ and $\Delta=x^+-x^-$. 

1D filtered values of quantities $z$ with filter width $\Delta$ positioned at $x_m$ are denoted as  $z_\Delta(x_m)$ and are calculated as
\begin{equation} 
	\overline{z}_\Delta(x_m)=\frac{1}{\Delta} \int_{x^-}^{x^+}z(x) dx=\frac{1}{\Delta} \int_{x_m-\frac{\Delta}{2}}^{x_m+\frac{\Delta}{2}}z(x) dx
	\label{eq:meanwc}
\end{equation}
The corresponding 3D (box) filter is obtained by convolution of three 1D filters.

\section{1D laminar flame pdf}
The filtered value of any quantity depending on progress variable $c$ can also be calculated using the corresponding pdf $p(c)$. In case of a 1D laminar flame profile, the pdf $p(c)$ is given by \cite{bray2006finite}, \cite{pfitzner2020pdf}:
\begin{equation} 
	p(c) = \frac{1}{\Delta}\frac{1}{dc/dx}H(c-c^-)H(c^+-c)
	\label{eq:pdfdef}
\end{equation}
where $H(...)$ is the Heaviside step function and $c^\pm=c(x^\pm)$ are the $c$ values of the profile at the edges of the filter interval. $dc/dx$ needs to be expressed in terms of $c$ using the inversion $x(c)$ of $c(x)$. The denominator $\Delta$ guarantees the correct normalisation of $p(c)$:
\begin{equation} 
	\int_0^1p(c)dc \stackrel{!}{=}1=\frac{1}{\Delta}\int_{c^-}^{c^+}\frac{1}{dc/dx}dc=\frac{1}{\Delta}\int_{x^-}^{x^+} dx=\frac{x^+-x^-}{\Delta}
	\label{eq:pdfnor}
\end{equation}
which is true for all flame profiles $c(x)$. 
The mean of any variable $z(c)$ can be calculated as:
\begin{equation} 
	\overline{z}=\int_0^1z(c)p(c)dc=\frac{1}{\Delta} \int_{c^-}^{c^+}\frac{z(c)}{dc/dx} dc=\frac{1}{\Delta} \int_{x^-}^{x^+}z(x) dx
	\label{eq:meanwcp}
\end{equation}
the last term showing the consistency with eq.(\ref{eq:meanwc}).  

\section{Planar surfaces moving through 3D filter volumes}
In 3D, LES filter volumes are often cubical or hexahedral. We first discuss here the case of a cubical filter of unit volume and generalize later to filter volumes of arbitrary size. 

\begin{figure} [ht]
	\centering
	\includegraphics[width=0.6\textwidth]{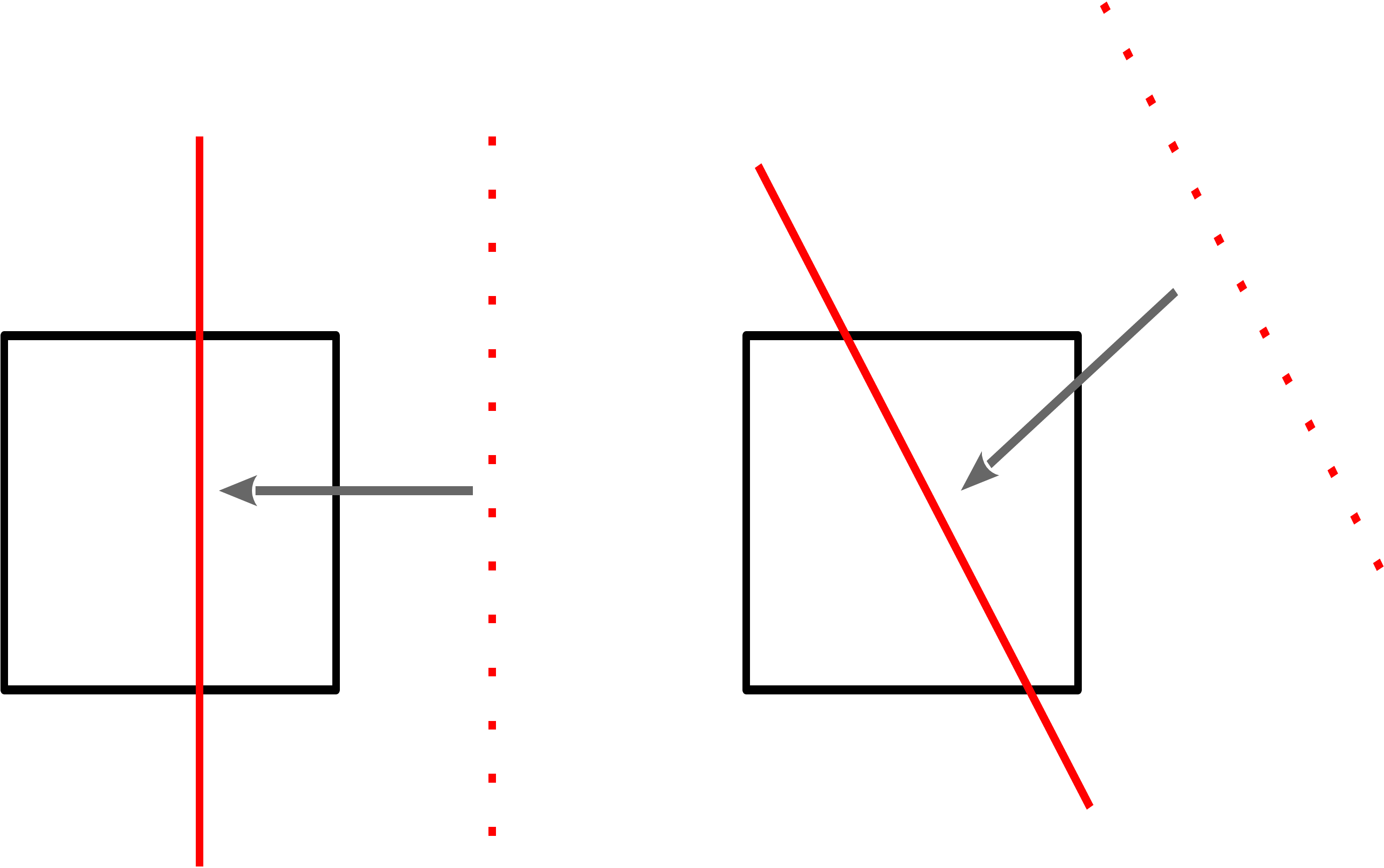}
	\caption{Linear flame moving through unit square filter}
	\label{fig:PlaneSquare}
\end{figure}

\begin{figure} [ht]
	\begin{minipage}[b]{.42\linewidth} 
		\begin{tikzpicture}
			\node[] (Grafik) at (0,0) {\includegraphics[width=1\textwidth]{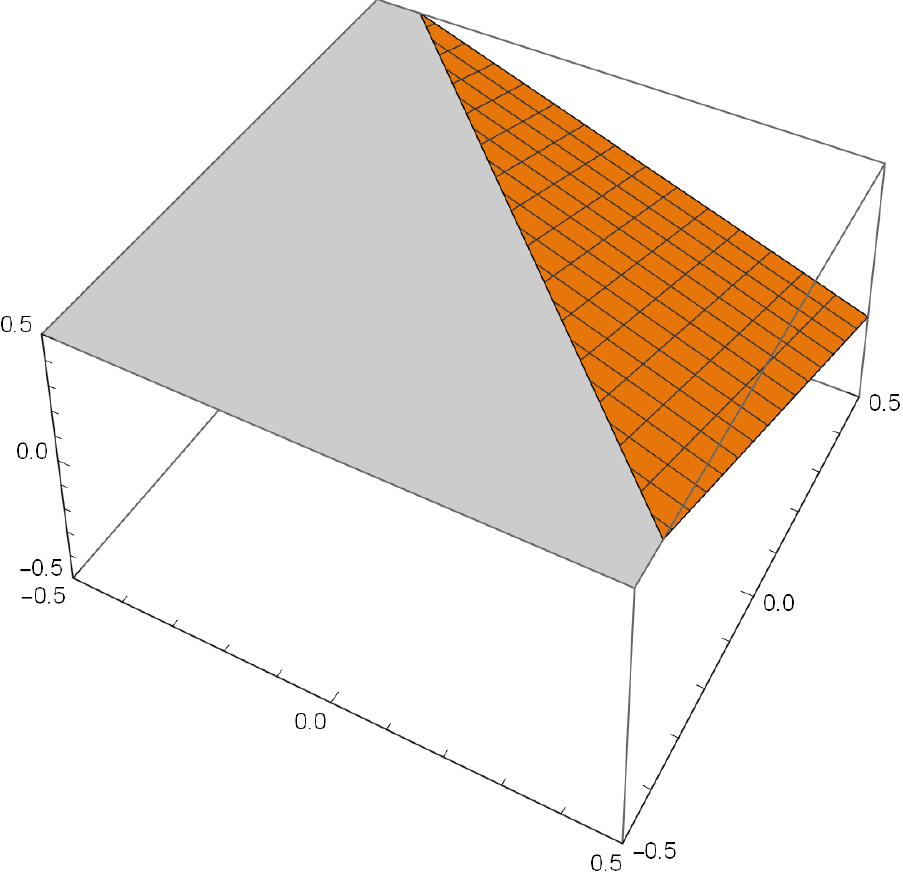}};
			\node[anchor=south,yshift=-10pt,xshift=10] at (Grafik.south) {$ $};
			\node[rotate=90,anchor=south,yshift=0,xshift=0pt] at (Grafik.west) {$ $};
		\end{tikzpicture}
	\end{minipage}
	\hspace{.1\linewidth}
	\begin{minipage}[b]{.42\linewidth} 
		\begin{tikzpicture}
			\node[] (Grafik) at (0,0) { \includegraphics[width=\linewidth]{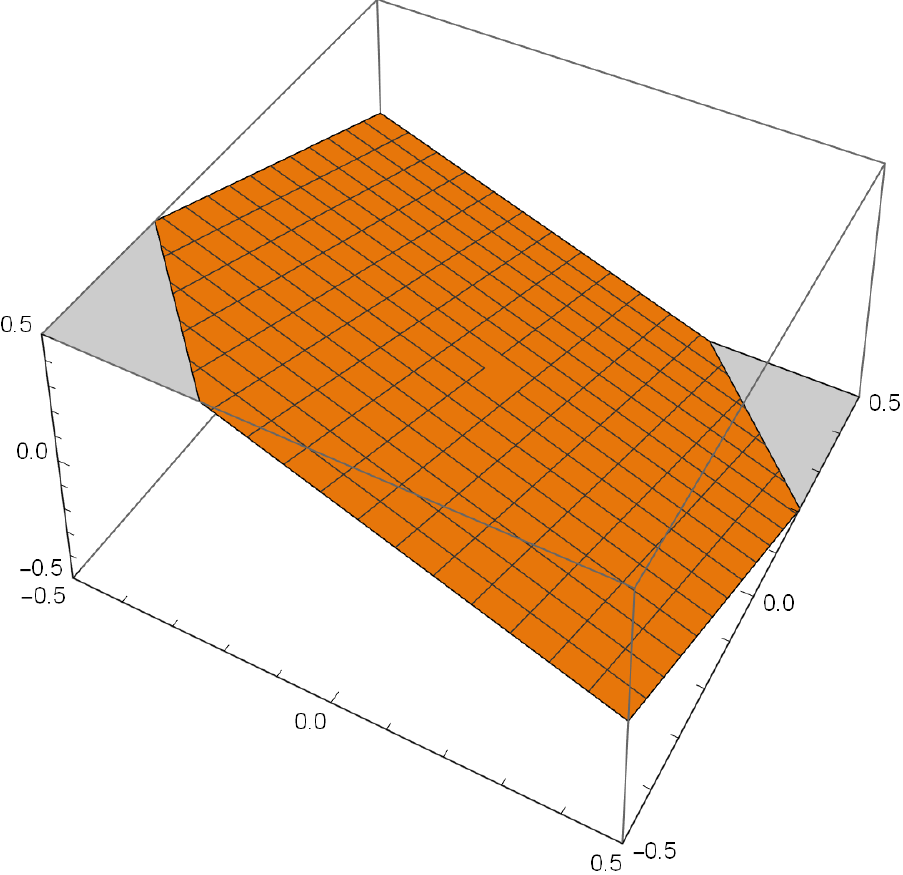}};
			\node[anchor=south,yshift=-10pt,xshift=10] at (Grafik.south) {$ $};
			\node[rotate=90,anchor=south,yshift=0,xshift=0pt] at (Grafik.west) {$ $};
		\end{tikzpicture} 
	\end{minipage}
	\caption{Slices of a plane contained within a unit cube}
	\label{fig:cubecut}
\end{figure}
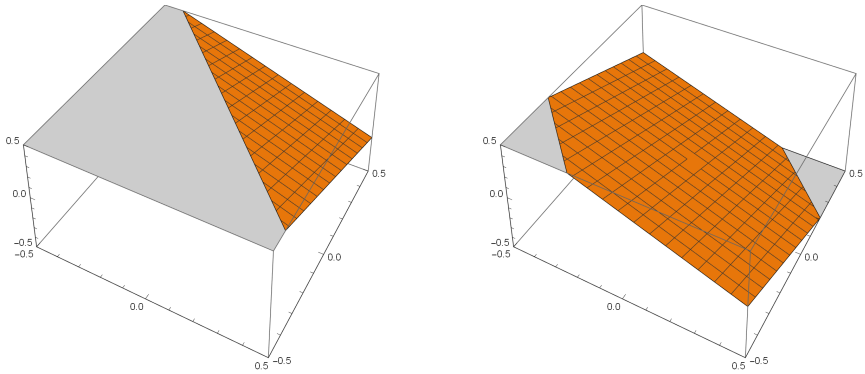

The 3D equivalent of a flame front moving through a 1D filter interval is a planar flame front moving through a cubical filter volume in a direction exactly normal to one of its faces. In 3D, however, flame fronts can move at arbitrary angles through the filter volume. Fig's (\ref{fig:PlaneSquare},\ref{fig:cubecut}) show 2D and 3D examples of flat surfaces slicing through a cubical filter volume at oblique angles. As the plane moves through the cube, the area $A$ of the plane contained within the filter volume will gradually increase from zero to a maximum and then decrease to zero again. 

In the quasi-1D case, the area of a planar surface contained within the unit cube will jump from zero to one when entering the filter volume and back to zero when leaving it. In the 2D and 3D cases with planar surfaces slicing through at oblique angles, the calculation of the area of the surface contained within the unit cube can be calculated by elementary geometry. 

\begin{figure} [ht]
	\begin{minipage}[b]{.42\linewidth} 
		\begin{tikzpicture}
			\node[] (Grafik) at (0,0) {\includegraphics[width=1\textwidth]{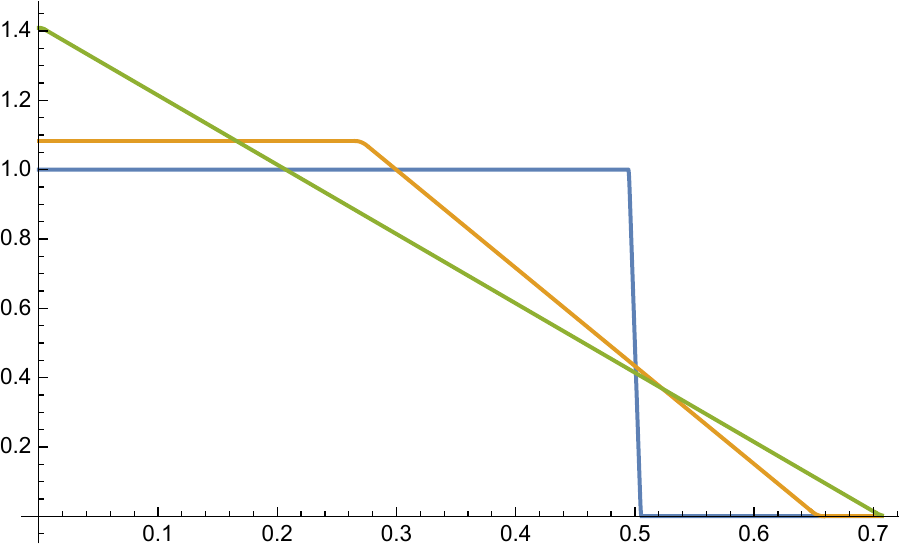}};
			\node[anchor=south,yshift=-10pt,xshift=10] at (Grafik.south) {$d$};
			\node[rotate=90,anchor=south,yshift=0,xshift=0pt] at (Grafik.west) {$A$};
		\end{tikzpicture}
	\end{minipage}
	\hspace{.1\linewidth}
	\begin{minipage}[b]{.42\linewidth} 
		\begin{tikzpicture}
			\node[] (Grafik) at (0,0) { \includegraphics[width=\linewidth]{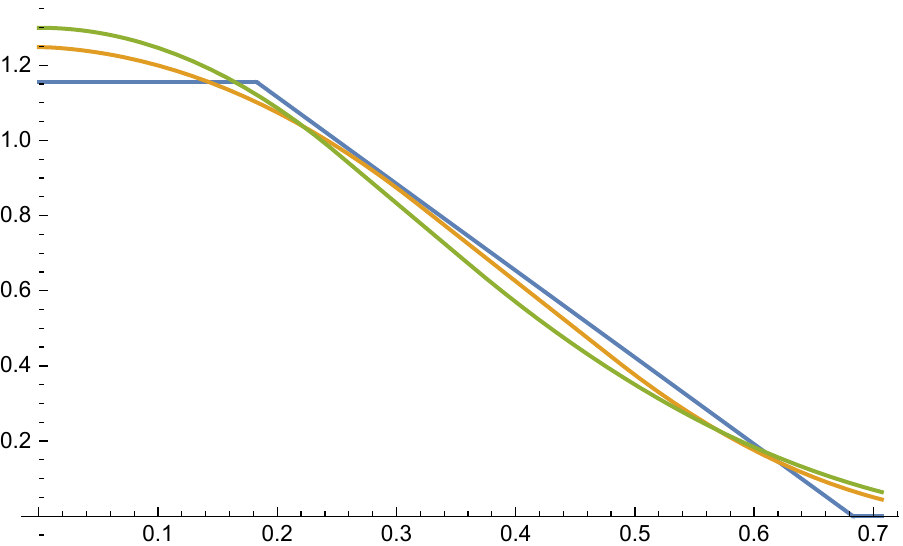}};
			\node[anchor=south,yshift=-10pt,xshift=10] at (Grafik.south) {$d$};
			\node[rotate=90,anchor=south,yshift=0,xshift=0pt] at (Grafik.west) {$A$};
		\end{tikzpicture} 
	\end{minipage}
	\caption{Area distributions of plane cutting through unit cube; $\phi=0$ (blue), $\phi=\pi/8$ (orange), $\phi=\pi/4$ (green) left: 2D case: $\theta=\pi/2$, ; right: 3D case: $\theta=\pi/3$}
	\label{fig:Aoblique}
\end{figure}
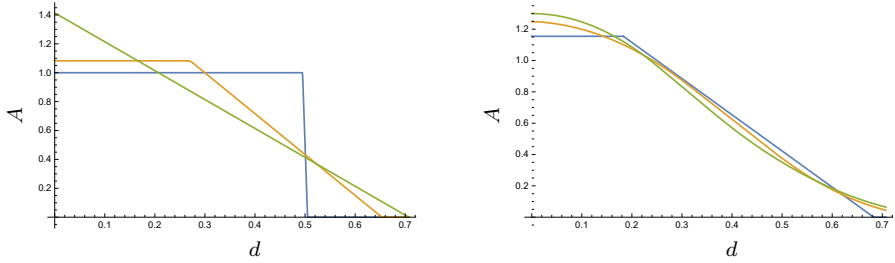

Characterizing this area  $A$ by two solid angles $\theta,\phi$ and the normal distance $d$ of the plane from the centre of the (unit) cube, we obtain distributions as shown in fig. (\ref{fig:Aoblique}). Fig. (\ref{fig:Aoblique} (left) shows examples of $A(d,\phi,\theta)$ for 1D and 2D cases, where the planar surface is still parallel to one of the coordinate axes. Fully 3D cases are shown in fig. (\ref{fig:Aoblique} (right). $A(d,\phi,\theta)$ is a piecewise constant, linear and parabolic function of $d$ at given $\phi,\theta$. Note that $A(d,\phi,\theta)$ features the same symmetry relationships as the filter volume. For a cubical volume, only angles $0<\theta<\pi/2$ and $0<\phi<\pi/4$ need to be considered.

It is evident that the maximum of the slicing area $A(d=0,\phi,\theta)$ is larger and it extends to larger $d$ values in the 2D and 3D cases than for the 1D case. However, genuinely 3D cases appear to generate quite similar $d$ dependences of $A$.

\begin{figure}
	\centering
	\begin{minipage}[b]{.40\linewidth} 
		\begin{tikzpicture}
			\node[] (Grafik) at (0,0) { \includegraphics[width=\linewidth]{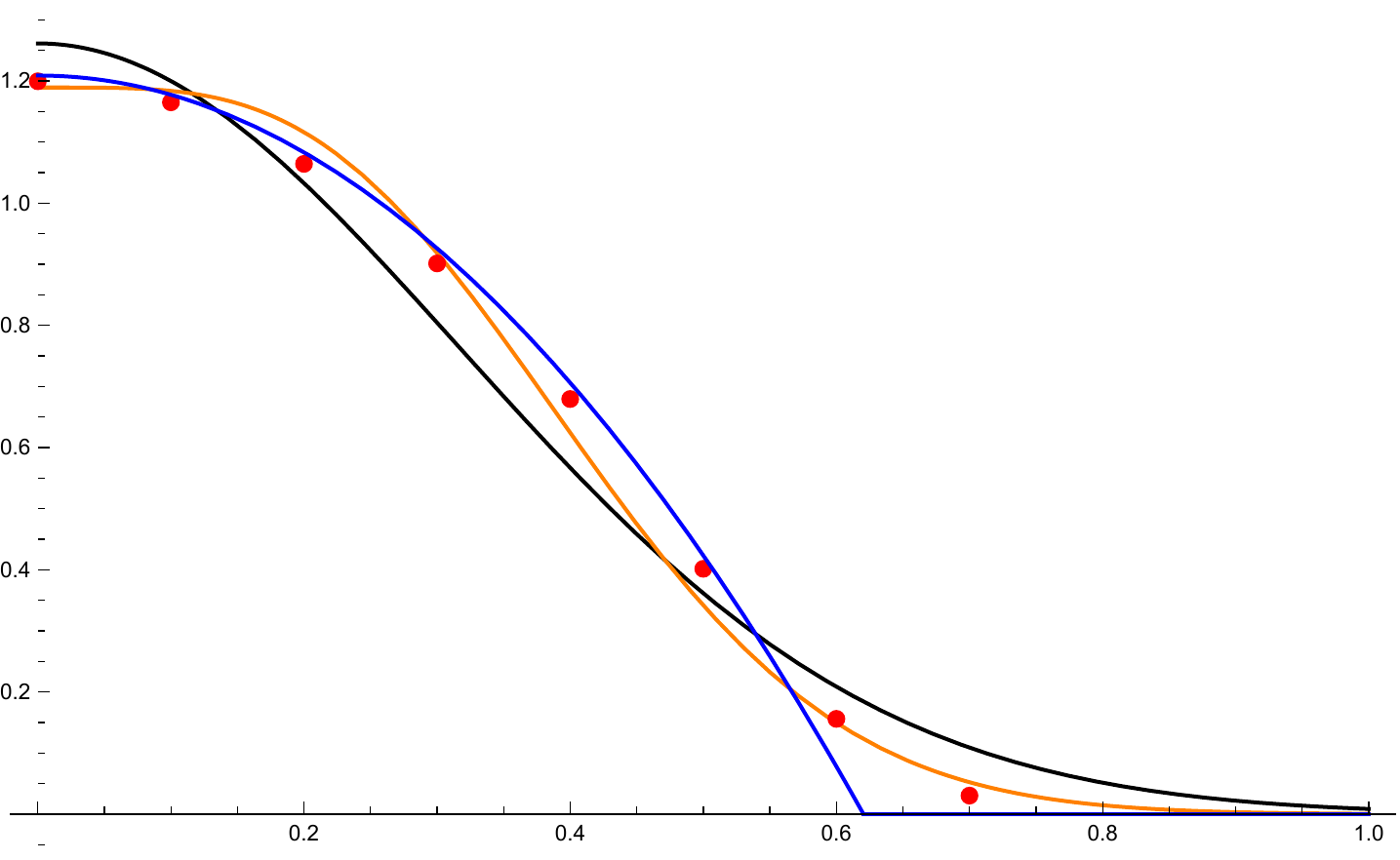}};
			\node[anchor=north,yshift=-0pt,xshift=10] at (Grafik.south) {$d$};
			\node[rotate=90,anchor=south,yshift=-0,xshift=0pt] at (Grafik.west) {$A$};
		\end{tikzpicture} 
	\end{minipage}
	\caption{Angle averaged $A_{av}(d)$ (dots) compared to $A_s(d)$, the area of a plane slicing through a unit volume sphere (blue) and Gaussian $A_g(d)$ (black) / modified Gaussian $A_{gm}(d)$ (orange) approximations}
	\label{fig:Aintsphere}
\end{figure}

In 3D LES, the angle of a flame front within the filter volume cannot be properly defined when there is subgrid flame folding. Assuming a random orientation of flame propagation angles one can average $A(d,\phi,\theta)$ over all angles, giving equal weight to each possible direction. For  spherical filter volumes, $A(d)$ is actually independent of propagation direction and a quadratic function of $d$. For a sphere of unit volume,the area is given by
\begin{equation}
	A_s(d)=\pi\left(\left(\frac{3}{4\pi}\right)^{\frac{2}{3}} - d^2\right)
	\label{eq:rsarea}
\end{equation}
Fig.(\ref{fig:Aintsphere}) shows $A_s(d)$ together with the angle-averaged $A_{av}(d)$ (dots) and with two approximations to $A(d)$ based on Gaussian functions:
\begin{equation}
	A_g(d)=\sqrt{\frac{5}{\pi }} e^{-5 d^2}
	\label{eq:Agauss2}
\end{equation}
\begin{equation}
	A_{gm}(d)=\frac{2}{3} \sqrt{\frac{10}{\pi }} e^{-10 d^2} \left(10 d^2+1\right)
	\label{eq:Agaussm}
\end{equation}
It can be seen that $A_{gm}(d)$ is a slightly better fit to $A_{av}(d)$ than $A_g(d)$. In the analysis of DNS data we found that both $A_g(d)$ and $A_{gm}(d)$ yielded very similar results in terms of the distributions of filtered reaction source term vs. Favre-filtered reaction progress. For simplicity, $A_g(d)$ has been used in the formulation of the combustion model.

\section{Filtering variables on oblique flame fronts}
We now show how to obtain filtered quantities for variables $z$ defined on flat laminar flame fronts crossing obliquely through cubical filter volumes. We first derive the relationships for the case of a unit cube filter and then generalize to larger filter volumes. In this section we use the canonical coordinate $\xi$ as spatial variable but results are valid also if the physical coordinate $x$ is used.

For a flat laminar flame front, we have the one-to-one relationship $c(\xi)$ between $c$ and the 1D coordinate $\xi$ normal to the flame front. Any variable $z(c)$ tabulated as function of progress variable $c$ can be calculated along $\xi$ as $z(\xi)=z(c(\xi))$. 

Let $A(d,\theta,\phi)$ be the area of a flat surface within the unit cube, positioned at a normal distance $d$ from its centre and $c_m$ the $c$ value of the flame in the centre of the unit cube (i.e. at $d=0$). Inverting the relationship between $c$ and spatial coordinate $\xi$ in the 1D profile yields the location $\xi_m=\xi(c_m)$ of the centre of the cube relative to the 1D $c(\xi)$ profile, see fig.(\ref{fig:FilterKernel}).

We can calculate the filtered value of a variable $z$ by multiplying its value $z(c^\ast)$ with the area of the (flat) $c^\ast$ isosurface contained in the filter volume and performing a 1D integration along a line normal to the flat flame and passing through the centre of the cube: 
\begin{equation}
	\overline{z}(\xi_m)=\int_{-\infty}^{\infty}r(\xi)z(\xi-\xi_m)d\xi
	\label{eq:zmean3D}
\end{equation}
where $r(\xi)$ is a filter kernel  which represents the area distribution along the integration path. 

The choice of $z(c) \equiv 1$ will yield the cell volume upon filtering, which is equal to one for the unit cube. This implies that $r(\xi)$ is a normalized filter kernel:
\begin{equation}
	\int_{-\infty}^{\infty}r(\xi)d\xi=1
	\label{eq:rnorm}
\end{equation}
Since $r(\xi)$ represents an area of a plane slicing through the filter volume, which always must be positive, we have $r(\xi) \geq 0$. Due to the symmetry of the filter volume, $r(\xi)=r(-\xi)$ and the integrability condition $r(\xi) \rightarrow 0$ fast enough for $\xi \rightarrow \pm \infty$, eq.(\ref{eq:zmean3D}) can also be written as
\begin{equation}
	\overline{z}(\xi_m)=\int_{-\infty}^{\infty}r(\xi-\xi_m)z(\xi)d\xi
	\label{eq:zmean3Da}
\end{equation}
Different filter kernels $r(\xi)=A(\xi,\phi,\theta)$ represent cases of planes moving at different oblique angles through the unit cube. In this work we choose $r_g(\xi)=A_g(\xi)$, i.e. a simple Gaussian approximation to the angle-averaged area distribution. As shown above, this filter 
is also a good approximation to the situation where planar surfaces move through a spherical filter unit volume and therefore also to e.g. polyhedral filter volumes.
\subsection{Favre filtering with temperature derived progress variable}
Favre filtered quantities are evaluated as
\begin{equation}
	\tilde{z}(\xi_m)=\frac{1}{\overline{\rho}}\int_{-\infty}^{\infty}r(\xi-\xi_m)\rho(\xi)z(\xi)d\xi
	\label{eq:ztilde3D}
\end{equation}
Note that for low Mach number, adiabatic and constant pressure premixed combustion and when using a normalized temperature progress variable, the density is a simple function of $c$:
\begin{equation}
	\rho(c)=\frac{\rho_u}{1 + c\tau}
	\label{eq:rhocrho1}
\end{equation}
where $\tau=\rho_u/\rho_b-1$ with $\rho_{u,b}$ being the unburnt and burnt gas density, respectively and therefore
\begin{equation}
	\rho(c)=\rho_u-\rho(c) c \tau
	\label{eq:rhocrho2}
\end{equation}
In this case, it is not necessary to calculate $\overline{\rho}$ for evaluation of $\tilde{c}=\overline{\rho c}/\overline{\rho}$, because
\begin{equation}
	\tilde{c}(\xi_m)=\frac{\overline{(\rho c)}(\xi_m)}{\overline{\rho}(\xi_m)}=\frac{\overline{(\rho c)}(\xi_m)}{\rho_u-\tau \overline{(\rho c)}(\xi_m)} 
	\label{eq:ctilde1}
\end{equation} \\

\subsection{Generalization to larger filter sizes}
\begin{figure}
	\centering
	\begin{minipage}[b]{.50\linewidth} 
		\begin{tikzpicture}
			\node[] (Grafik) at (0,0) { \includegraphics[width=\linewidth]{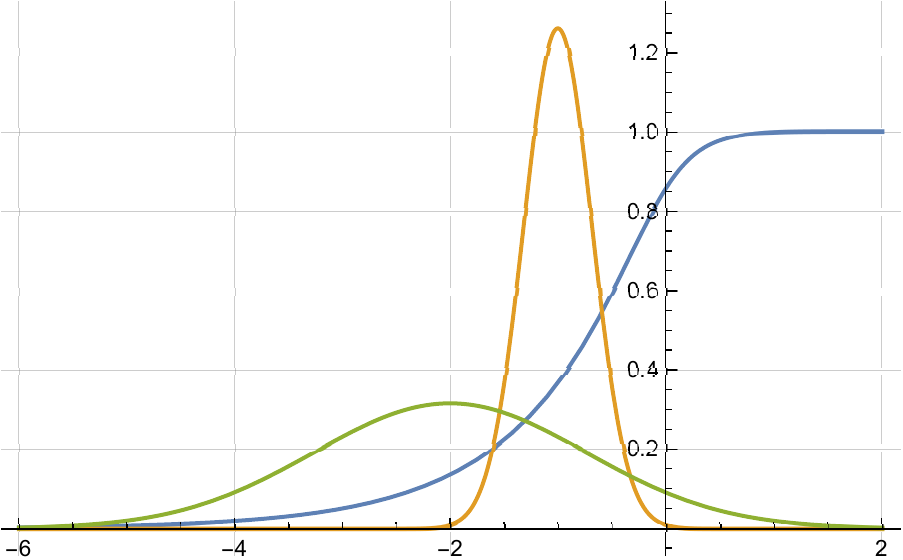}};
			\node[anchor=north,yshift=-0pt,xshift=10] at (Grafik.south) {$\xi$};
			\node[rotate=90,anchor=south,yshift=-0,xshift=0pt] at (Grafik.west) {$c(\xi),r(\xi)$};
		\end{tikzpicture} 
	\end{minipage}
	\caption{$c(\xi)$ (blue) with Gaussian filter at $\Delta_\xi=1,\xi_m=-1$ (orange) and $\Delta_\xi=4,\xi_m=-2$, where $\Delta_\xi$ denotes the filter size in $\xi$ space}
	\label{fig:FilterKernel}
\end{figure}

In the generalization of the filter kernel concept to larger cubical filter volumes of side length $\Delta$, we note that a geometrically similar situation exists to the unit cube case. However, the filter volume $\Omega$ will be multiplied by $\Delta^3$ and the area of isosurfaces within the cell by $\Delta^2$. Distances from the cube center will scale with $\Delta$. When calculating mean values in the larger cube, the filter kernel $r(\xi)$ is therefore replaced by $r_\Delta(\xi)=\frac{1}{\Delta}r(\xi/\Delta)$, yielding
\begin{equation}
	\overline{z}_\Delta(\xi_m)=\int_{-\infty}^{\infty}r_\Delta(\xi-\xi_m)z(\xi)d\xi=\frac{1}{\Delta}\int_{-\infty}^{\infty}r\left(\frac{\xi-\xi_m}{\Delta}\right)z(\xi)d\xi
	\label{eq:zmean3DDelta}
\end{equation}
Examples of filter kernels located at different $\xi$ positions and for different filter sizes $\Delta$ are shown in fig.(\ref{fig:FilterKernel}). $\Delta_\xi$ here is the filter width evaluated in canonical coordinates, where in the present case the thermal flame thickness is $\delta_{th} \approx 1.79$.

\section{Effect of filter kernel form on $\overline{c}_\Delta$, $\tilde{c}_\Delta$ and $\overline{\omega}_\Delta$}
It is interesting to investigate how much the oblique flame filter kernel changes $\overline{c}_\Delta$, $\tilde{c}_\Delta$ and $\overline{\omega}_\Delta$ from the quantities calculated with the 1D step filter kernel.
Before we present results, we note that for small filter size $\Delta \ll \delta_{th}$, all filters approach regularisations of the delta function, so in this limit $\overline{z}_\Delta(\xi_m)=\tilde{z}_\Delta(\xi_m)=z(\xi_m)$.

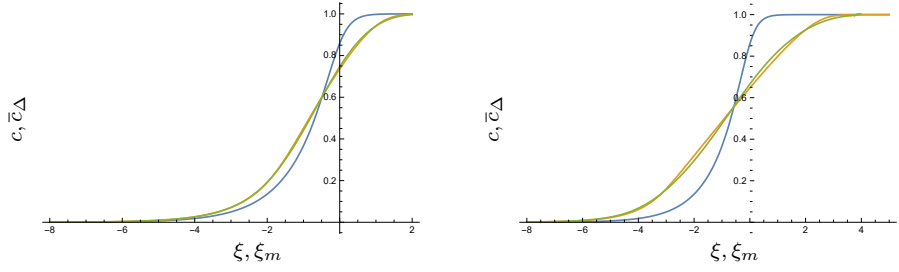
\begin{figure} [ht]
	\begin{minipage}[b]{.42\linewidth} 
		\begin{tikzpicture}
			\node[] (Grafik) at (0,0) {\includegraphics[width=1\textwidth]{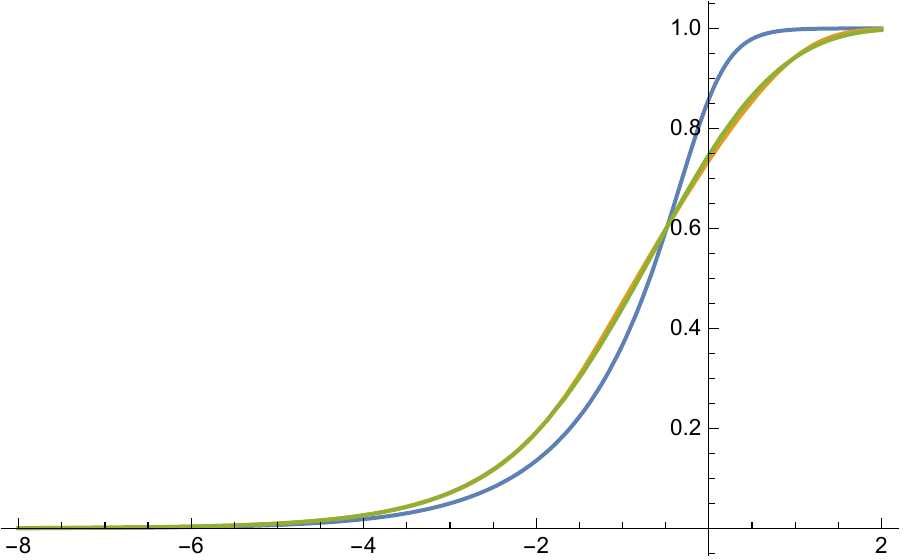}};
			\node[anchor=south,yshift=-10pt,xshift=10] at (Grafik.south) {$\xi,\xi_m$};
			\node[rotate=90,anchor=south,yshift=0,xshift=0pt] at (Grafik.west) {$c,\overline{c}_\Delta$};
		\end{tikzpicture}
	\end{minipage}
	\hspace{.1\linewidth}
	\begin{minipage}[b]{.42\linewidth} 
		\begin{tikzpicture}
			\node[] (Grafik) at (0,0) { \includegraphics[width=\linewidth]{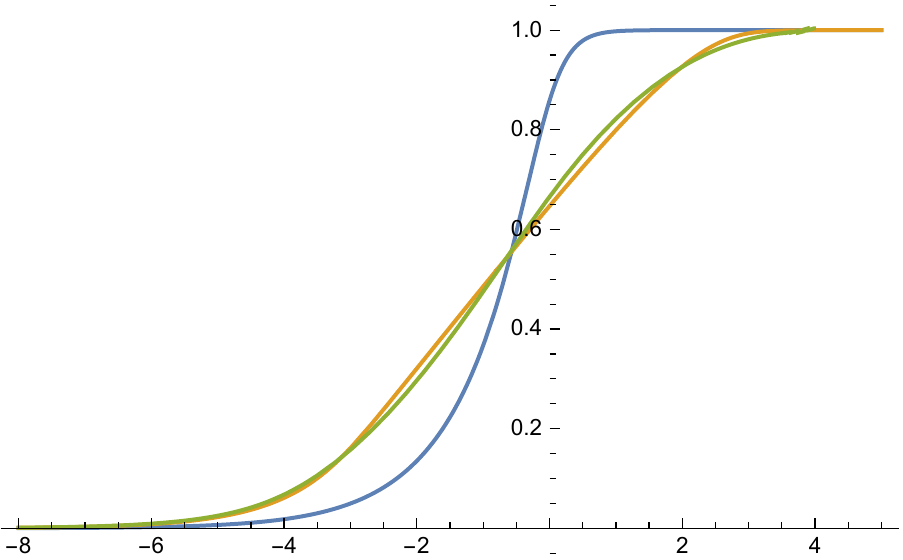}};
			\node[anchor=south,yshift=-10pt,xshift=10] at (Grafik.south) {$\xi,\xi_m$};
			\node[rotate=90,anchor=south,yshift=0,xshift=0pt] at (Grafik.west) {$c,\overline{c}_\Delta$};
		\end{tikzpicture} 
	\end{minipage}
	\caption{$c(\xi),\overline{c}_\Delta(\xi_m)$; blue: $c(\xi)$, orange: $\overline{c}_\Delta(\xi_m)$ evaluated with 1D filter, green: $\overline{c}_\Delta(\xi_m)$ evaluated with spherical filter left: $\Delta=3$, right: $\Delta=6$}
	\label{fig:cmean}
\end{figure}

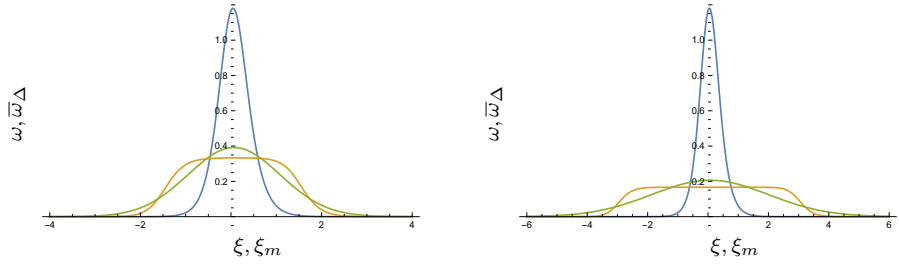
\begin{figure} [ht]
	\begin{minipage}[b]{.42\linewidth} 
		\begin{tikzpicture}
			\node[] (Grafik) at (0,0) {\includegraphics[width=1\textwidth]{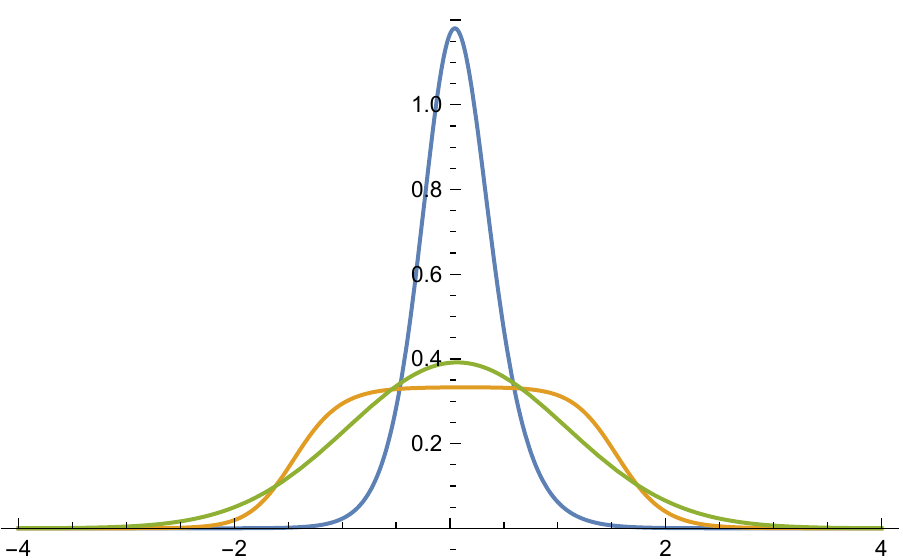}};
			\node[anchor=south,yshift=-10pt,xshift=10] at (Grafik.south) {$\xi,\xi_m$};
			\node[rotate=90,anchor=south,yshift=0,xshift=0pt] at (Grafik.west) {$\omega,\overline{\omega}_\Delta$};
		\end{tikzpicture}
	\end{minipage}
	\hspace{.1\linewidth}
	\begin{minipage}[b]{.42\linewidth} 
		\begin{tikzpicture}
			\node[] (Grafik) at (0,0) { \includegraphics[width=\linewidth]{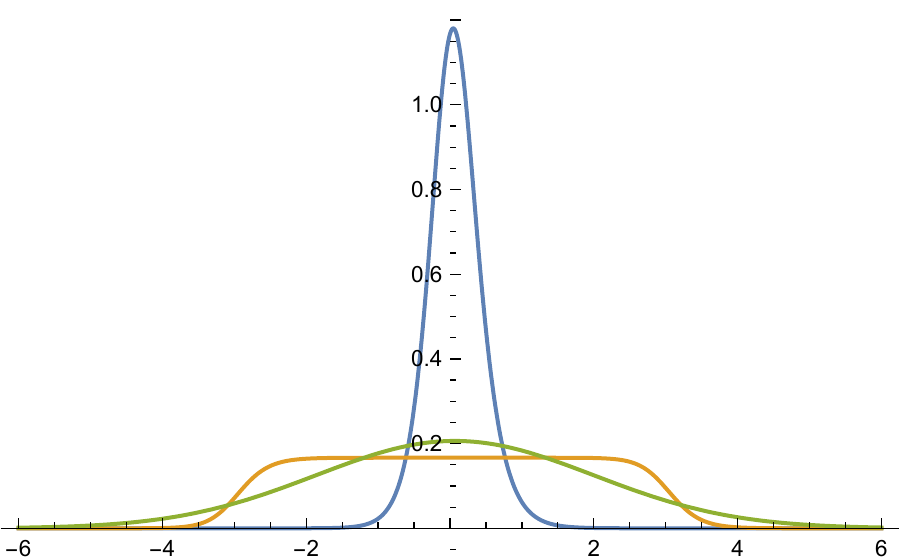}};
			\node[anchor=south,yshift=-10pt,xshift=10] at (Grafik.south) {$\xi,\xi_m$};
			\node[rotate=90,anchor=south,yshift=0,xshift=0pt] at (Grafik.west) {$\omega,\overline{\omega}_\Delta$};
		\end{tikzpicture} 
	\end{minipage}
	\caption{$\omega(\xi),\overline{\omega}_\Delta(\xi_m)$; blue: $\omega(\xi)$, orange: $\overline{\omega}_\Delta(\xi_m)$ evaluated with 1D filter, green: $\overline{\omega}_\Delta(\xi_m)$ evaluated with  spherical filter; left: $\Delta=3$, right: $\Delta=6$}
	\label{fig:ommean}
\end{figure}

For moderate to large filter sizes, of the order or larger than $\delta_{th}$, $\overline{c}_\Delta(\xi_m)$ and $\overline{\omega}_\Delta(\xi_m)$ will smooth out. Fig.'s (\ref{fig:cmean},\ref{fig:ommean}) show comparisons of $c(\xi)$ with $\overline{c}_\Delta(\xi_m)$ and the laminar $\omega_n(\xi)$ with $\overline{\omega}_\Delta(\xi_m)$ for $\Delta_\xi=3,6$. While the form of $\overline{c}_\Delta(\xi_m)$ is not particularly sensible to the particular filter $r(\xi)$ (the same is true for $\tilde{c}_\Delta(\xi_m)$), the strong dependence of the form of  $\overline{\omega}_\Delta(\xi_m)$ on the filter kernel is apparent. 

For very large filter sizes  $\Delta \rightarrow \infty$, $\overline{c}$ and $\overline{(\rho c)}$ will be similar to those evaluated with step functions  $H(\xi)$, i.e. they will approach linear ramp functions. In contrast, $\omega(\xi)$ will behave like delta function $\delta(\xi)$ and $\overline{\omega}$ will mimic the (scaled and stretched) filter kernel:  
\begin{equation*}
	\overline{c}_\infty(\xi_m,\Delta)= H(\xi_m+\Delta/2)H(\Delta/2-\xi_m)(\xi_m/\Delta+1/2) +H(\xi_m-\Delta/2)
\end{equation*}
\begin{equation*}
	\tilde{c}_\infty(\xi_m,\Delta) =\frac{\overline{c}_\infty(\xi_m,\Delta)}{1+\tau\left(1-\overline{c}_\infty(\xi_m,\Delta)\right)}=
\end{equation*}
\begin{equation*}
	 = H(\xi_m+\Delta/2)H(\Delta/2-\xi_m)\frac{\xi_m/\Delta+1/2}{1+\tau\left(1/2-\xi_m/\Delta)\right)}+H(\xi_m-\Delta/2)
\end{equation*}
\begin{equation}
	\overline{\omega}_\Delta(\xi_m) \approx \frac{1}{\Delta}r(\xi_m/\Delta)
	\label{eq:Deltalim}
\end{equation}
Due to their low sensitivity to the form of the filter kernel, the 1D step filter can be used in the evaluation of $\overline{c}_\Delta$,  $\overline{\rho}_\Delta$ and $\overline{\rho c}_\Delta$. When calculating $\overline{\omega}_\Delta$, however, it 
is important to use a suitable filter kernel $r(\xi)$. In the proposed combustion model, the step filter is used to calculate $\overline{c}_\Delta$,  $\overline{\rho}_\Delta$ and $\overline{\rho c}_\Delta$ while the filter kernel $r_g(\xi)$ is used in the evaluation of $\overline{\omega}_\Delta$.

\section{Relation between  $\overline{c}_\Delta$, $\tilde{c}_\Delta$ and  $\overline{\omega}_\Delta$}
Since $\overline{c}_\Delta(\xi_m)$ and $\tilde{c}_\Delta(\xi_m)$ calculated from the filtered 1D laminar flame are monotonic functions of $\xi_m$, one can generate parametric functions $\overline{\omega}_\Delta(\overline{c})$ and $\overline{\omega}_\Delta(\tilde{c})$. These relationships are not analytical but can be calculated and tabulated numerically, varying $\xi_m$ simultaneously in $\overline{c}_\Delta(\xi_m)$, $\tilde{c}_\Delta(\xi_m)$  and $\overline{\omega}_\Delta(\xi_m)$ at given $\Delta$. 

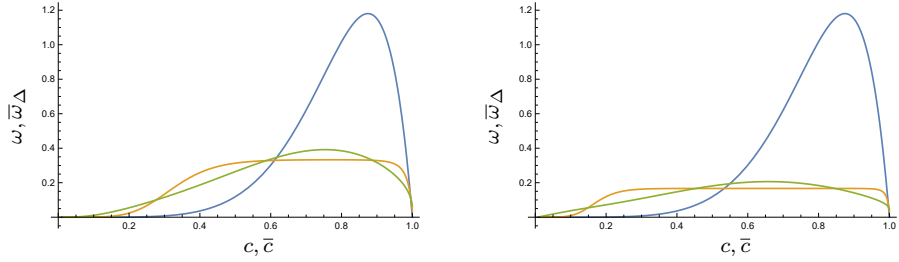
\begin{figure} [ht]
	\begin{minipage}[b]{.42\linewidth} 
		\begin{tikzpicture}
			\node[] (Grafik) at (0,0) {\includegraphics[width=1\textwidth]{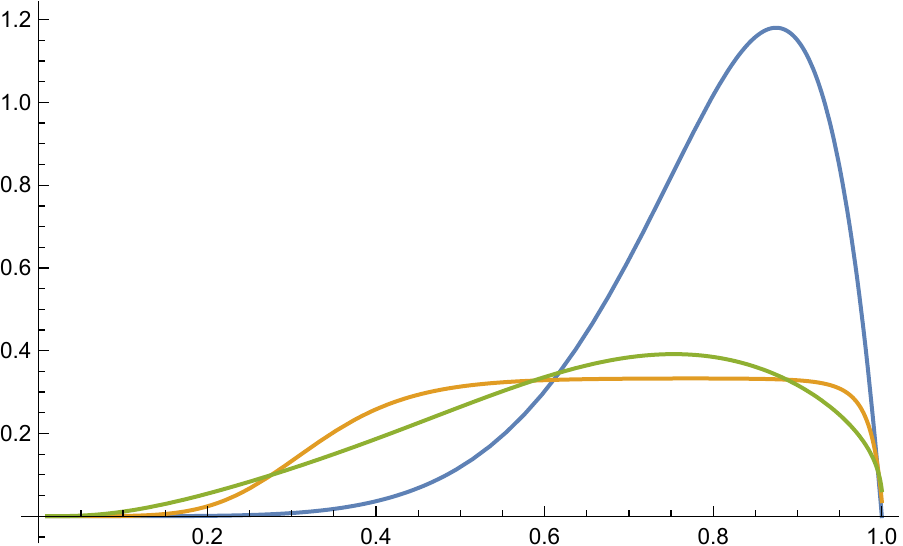}};
			\node[anchor=south,yshift=-10pt,xshift=10] at (Grafik.south) {$c,\overline{c}$};
			\node[rotate=90,anchor=south,yshift=0,xshift=0pt] at (Grafik.west) {$\omega,\overline{\omega}_\Delta$};
		\end{tikzpicture}
	\end{minipage}
	\hspace{.1\linewidth}
	\begin{minipage}[b]{.42\linewidth} 
		\begin{tikzpicture}
			\node[] (Grafik) at (0,0) { \includegraphics[width=\linewidth]{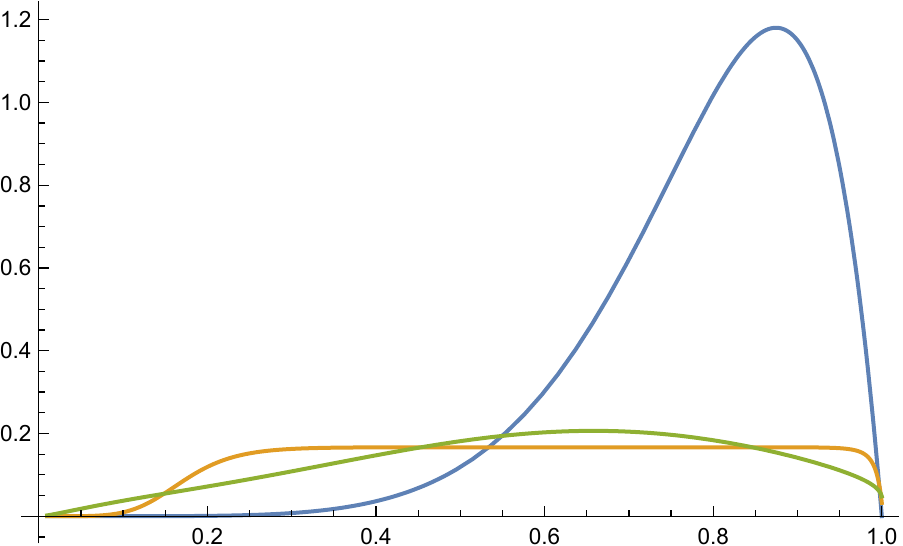}};
			\node[anchor=south,yshift=-10pt,xshift=10] at (Grafik.south) {$c,\overline{c}$};
			\node[rotate=90,anchor=south,yshift=0,xshift=0pt] at (Grafik.west) {$\omega,\overline{\omega}_\Delta$};
		\end{tikzpicture} 
	\end{minipage}
	\caption{$\omega(c)$, $\overline{\omega}_\Delta(\overline{c})$; blue: $\omega(c)$, orange: 1D filter, green: Gaussian filter; left: $\Delta_\xi=3$, right: $\Delta_\xi=6$}
	\label{fig:commean}
\end{figure}

Fig.(\ref{fig:commean}) shows $\overline{\omega}_\Delta(\overline{c})$ for $\Delta_\xi=3,6$ together with the unfiltered source term. Fig.(\ref{fig:commeantilde}) compares $\overline{\omega}_\Delta(\overline{c})$ and $\overline{\omega}_\Delta(\tilde{c})$ for $\Delta_\xi=0.5,3,10$. 

Fig.(\ref{fig:commeanlim}) presents $\overline{\omega}_\Delta(\overline{c})$ at large filter sizes $\Delta_\xi=10,20$ together with the limiting form evaluated from eq.(\ref{eq:Deltalim}) and using the Gaussian filter $r_g(\xi)$. Convergence towards the limiting form is apparent but the limit is not yet fully achieved even at $\Delta_\xi=20$, a filter size which is 11 times larger than the laminar flame thickness.
\begin{figure} [ht]
	\begin{minipage}[b]{.42\linewidth} 
		\begin{tikzpicture}
			\node[] (Grafik) at (0,0) {\includegraphics[width=1\textwidth]{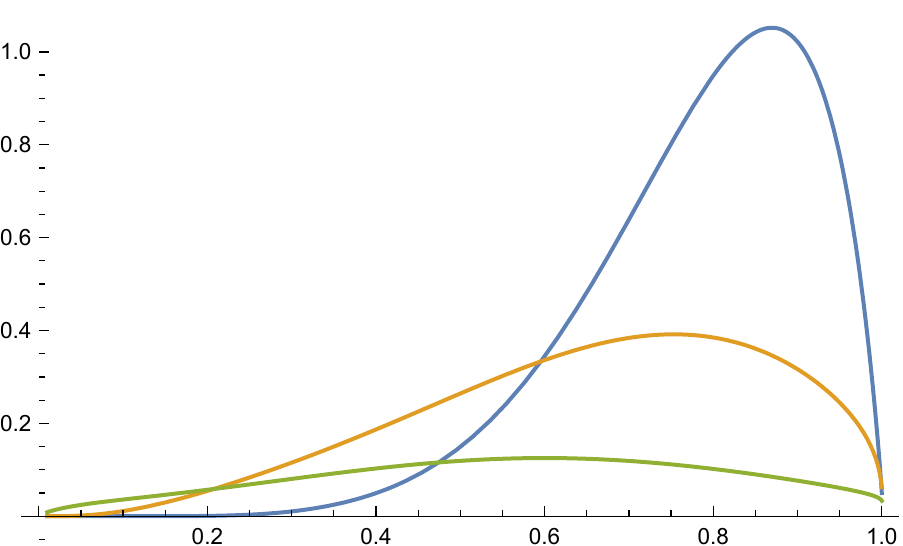}};
			\node[anchor=south,yshift=-10pt,xshift=10] at (Grafik.south) {$\overline{c}$};
			\node[rotate=90,anchor=south,yshift=0,xshift=0pt] at (Grafik.west) {$\overline{\omega}_\Delta$};
		\end{tikzpicture}
	\end{minipage}
	\hspace{.1\linewidth}
	\begin{minipage}[b]{.42\linewidth} 
		\begin{tikzpicture}
			\node[] (Grafik) at (0,0) { \includegraphics[width=\linewidth]{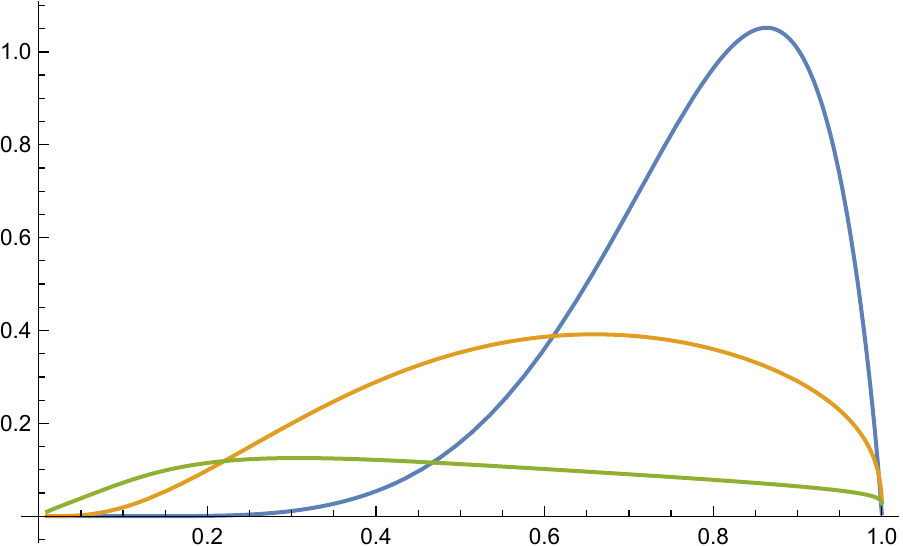}};
			\node[anchor=south,yshift=-10pt,xshift=10] at (Grafik.south) {$\tilde{c}$};
			\node[rotate=90,anchor=south,yshift=0,xshift=0pt] at (Grafik.west) {$\overline{\omega}_\Delta$};
		\end{tikzpicture} 
	\end{minipage}
	\caption{$\overline{\omega}_\Delta(\overline{c})$ (left), $\overline{\omega}_\Delta(\tilde{c})$ (right); blue: $\Delta_\xi=0.5$, orange: $\Delta_\xi=3$, green: $\Delta_\xi=10$}
	\label{fig:commeantilde}
\end{figure}
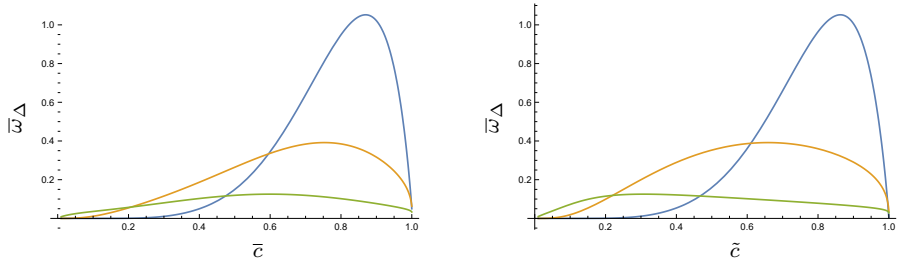
\begin{figure} [ht]
	\begin{minipage}[b]{.42\linewidth} 
		\begin{tikzpicture}
			\node[] (Grafik) at (0,0) {\includegraphics[width=1\textwidth]{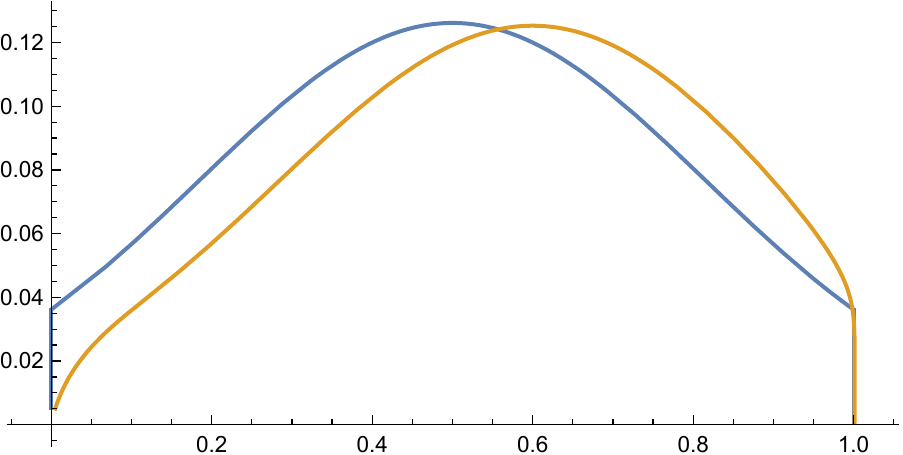}};
			\node[anchor=south,yshift=-10pt,xshift=10] at (Grafik.south) {$\overline{c}$};
			\node[rotate=90,anchor=south,yshift=0,xshift=0pt] at (Grafik.west) {$\overline{\omega}$};
		\end{tikzpicture}
	\end{minipage}
	\hspace{.1\linewidth}
	\begin{minipage}[b]{.42\linewidth} 
		\begin{tikzpicture}
			\node[] (Grafik) at (0,0) { \includegraphics[width=\linewidth]{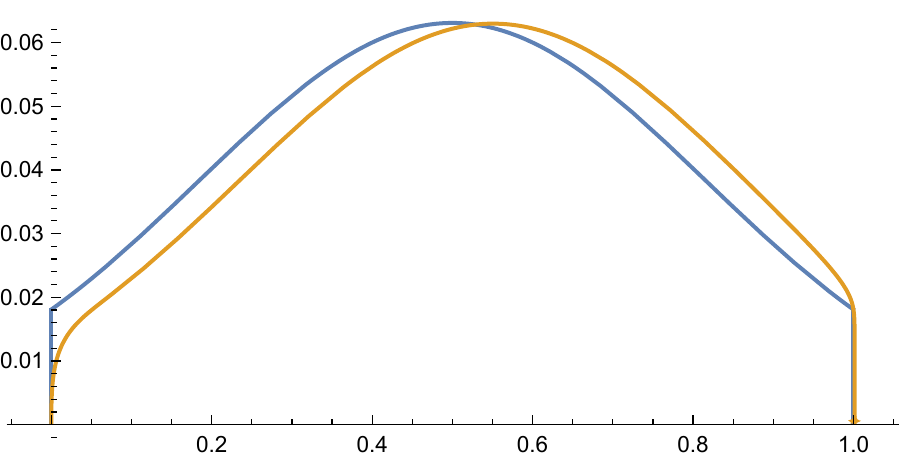}};
			\node[anchor=south,yshift=-10pt,xshift=10] at (Grafik.south) {$\overline{c}$};
			\node[rotate=90,anchor=south,yshift=0,xshift=0pt] at (Grafik.west) {$\overline{\omega}$};
		\end{tikzpicture} 
	\end{minipage}
	\caption{$\overline{\omega}_\Delta(\overline{c})$ with Gaussian filter $r_g(\xi)$; blue: limiting form, orange:  at  $\Delta_\xi=10$ (left), $\Delta_\xi=20$ (right)}
	\label{fig:commeanlim}
\end{figure}
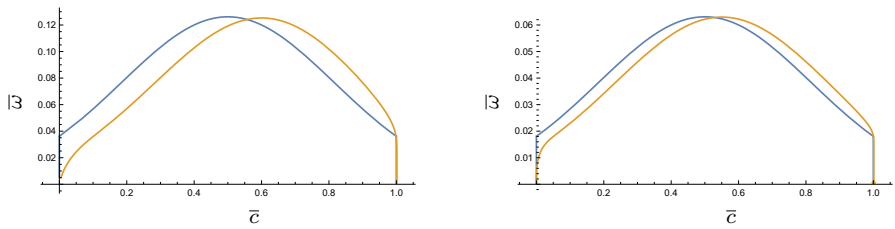

\section{Effect of oblique flame propagation on pdf}
Filtered quantities depending on $c$ only can be evaluated in $c$ space using the pdf $p(c)$ according to eq.(\ref{eq:meanwc}). The multidimensional pdf can be decomposed as \cite{pfitzner2021near} 
\begin{equation}
	p(c^\ast)=\frac{1}{\Omega}\frac{\Sigma(c^\ast)I(c^\ast)}{\mid dc/d\xi_{1D,c^\ast} \mid}
	\label{pdecomp}
\end{equation}
where $\Omega$ is the cell volume and $\Sigma(c^\ast)$ is the area of the $c^\ast$ isosurface within the filter volume.  The $c$ derivative in the denominator represents the pdf of the 1D flame profile. $I(c^\ast)$ is the ratio of local $c$ derivative in the flame front and in the 1D flame at $c=c^\ast$ and therefore sensitive to aberrations of the local flame structure from 1D flamelet profile. In \cite{pfitzner2021near} it was found that $I(c^\ast) \approx 1$ for the DNS data investigated here even for the large $u'/s_L$ cases.

It is interesting to see how the spatial filter $r(\xi)$ derived for planar obliquely propagating flame fronts translates into area  distributions in $c$ space, $\Sigma(c)$. Due to the strongly non linear transformation between $\xi$ and $c$, the relatively simple filter kernel $r(\xi-\xi_m)$ in $\xi$ space translate into a complex functional behaviour in $c$ space. 

\begin{figure} [ht]
	\begin{minipage}[b]{.42\linewidth} 
		\begin{tikzpicture}
			\node[] (Grafik) at (0,0) {\includegraphics[width=1\textwidth]{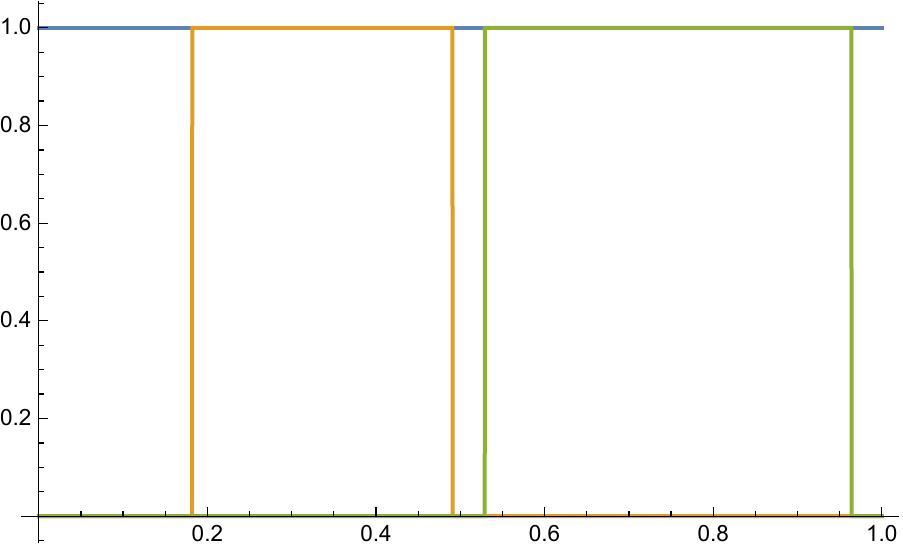}};
			\node[anchor=south,yshift=-10pt,xshift=10] at (Grafik.south) {$c$};
			\node[rotate=90,anchor=south,yshift=0,xshift=0pt] at (Grafik.west) {$\Sigma(c)$};
		\end{tikzpicture}
	\end{minipage}
	\hspace{.1\linewidth}
	\begin{minipage}[b]{.42\linewidth} 
		\begin{tikzpicture}
			\node[] (Grafik) at (0,0) { \includegraphics[width=\linewidth]{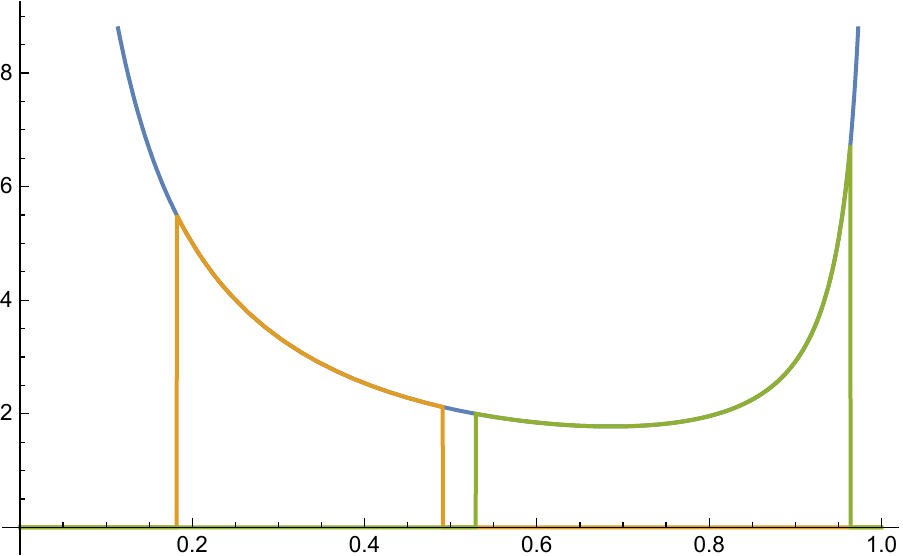}};
			\node[anchor=south,yshift=-10pt,xshift=10] at (Grafik.south) {$c$};
			\node[rotate=90,anchor=south,yshift=0,xshift=0pt] at (Grafik.west) {$p(c)$};
		\end{tikzpicture} 
	\end{minipage}
	\caption{Slicing area in unit cube and $p(c)$ using 1D filter kernel $H(\frac{1}{2}- \mid \xi \mid)$ for $\Delta_\xi=1$ and  $\overline{c} \approx 0.3$  (orange),  $\overline{c} \approx 0.8$  (green)  together with $\frac{1}{\Delta}\frac{1}{dc/d\xi}$ (blue); left: slicing area $\Sigma(c)$ in unit cube, right: $p(c)$}
	\label{fig:sigmap1d1}
\end{figure}

Fig.(\ref{fig:sigmap1d1}) shows the pdf's for situations for cases of $\overline{c} \approx 0.3,0.8$ in the case of the 1D flame filter.The Heaviside function cutoff of the 1D filter in $\xi$ space translates into a Heaviside function cutoff for $\Sigma(c)$ also in $c$ space. 
\begin{figure} [ht]
	\begin{minipage}[b]{.42\linewidth} 
		\begin{tikzpicture}
			\node[] (Grafik) at (0,0) {\includegraphics[width=1\textwidth]{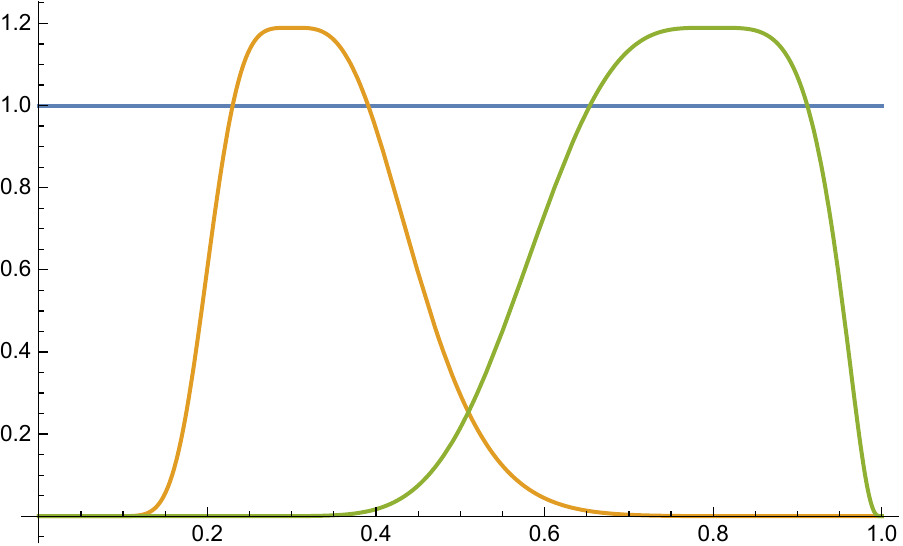}};
			\node[anchor=south,yshift=-10pt,xshift=10] at (Grafik.south) {$c$};
			\node[rotate=90,anchor=south,yshift=0,xshift=0pt] at (Grafik.west) {$\Sigma(c)$};
		\end{tikzpicture}
	\end{minipage}
	\hspace{.1\linewidth}
	\begin{minipage}[b]{.42\linewidth} 
		\begin{tikzpicture}
			\node[] (Grafik) at (0,0) { \includegraphics[width=\linewidth]{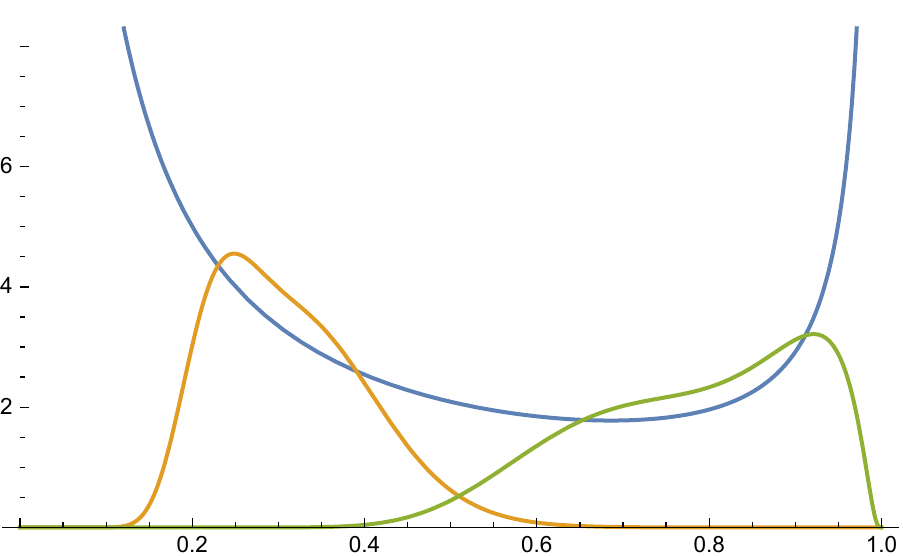}};
			\node[anchor=south,yshift=-10pt,xshift=10] at (Grafik.south) {$c$};
			\node[rotate=90,anchor=south,yshift=0,xshift=0pt] at (Grafik.west) {$p(c)$};
		\end{tikzpicture} 
	\end{minipage}
	\caption{Slicing area in unit cube and $p(c)$ using 1D filter kernel $r_{gm}(\xi)$ for $\Delta_\xi=1$ and  $\overline{c} \approx 0.3$  (orange),  $\overline{c} \approx 0.8$  (green)  together with $\frac{1}{\Delta}\frac{1}{dc/d\xi}$ (blue); left: slicing area $\Sigma(c)$ in unit cube, right: $p(c)$}
	\label{fig:sigmap3d1}
\end{figure}
\begin{figure} [ht]
	\begin{minipage}[b]{.42\linewidth} 
		\begin{tikzpicture}
			\node[] (Grafik) at (0,0) {\includegraphics[width=1\textwidth]{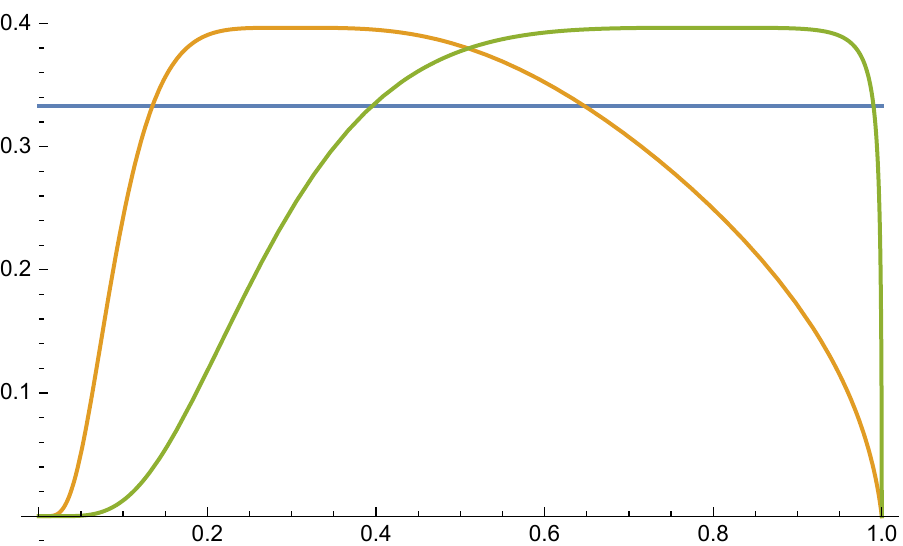}};
			\node[anchor=south,yshift=-10pt,xshift=10] at (Grafik.south) {$c$};
			\node[rotate=90,anchor=south,yshift=0,xshift=0pt] at (Grafik.west) {$\Sigma(c)$};
		\end{tikzpicture}
	\end{minipage}
	\hspace{.1\linewidth}
	\begin{minipage}[b]{.42\linewidth} 
		\begin{tikzpicture}
			\node[] (Grafik) at (0,0) { \includegraphics[width=\linewidth]{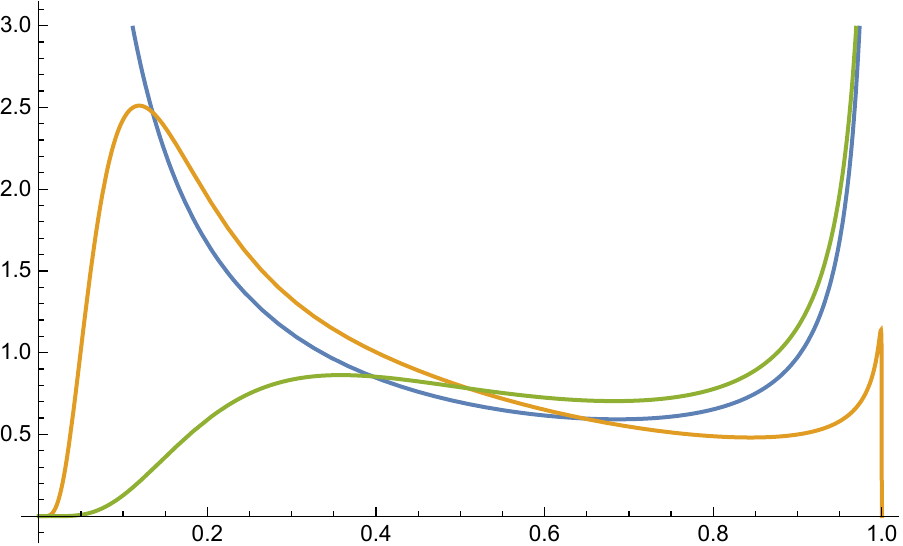}};
			\node[anchor=south,yshift=-10pt,xshift=10] at (Grafik.south) {$c$};
			\node[rotate=90,anchor=south,yshift=0,xshift=0pt] at (Grafik.west) {$p(c)$};
		\end{tikzpicture} 
	\end{minipage}
	\caption{Slicing area in unit cube and $p(c)$ using 1D filter kernel $r_{gm}(\xi)$ for $\Delta_\xi=3$ and  $\overline{c} \approx 0.3$  (orange),  $\overline{c} \approx 0.8$  (green)  together with $\frac{1}{\Delta}\frac{1}{dc/d\xi}$ (blue); left: slicing area $\Sigma(c)$  in unit cube, right: $p(c)$}
	\label{fig:sigmap3d3}
\end{figure}\begin{figure} [ht]
	\begin{minipage}[b]{.42\linewidth} 
		\begin{tikzpicture}
			\node[] (Grafik) at (0,0) {\includegraphics[width=1\textwidth]{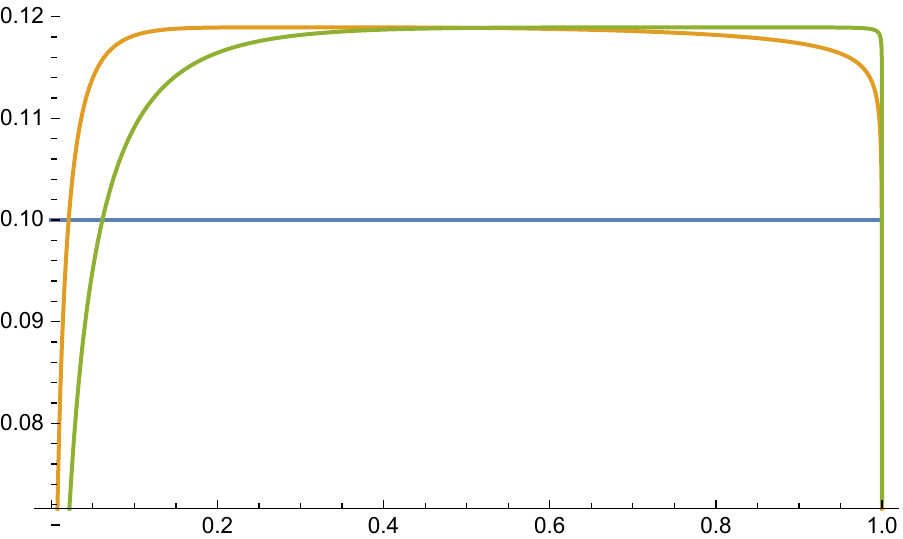}};
			\node[anchor=south,yshift=-10pt,xshift=10] at (Grafik.south) {$c$};
			\node[rotate=90,anchor=south,yshift=0,xshift=0pt] at (Grafik.west) {$\Sigma(c)$};
		\end{tikzpicture}
	\end{minipage}
	\hspace{.1\linewidth}
	\begin{minipage}[b]{.42\linewidth} 
		\begin{tikzpicture}
			\node[] (Grafik) at (0,0) { \includegraphics[width=\linewidth]{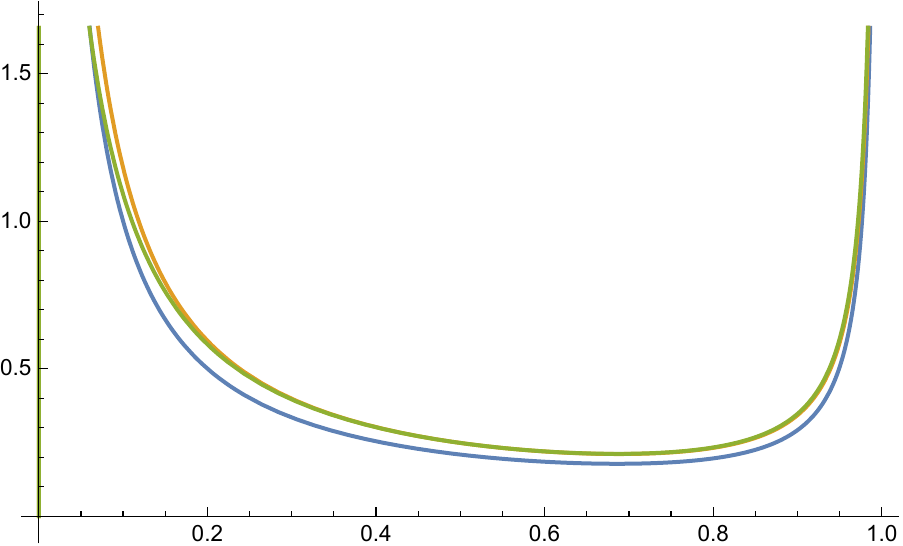}};
			\node[anchor=south,yshift=-10pt,xshift=10] at (Grafik.south) {$c$};
			\node[rotate=90,anchor=south,yshift=0,xshift=0pt] at (Grafik.west) {$p(c)$};
		\end{tikzpicture} 
	\end{minipage}
	\caption{Slicing area in unit cube and $p(c)$ using 1D filter kernel $r_{gm}(\xi)$ for $\Delta_\xi=10$ and  $\overline{c} \approx 0.3$  (orange),  $\overline{c} \approx 0.8$  (green)  together with $\frac{1}{\Delta}\frac{1}{dc/d\xi}$ (blue); left: slicing area $\Sigma(c)$ in unit cube, right: $p(c)$}
	\label{fig:sigmap3d10}
\end{figure}
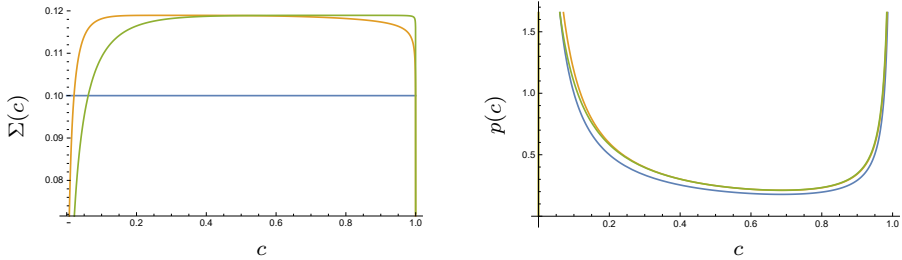

Fig.'s (\ref{fig:sigmap3d1},\ref{fig:sigmap3d3},\ref{fig:sigmap3d10}) show the corresponding $\Sigma(c)$ and $p(c)$'s when using the modified Gaussian filter $r_g(\xi)$ for $\Delta_\xi=1,3,10$. We see that while for intermediate filter sizes $\Delta_\xi \approx 3$ quite complicated shapes of $\Sigma(c)$ and $p(c)$ are produced, for large filters $\Sigma(c)$  becomes flat again over most of the $c$ range while the pdf is just increased by a nearly constant value (however dropping earlier near the ends earlier than the 1D pdf). 

Due to the relatively complex forms of $\Sigma(c)$ and $p(c)$ generated by the multidimensional filter kernels even in the case of flat oblique flame fronts, it can be assumed that accurate models of such  multidimensional effects would be quite difficult to formulate directly in $c$ space. Complex shapes of the subgrid pdf for moderate filter sizes were also noted in \cite{pfitzner2020pdf} when analysing DNS data filter sizes typical of LES. 

In \cite{pfitzner2020pdf}, $\overline{\omega}$ was modelled as function of $\overline{c}$ and $\Delta/\delta_{th}$ using the 1D step filter. Since in the LES a transport equation is solved for $\tilde{c}$, an application of that model would require an additional (model) relationship between $\overline{c}$ and $\tilde{c}$. To avoid this, in the present approach we formulate the  $\overline{\omega}$ model directly as function of $\tilde{c}$ (and $\Delta/\delta_{th}$). The applicability of this model concept will now be investigated by comparison with filtered DNS data.

\section{Description of DNS datasets}
\label{sec:DNS}
The DNS datasets are chosen from an existing database created using single step Arrhenius chemistry with  unity Lewis number, heat release parameter $\tau=(3,4.5)$ and  Arrhenius parameters $\alpha=(3/4,9/11)$, $\beta=6$, $\beta_1=0$ and constant $\lambda/c_p$. To remain in the wrinkled / corrugated flame regime, these DNS datasets feature moderate Reynolds and Karlovitz numbers, see table \ref{tab:database}. The parameter in the analytical model relations $c_n(\xi)$ and $\omega_n(c)$ for evaluation of the DNS data is chosen as $n=(4.4,4.45)$ for $\tau=(3,4.5)$ using the correlations provided in \cite{pfitzner2021analytic}. 

Single-step Arrhenius chemistry cannot represent realistic chemistry for a large range of operating conditions (fuel, stoichiometry, educt temperature, pressure) with a single set of parameters, but it is often able to reproduce the major characteristics of turbulence chemistry interaction with Arrhenius parameters adapted to the specific operating conditions. This is supported by a recent detailed comparison of DNS with single-step and detailed chemistry was presented by Keil et al. \cite{keil2021comparison}.

Five statistically planar flames with moderate and higher turbulence intensity with $\tau=4.5$ and four cases with $\tau=3$ were selected from the database \cite{klein2019priori,KeKl2019}, which was generated with the compressible DNS code SENGA \cite{Jenkins:1999}. The governing equations are solved in non-dimensional form using tenth order finite differences and a third order Runge Kutta scheme. A detailed description of the numerical methodology, the boundary conditions and initialisation procedure can be found in \cite{Gao:2015a,Klein:2016}, but the present database \cite{klein2019priori,KeKl2019} features higher turbulence level and larger scale separation. 

The flame turbulence interaction takes place under decaying turbulence with a simulation time 
$t_{sim}$  larger than  $max(2 t_f,t_c)$, where $t_f=l/u'$ is the initial eddy turn over time and $t_c=\delta_{th}/S_L$ is the chemical time scale. Here $l$ is the integral length scale and $\delta_{th}$ denotes the thermal flame thickness. 
It was demonstrated there that the results remain qualitatively similar halfway through the simulation and that a satisfactory level of convergence of statistics has been achieved. Hence, results will not change qualitatively if a different snapshot was considered. The filtering was done on flow field snapshots at a single time step of the DNS evolution, where turbulent kinetic energy and the global burning rate were not changing rapidly with time as shown in \cite{chakraborty2011effects}.

A detailed discussion of the specific advantages and disadvantages of this planar flame configuration can be found in \cite{Klein:2017a}. There three different methods to introduce turbulence in the computational domain of DNS of statistically planar turbulent premixed flame configurations have been reviewed and their advantages and disadvantages in terms of run time, natural flame development, control of turbulence parameters and convergence of statistics extracted from the simulations have been discussed in detail. No clearly superior method could be identified in these studies.

\begin{table}   
	\caption{The turbulence initial flow parameters for the considered cases.}
	\begin{tabular}{l l l l l l l l l l}
		\hline\noalign{\smallskip}
		$u^\prime /S_L $        &$Re_t$ &  $\tau$   & $l/\delta_{th}$           &$Ka$    &$Da$\\
		\noalign{\smallskip}\hline\noalign{\smallskip}
		$1.0      $     & $11.7$        &$3.0      $                            & $4.58$                         &$4.58$ &$0.47$\\
		$5.0      $     & $58.3$        &$3.0, 4.5$                             & $4.58$                         &$0.93$ &$5.23$\\      
		$7.5      $     & $87.5$        &$3.0, 4.5$                             & $4.58$                        &$0.62$ &$9.60$\\      
		$9.0      $     & $105.0$       &$3.0, 4.5$                             & $4.58$                        &$0.51$ &$12.62$\\
		$15.0      $    & $175.0$       &$3.0, 4.5$                             & $4.58$                        &$0.31$ &$27.16$\\
		\noalign{\smallskip}\hline
	\end{tabular}
	\label{tab:database}
\end{table}

\begin{figure} [ht]
	\hspace{-1.75cm}
	\includegraphics[width=0.7\linewidth]{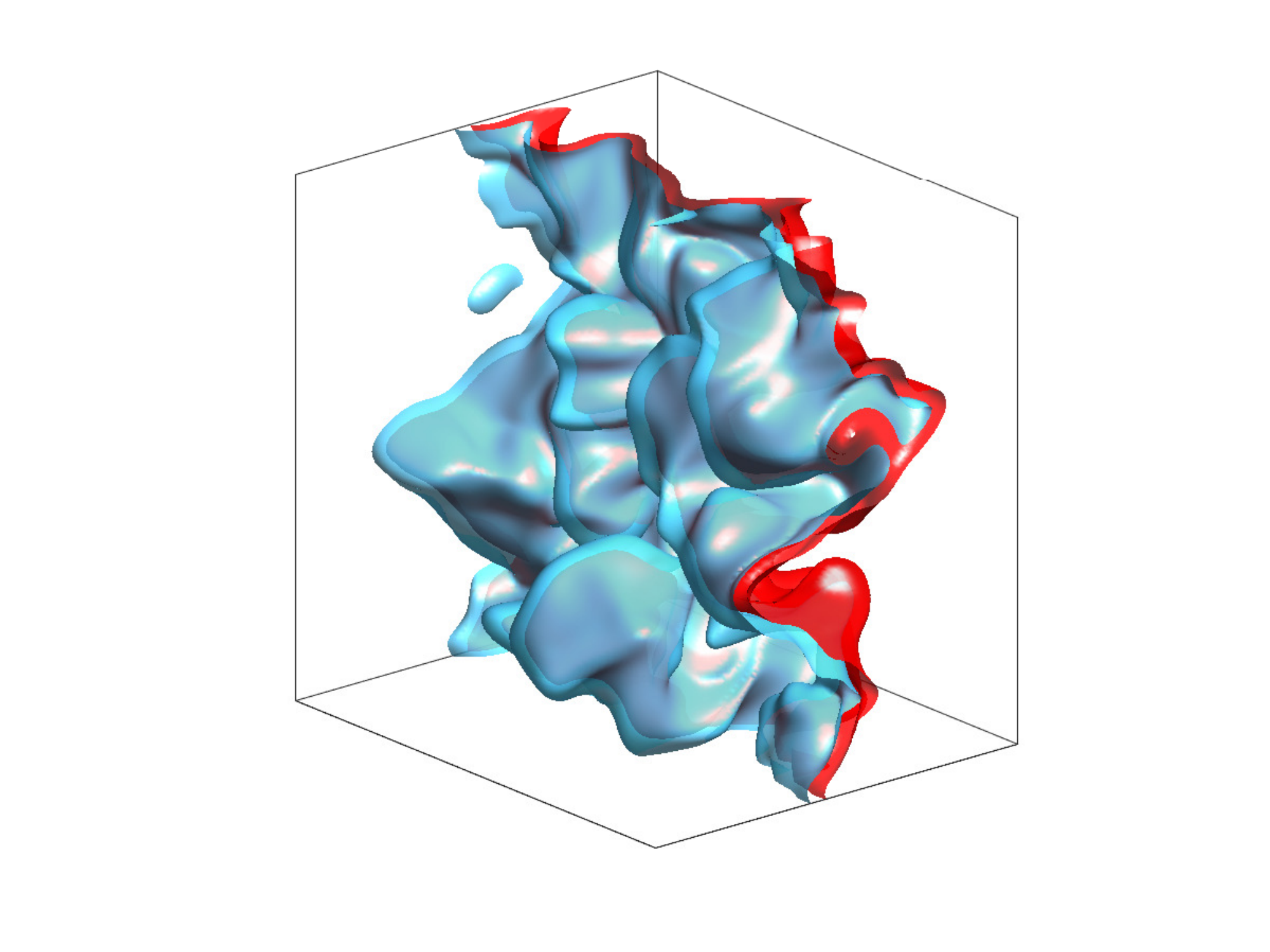}
	\hspace{-1.75cm}
	\includegraphics[width=0.7\linewidth]{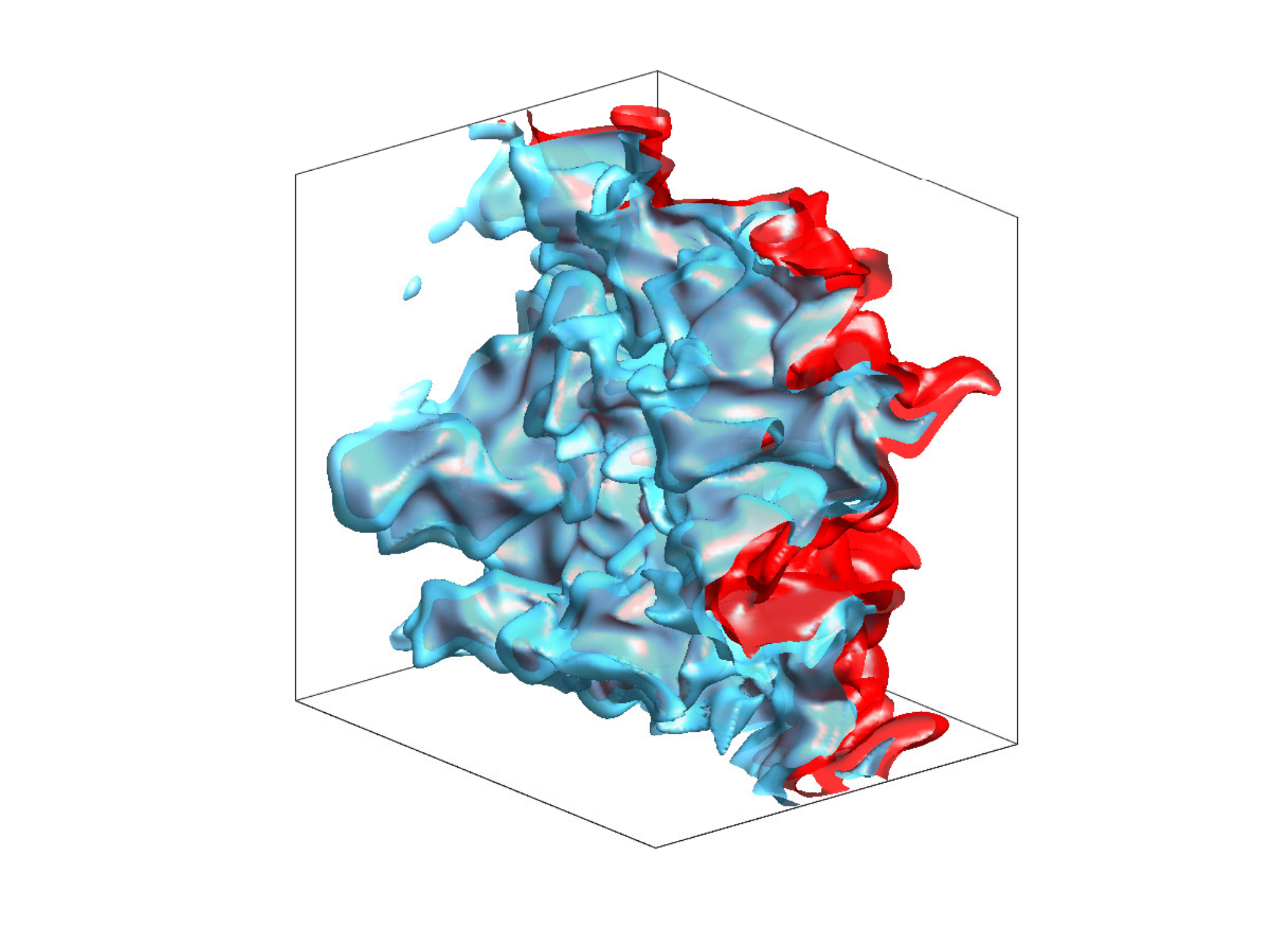}
	\caption{Instantaneous view of $c$-isosurfaces for cases  $u'/s_L=5,15$ and $\tau=4.5$. The value of $c$ increases from 0.1 to 0.9 from light blue to red.}
	\label{fig:DNS}
\end{figure}

Simulations were done on $512^3$ cartesian grids with approximately 11 DNS cells resolving the laminar flame thickness $\delta_{th}$. 
The turbulence level at the start of the DNS was $u'/s_L=1,5,7.5,9,15$. At the time of the evaluated snapshots, the turbulence level had decreased to approximately half of this value. The thermophysical properties
such as dynamic viscosity, thermal conductivity, and density-weighted mass diffusivity
are taken to be constant and independent of temperature. Standard values for (constant) Schmidt $Sc=0.7$ and Prandtl $Pr=0.7$ number as well as for the ratio of specific heats $\gamma=1.4$ were used. The integral length scale is of the order of 4-5 times the laminar flame thickness. 
Instantaneous views of isosurfaces for cases $u'/s_L=5,15$, $\tau=4.5$ are shown in fig.(\ref{fig:DNS}). 

\section{Filtering of DNS data}
The local values of the source term were evaluated as $\omega_n(c)$ with $n=4.4,4.45$ for $\tau=3,4.5$ from the DNS $c$ snapshot fields. Due to the constancy of $c_p/\lambda$, the spatial coordinates of the DNS could be rescaled by a constant factor into the canonical $\xi$ coordinates. For $\tau=(3,4.5)$ the laminar flame thickness in $\xi$ coordinates is $\delta_{th}=(1.8,1.79)$.

The DNS data were filtered using a simple box filter with side lengths which are multiples of the DNS cell length. For small filter sizes, the DNS domain was partitioned into non-overlapping LES box filters while for larger filter sizes  additional filtered data were generated by moving the box filter across the DNS dataset with pivots larger than the laminar flame thickness. Evaluated quantities were $\overline{c}$, $\tilde{c}$ and their gradients on the LES grid, $\overline{\omega}$ and $u'_\Delta=\sqrt{\left(\overline{(u')^2}+\overline{(v')^2}+\overline{(w')^2}\right)/3}$. 

For comparison of filtered DNS data with the presently proposed combustion model, which parametrizes $\overline{\omega}$ as function of $\tilde{c}$ and $\Delta/\delta_{th}$, the filtered $\overline{\omega}$ data were sorted into $\tilde{c}$ bins of width 0.05 and mean values were formed in each $\tilde{c}$ bin. These data are plotted as points in the following figures.

Note that the thickness of the turbulent flame brush in the DNS snapshots is much larger than the integral length scale. This 
motivated the use of filter sizes up to half the turbulent flame brush thickness.

\section{Model validation using filtered DNS data}
\begin{figure} [ht]
	\begin{minipage}[b]{.42\linewidth} 
		\begin{tikzpicture}
			\node[] (Grafik) at (0,0) {\includegraphics[width=1\textwidth]{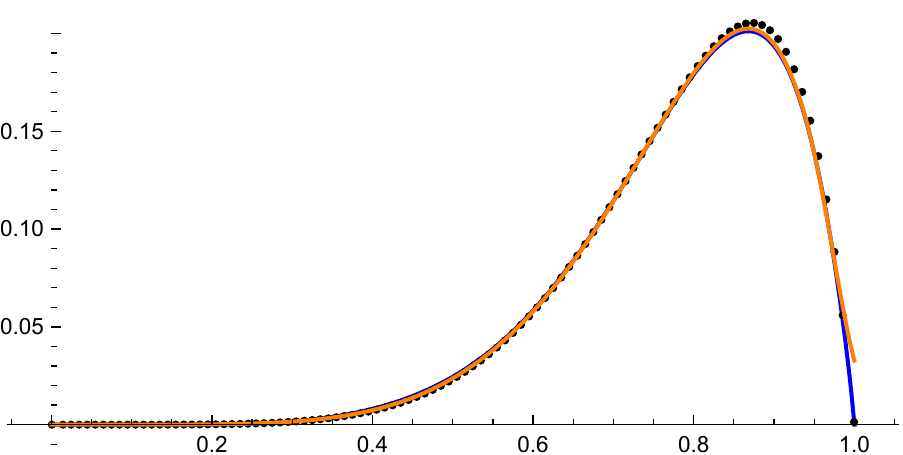}};
			\node[anchor=south,yshift=-10pt,xshift=10] at (Grafik.south) {$\tilde{c}$};
			\node[rotate=90,anchor=south,yshift=0,xshift=0pt] at (Grafik.west) {$\overline{\omega} \cdot \Delta$};
		\end{tikzpicture}
	\end{minipage}
	\hspace{.1\linewidth}
	\begin{minipage}[b]{.42\linewidth} 
		\begin{tikzpicture}
			\node[] (Grafik) at (0,0) { \includegraphics[width=\linewidth]{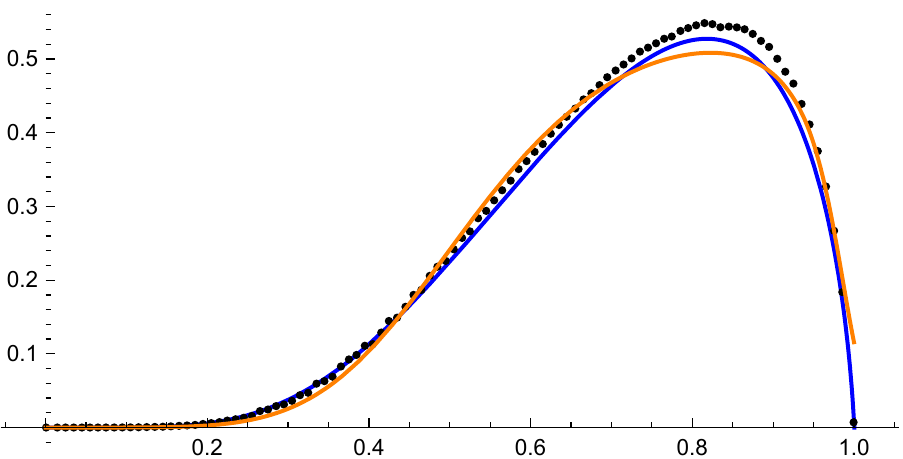}};
			\node[anchor=south,yshift=-10pt,xshift=10] at (Grafik.south) {$\tilde{c}$};
			\node[rotate=90,anchor=south,yshift=0,xshift=0pt] at (Grafik.west) {$\overline{\omega} \cdot \Delta$};
		\end{tikzpicture} 
	\end{minipage}
	\caption{$\overline{\omega} \cdot \Delta$ as function $\tilde{c}$ for the $u'/sL=1$ case; dots: DNS, blue: Gaussian filter, orange: 1D model;  left: $\Delta/\Delta_{DNS}=2$, right: $\Delta/\Delta_{DNS}=8$}
	\label{fig:commeanDNSu1}
\end{figure}
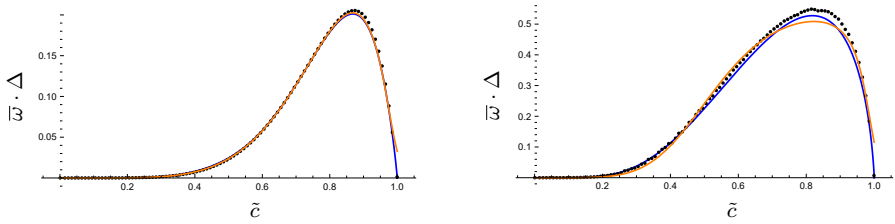

We first investigate the validity of the filter kernel approach for cases with little or no subgrid flame wrinkling by comparison of DNS scatter plots of binned $\overline{\omega}$ vs. $\tilde{c}$. Fig.(\ref{fig:commeanDNSu1}) shows this evaluation for filter sizes of $\Delta/\Delta_{DNS}$ of 2 and 8 for the 1D step filter and the Gaussian filter kernel. 

For the very small filter width $\Delta/\Delta_{DNS}=2$, the laminar reaction source term is reproduced by both filter kernels as expected. For $\Delta/\Delta_{DNS}=8$, which is a little smaller than the laminar flame thickness, the Gaussian filter results are already seen to better reproduce the shape of the DNS filtered $\overline{\omega}(\tilde{c})$ data than the results using the 1D step filter. Also, the maximum of $\overline{\omega}$ from the filtered DNS  is already seen to be slightly larger than the model predictions due to subgrid flame folding.

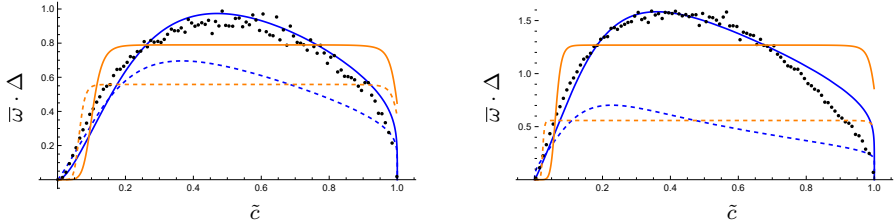
\begin{figure} [ht]
	\begin{minipage}[b]{.42\linewidth} 
		\begin{tikzpicture}
			\node[] (Grafik) at (0,0) {\includegraphics[width=1\textwidth]{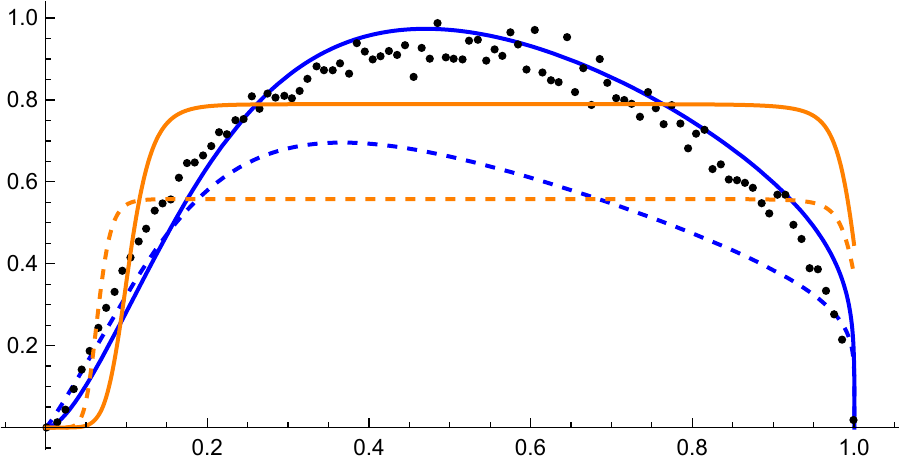}};
			\node[anchor=south,yshift=-10pt,xshift=10] at (Grafik.south) {$\tilde{c}$};
			\node[rotate=90,anchor=south,yshift=0,xshift=0pt] at (Grafik.west) {$\overline{\omega} \cdot \Delta$};
		\end{tikzpicture}
	\end{minipage}
	\hspace{.1\linewidth}
	\begin{minipage}[b]{.42\linewidth} 
		\begin{tikzpicture}
			\node[] (Grafik) at (0,0) { \includegraphics[width=\linewidth]{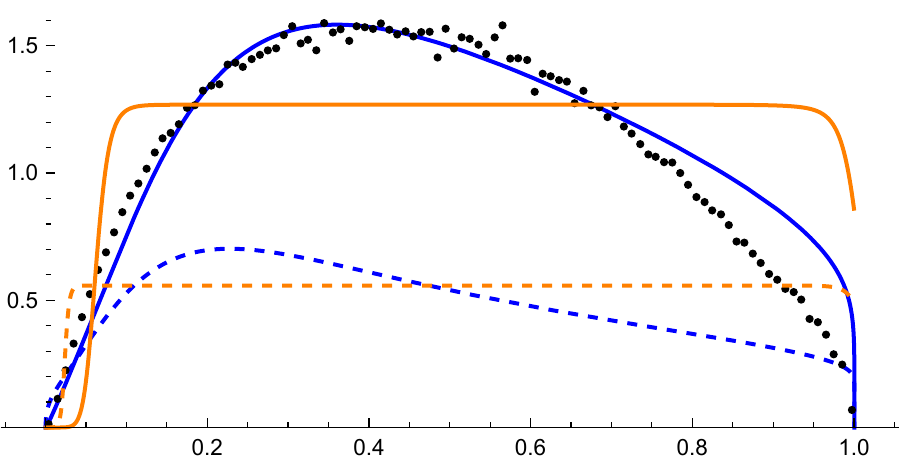}};
			\node[anchor=south,yshift=-10pt,xshift=10] at (Grafik.south) {$\tilde{c}$};
			\node[rotate=90,anchor=south,yshift=0,xshift=0pt] at (Grafik.west) {$\overline{\omega} \cdot \Delta$};
		\end{tikzpicture} 
	\end{minipage}
	\caption{$\overline{\omega} \cdot \Delta$ (solid lines) as function $\tilde{c}$ with fitted wrinkling factor $\Xi$; left: $u'/sL=7.5, \Delta/\Delta_{DNS}=48$, right: $u'/s_L=15, \Delta/\Delta_{DNS}=112$; dots: DNS, blue: Gaussian filter, orange: 1D model;  dashed curves: $\Xi=1$}
	\label{fig:commeanDNShigh}
\end{figure}

For larger filters and at higher freestream turbulence levels effects of subgrid flame folding need to be taken into
account. Using the ratio of the model and filtered DNS $\overline{\omega}$ as estimate of a wrinkling factor $\Xi$ and
performing the filtering operation at $\Delta'=\Delta/\Xi$ yields a very good agreement between the filtered DNS data and the model $\overline{\omega}$. Fig.(\ref{fig:commeanDNShigh}) shows the conditionally filtered DNS data with $\overline{\omega}_\Delta(\tilde{c})$ and $\overline{\omega}_{\Delta/\Xi}(\tilde{c})$ using the Gaussian filter kernel $r_g(\xi)$ and the 1D step filter. 
It is apparent that the 1D box filter does not reproduce well the qualitative distribution of $\overline{\omega}(\tilde{c})$ at larger filter sizes while the Gaussian filter appears
to give a good fit if an appropriate $\Xi$ is chosen.

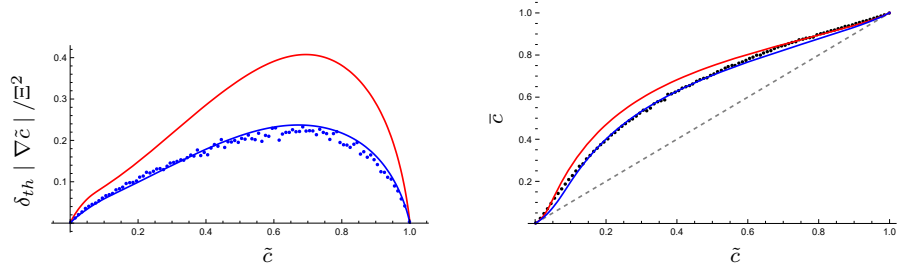
\begin{figure} [ht]
	\begin{minipage}[b]{.42\linewidth} 
		\begin{tikzpicture}
			\node[] (Grafik) at (0,0) {\includegraphics[width=1\textwidth]{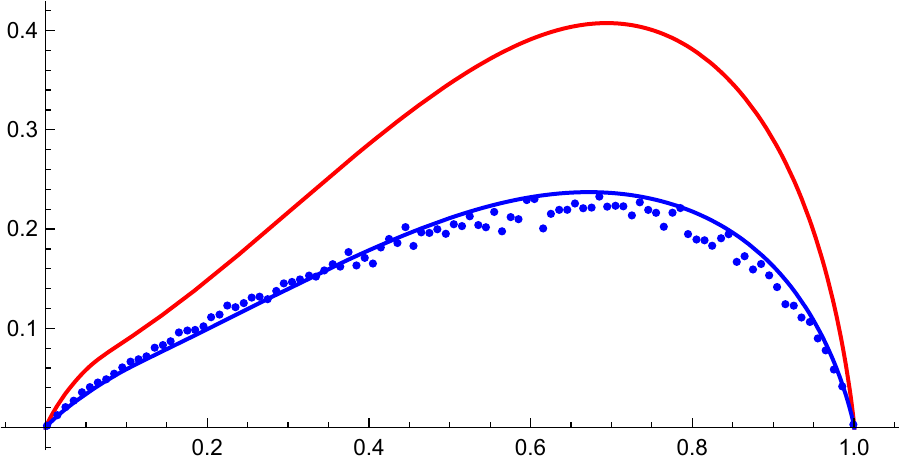}};
			\node[anchor=south,yshift=-10pt,xshift=10] at (Grafik.south) {$\tilde{c}$};
			\node[rotate=90,anchor=south,yshift=0,xshift=0pt] at (Grafik.west) {$\delta_{th}\mid \nabla \tilde{c} \mid/\Xi^2 $};
		\end{tikzpicture}
	\end{minipage}
	\hspace{.1\linewidth}
	\begin{minipage}[b]{.42\linewidth} 
		\begin{tikzpicture}
			\node[] (Grafik) at (0,0) { \includegraphics[width=\linewidth]{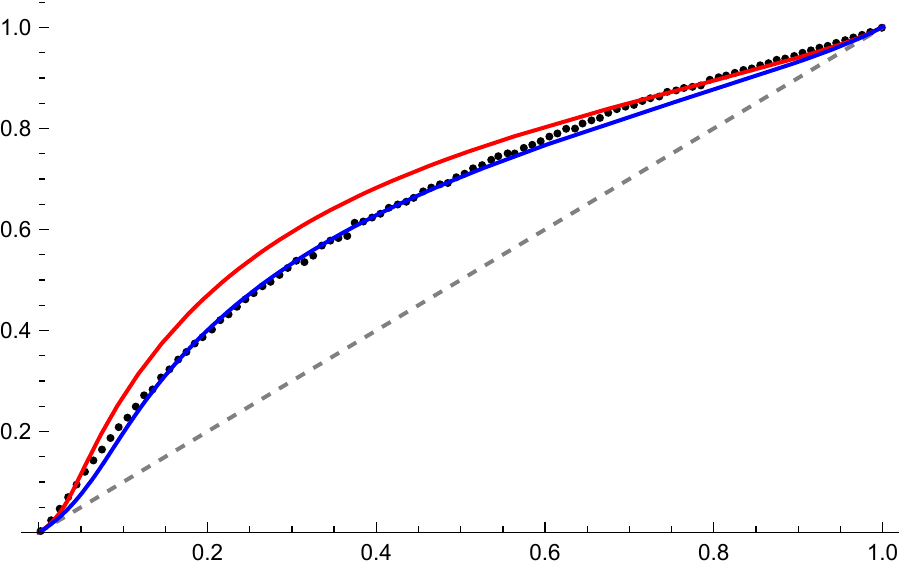}};
			\node[anchor=south,yshift=-10pt,xshift=10] at (Grafik.south) {$\tilde{c}$};
			\node[rotate=90,anchor=south,yshift=0,xshift=0pt] at (Grafik.west) {$\overline{c}$};
		\end{tikzpicture} 
	\end{minipage}
	\caption{left:  ($\delta_{th}\mid \nabla c \mid /\Xi^2) $ vs. $\tilde{c}$, $u'/s_L=15, \Delta/\Delta_{DNS}=48$; 
	right:  $\overline{c}$ vs. $\tilde{c}$, $u'/s_L=15, \Delta=64$; dots: DNS; red line: model with $\Xi=1$, blue line: model with adapted $\Xi$}
	\label{fig:gradccbar}
\end{figure}

We have found that besides $\overline{\omega}$, also other quantities filtered conditionally on $\tilde{c}$ seem to be reproduced quite well (using the $\Xi$ from the fit to the $\overline{\omega}$ maximum). Fig.(\ref{fig:gradccbar} (right) shows $\overline{c}$ plotted vs. $\tilde{c}$. Fig.(\ref{fig:gradccbar} (left) compares the magnitude of the LES gradient of $\tilde{c}$ evaluated from DNS data with the gradient of $\tilde{c}$ of the 1D profile filtered at $\Delta'=\Delta/\Xi$ and divided by $\Xi^2$. Both gradients are non-dimensionalized with the laminar flame thickness. 

The reason for having to scale the $\tilde{c}$ gradient by $1/\Xi^2$ is the following: the filtered gradient increases with decreasing filter size. So the filtering at a smaller  $\Delta'=\Delta/\Xi$ needs to be compensated by the first factor $\Xi$ to bring the filtered gradient back to the level with no subgrid flame folding. The fact that the magnitude of the LES gradient of $\tilde{c}$ of a flame which features subgrid flame folding is smaller than in the case of no subgrid folding is taken into account by the second $\Xi$ factor. The level of agreement is similar for other values of $u'/s_L$ and $\Delta$. 

In our recent work trying to fit $\overline{\omega}$ using neural networks \cite{shin2021apriori}, we found that $\overline{\omega}_\Delta$ could be correlated well using ($\tilde{c}$, $\Delta/\delta_{th}$) and in addition one of the variables $u'/s_L$, $\mid \nabla \tilde{c} \mid_{LES}$ or $\overline{c}$. We were quite surprised since we did not expect that the effect of subgrid flame wrinkling would be apparent e.g. in the relation between $\tilde{c}$ and $\overline{c}$ or in the magnitude of the LES gradient of $\tilde{c}$.

This finding can now be explained as follows: if there exist definite relationships between $\Xi$, $u'_\Delta$, $\mid \nabla \tilde{c} \mid_{LES}$ and $\overline{c}$ (at given $\tilde{c}$ and $\Delta$), one can expect that a deep learning algorithm operating on a large enough network can find and represent such relationships even without having any clue of the underlying physics.

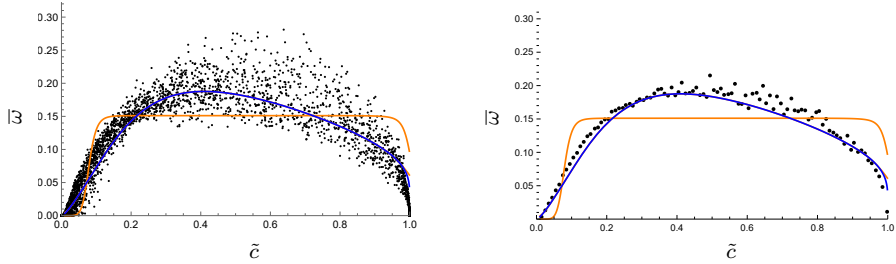
\begin{figure} [ht]
	\begin{minipage}[b]{.42\linewidth} 
		\begin{tikzpicture}
			\node[] (Grafik) at (0,0) {\includegraphics[width=1\textwidth]{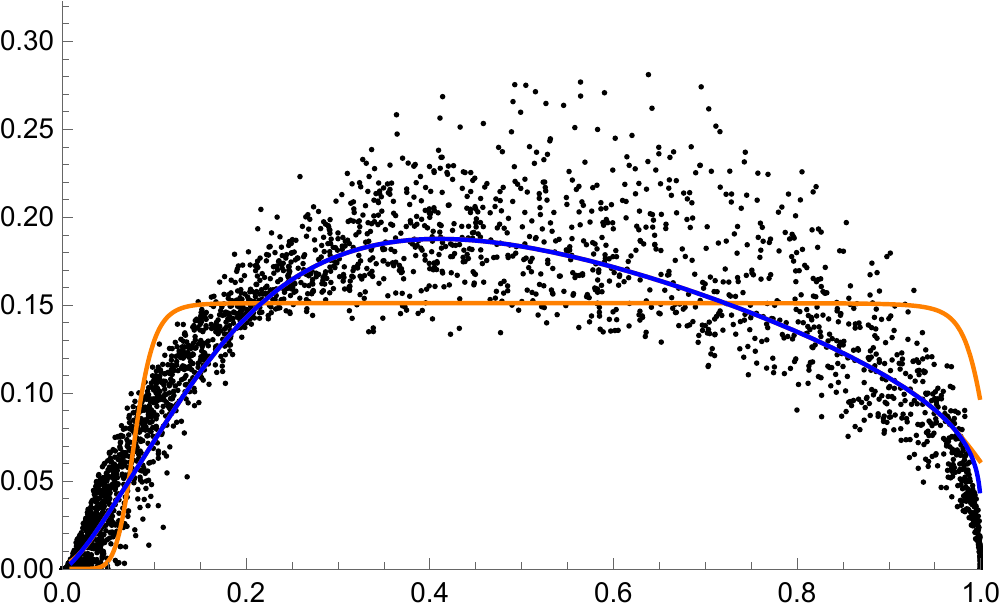}};
			\node[anchor=south,yshift=-10pt,xshift=10] at (Grafik.south) {$\tilde{c}$};
			\node[rotate=90,anchor=south,yshift=0,xshift=0pt] at (Grafik.west) {$\overline{\omega}$};
		\end{tikzpicture}
	\end{minipage}
	\hspace{.1\linewidth}
	\begin{minipage}[b]{.42\linewidth} 
		\begin{tikzpicture}
			\node[] (Grafik) at (0,0) { \includegraphics[width=\linewidth]{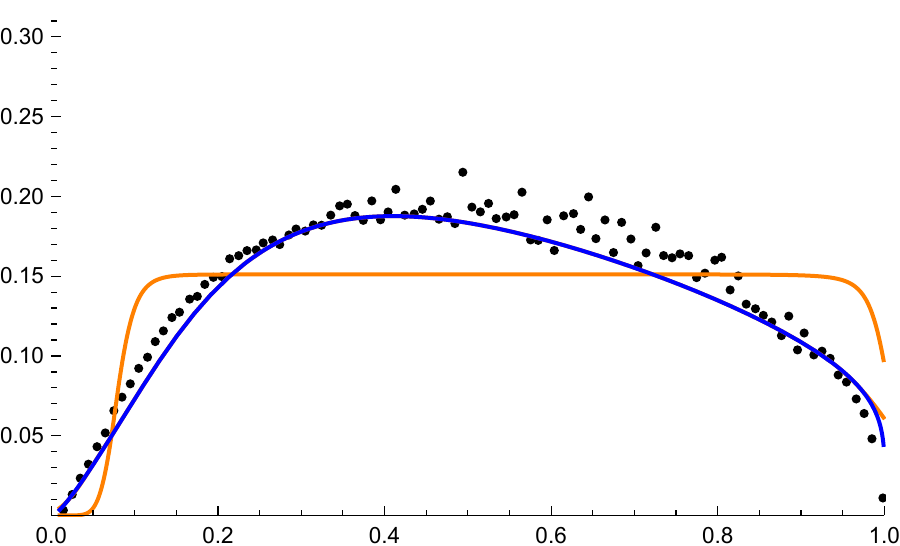}};
			\node[anchor=south,yshift=-10pt,xshift=10] at (Grafik.south) {$\tilde{c}$};
			\node[rotate=90,anchor=south,yshift=0,xshift=0pt] at (Grafik.west) {$\overline{\omega}$};
		\end{tikzpicture} 
	\end{minipage}
	\caption{Comparison of raw and binned $\overline{\omega}$ for $u'/s_L=5,\tau=4.5,\Delta/\Delta_{DNS}=48$; black: DNS; Lines: model with adapted $\Xi$; blue: Gaussian filter, orange: 1D step filter}
	\label{fig:omcbarunbinned}
\end{figure}

We would also like to emphasize that the raw filtered DNS data at larger filter sizes show a considerable scatter around the conditionally averaged ones. Fig.(\ref{fig:omcbarunbinned}) compares the raw (left) and the conditionally averaged $\overline{\omega}_\Delta$ for $u'/s_L=5,\tau=4.5$ and $\Delta/\Delta_{DNS}=48$. Also shown are the model results using $\Delta'=\Delta/\Xi$ with $\Xi=1.2$. Future investigations will show whether at least some of the scatter can be captured by an extension of model using additional LES variables.

\section{Development of wrinkling factor model}
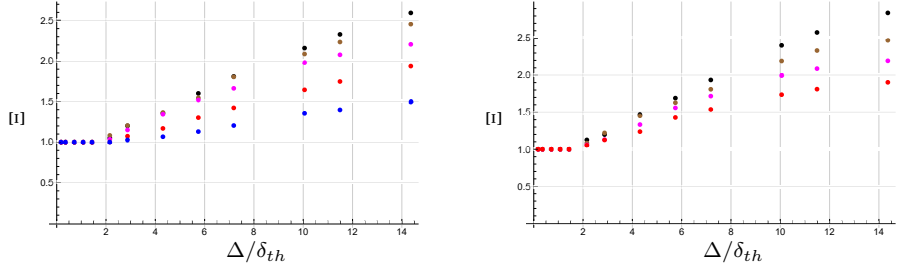
\begin{figure} [ht]
	\begin{minipage}[b]{.42\linewidth} 
		\begin{tikzpicture}
			\node[] (Grafik) at (0,0) {\includegraphics[width=1\textwidth]{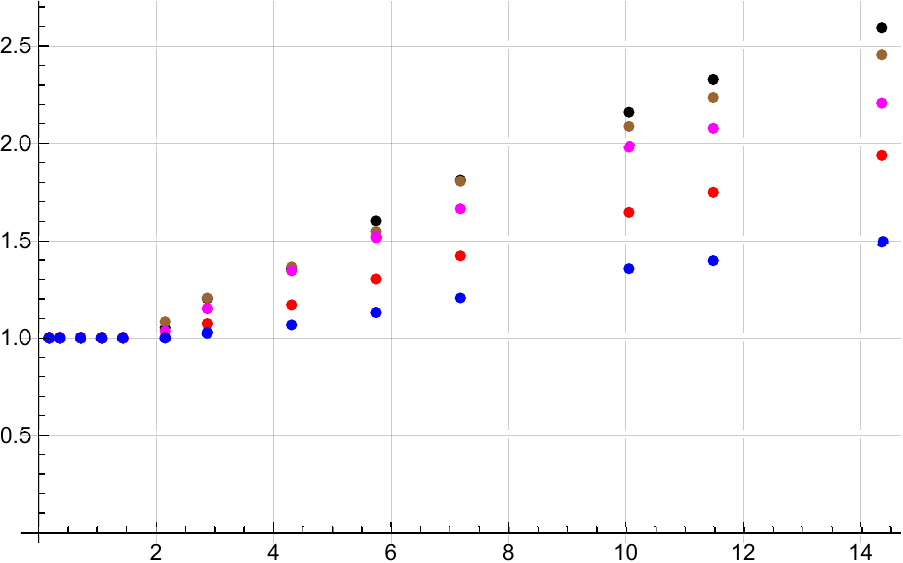}};
			\node[anchor=south,yshift=-10pt,xshift=10] at (Grafik.south) {$\Delta/\delta_{th}$};
			\node[rotate=90,anchor=south,yshift=0,xshift=0pt] at (Grafik.west) {$\Xi$};
		\end{tikzpicture}
	\end{minipage}
	\hspace{.1\linewidth}
	\begin{minipage}[b]{.42\linewidth} 
		\begin{tikzpicture}
			\node[] (Grafik) at (0,0) { \includegraphics[width=\linewidth]{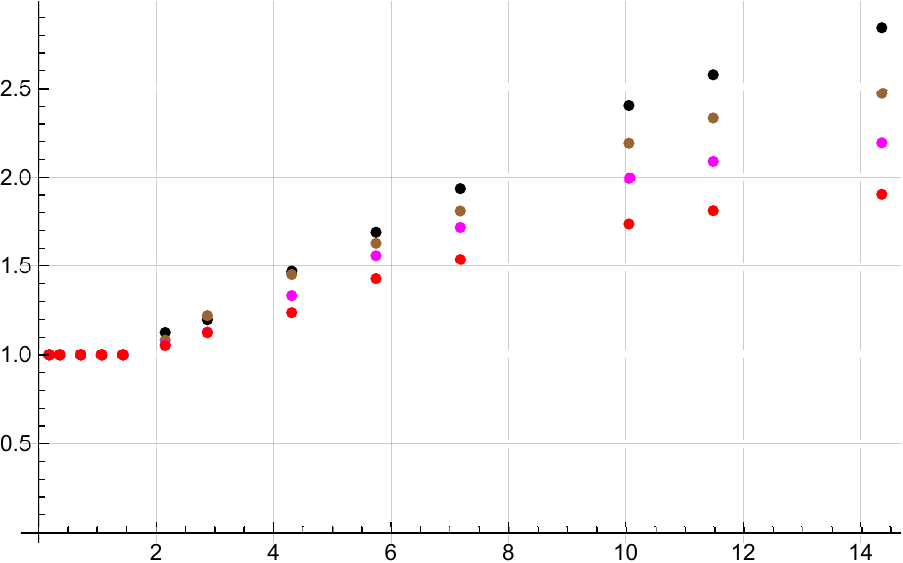}};
			\node[anchor=south,yshift=-10pt,xshift=10] at (Grafik.south) {$\Delta/\delta_{th}$};
			\node[rotate=90,anchor=south,yshift=0,xshift=0pt] at (Grafik.west) {$\Xi$};
		\end{tikzpicture} 
	\end{minipage}
	\caption{Effective wrinkling factor $\Xi$ as a function $\Delta/\delta_{th}$ for $\tau=4.5$ (left) and $\tau=3$ (right); blue: $u'/s_L=1$, red: $u'/s_L=1$, magenta: $u'/s_L=7.5$, brown: $u'/s_L=9$, black: $u'/s_L=15$}
	\label{fig:wrinklingfactorXi}
\end{figure}

As a next step, we formulate models for the effective wrinkling factor $\Xi$ based on analysis of filtered DNS data.
Fig. (\ref{fig:wrinklingfactorXi}) shows wrinkling factors $\Xi$ extracted from the filtered DNS data at $\tau=4.5$ and $\tau=3$. $\Xi$'s were calculated such that the maximum of $\overline{\omega}$ filtered at $\Delta'=\Delta/\Xi$ reproduces the maximum of the filtered and conditionally averaged DNS data. At large $\Delta$, $\Xi$ is fitted to 95\% of the maximum of binned DNS data to compensate their slight scatter.

Evidently, up to $\Delta/\delta_{th} \approx 2$, there is no effect from subgrid wrinkling (i.e. $\Xi=1$). For larger filter sizes, $\Xi$ increases roughly linearly with filter size $\Delta/\delta_{th}$ at slopes which increase with freestream turbulence level  $u'/s_L$. The heat release parameter $\tau$ appears to have little effect.
 
Since subgrid wrinkling is a genuine multidimensional effect, it cannot be extracted from the quasi-1D model and additional LES variables in addition to $\tilde{c}$ and $\Delta/\delta_{th}$ are needed for a wrinkling factor model. 

Many such models have been formulated in the framework of flame surface density theory. Some of which use the concept of fractal flame folding in the form of $\Xi=\Gamma^{D-2}$, where $D$ is a fractal dimension $2 \le D \le D_{max}$ and $\Gamma$ denoting the ratio of outer to inner cutoff scales. The classical value of $D_{max}$ is 7/3, but some authors (e.g. \cite{keppeler2014low}) have advocated a larger value of $D_{max}=8/3$ based on analyses of scalar mixing inside of turbulent jets. In the analysis of hydrogen-air flame fronts developing in the wake of a shock wave, Bambauer et al. \cite{bambauer2021vortex} observed a large range of fractal dimensions between $D=2$ and even near to the theoretical limit of $D=3$, which settled to values of $D \approx 7/3$ for $\phi=1$ and $D \approx 8/3$ for $\phi=0.5$. 

In most models of the literature, both $\Gamma$ and $D$ are functions of non dimensional filter size $\Delta/\delta_{th}$ and of $u'_\Delta/s_L$ or subgrid Karlovitz number $Ka_\Delta$. We find that the $\Xi$ distributions evaluated from the DNS can be reproduced quite well using the fractal dimension approach with fractal dimensions based on $u'_\Delta/s_L$ as well as based on $Ka_\Delta$. The approximately linear dependence of $\Xi$ with filter size starting at $\Delta/\delta_{th} \approx 1..2$ can be modelled using a lower cutoff at $\Xi=1$.

In the process of fitting $\Xi$ model parameter values to data shown in fig.(\ref{fig:wrinklingfactorXi}), we used mean values of $u'_\Delta/s_L$. In the application of the model in LES, the local $u'_\Delta/s_L$ from the LES subgrid turbulence model would obviously be used.

\subsection{Model based on Fureby fractal dimension}
The first model uses the form of the fractal dimension proposed by Fureby \cite{fureby2005fractal}, however with a subgrid turbulence level increased by a constant factor and an adapted model constant in $\Gamma$: 
\begin{equation*}
	u_F=2.4 \left(\frac{u'_\Delta}{s_{L0}}\right)
	\label{eq:fu}
\end{equation*}
\begin{equation*}
	D_F=\frac{2}{u_F+1}+\frac{7/3}{1/u_F+1}
	\label{eq:dfur}
\end{equation*}
\begin{equation*}
	\Gamma_F=0.3 \cdot u_F \cdot\left(\frac{\Delta}{\delta_{th}}\right)
	\label{eq:fgam}
\end{equation*}
\begin{equation}
	\Xi_F=Max\left[1,\Gamma_F^{D_F-2}\right]
	\label{eq:xifmod}
\end{equation}

\subsection{Model based on Keppeler fractal dimension}
In a second $\Xi$ model, the fractal dimension of the Keppeler model \cite{keppeler2014low} is used, where the subgrid Karlovitz number $Ka_\Delta$ instead of $u'/s_L$ scales the effect of turbulence on subgrid flame wrinkling. Also in this case, constants in the definition of fractal dimension and the model constant had to be adapted to fit the DNS data.
\begin{equation*}
	Ka_\Delta=\left(\frac{u'_\Delta}{s_{L}}\right)^{3/2}\left(\frac{\Delta}{\delta_{th}}\right)^{-1/2}
	\label{eq:karlovitz}
\end{equation*}
\begin{equation*}
	D_K=\frac{8/3 \cdot Ka_\Delta +3.11}{Ka_\Delta +1.42}
	\label{eq:dkep}
\end{equation*}
\begin{equation}
	\Xi_K=Max\left[1,0.69\left(\frac{\Delta}{\delta_{th}}\right)^{D_K-2}\right]
	\label{eq:xikmod}
\end{equation}
\begin{figure} [ht]
	\begin{minipage}[b]{.42\linewidth} 
		\begin{tikzpicture}
			\node[] (Grafik) at (0,0) {\includegraphics[width=1\textwidth]{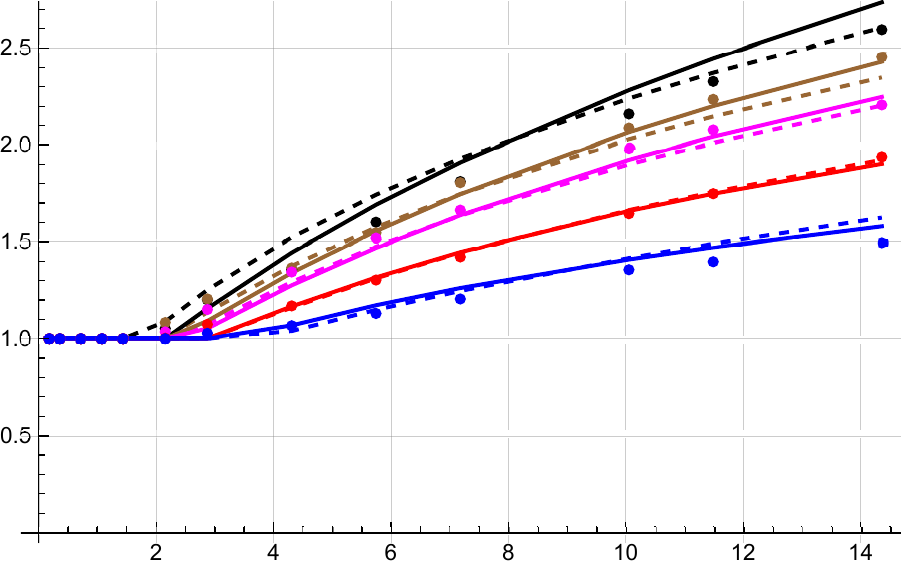}};
			\node[anchor=south,yshift=-10pt,xshift=10] at (Grafik.south) {$\Delta/\delta_{th}$};
			\node[rotate=90,anchor=south,yshift=0,xshift=0pt] at (Grafik.west) {$\Xi$};
		\end{tikzpicture}
	\end{minipage}
	\hspace{.1\linewidth}
	\begin{minipage}[b]{.42\linewidth} 
		\begin{tikzpicture}
			\node[] (Grafik) at (0,0) { \includegraphics[width=\linewidth]{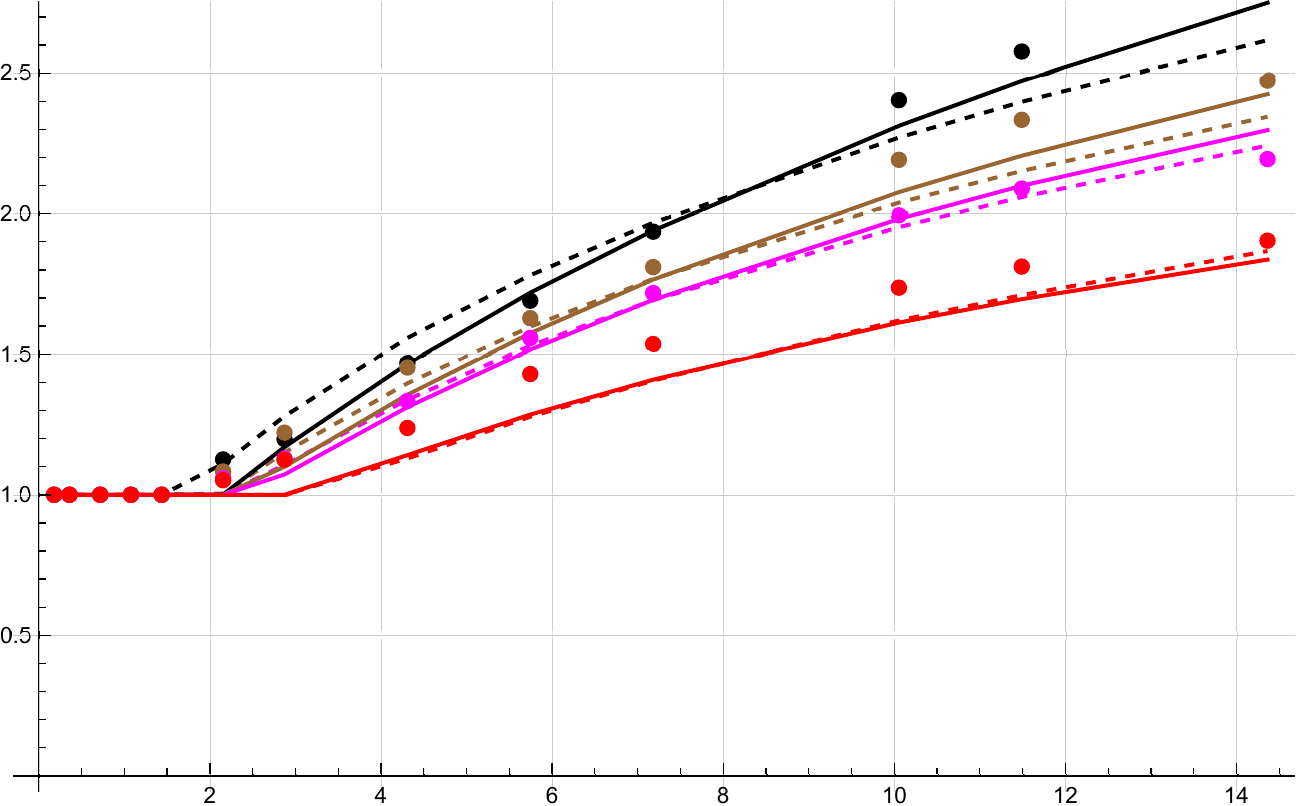}};
			\node[anchor=south,yshift=-10pt,xshift=10] at (Grafik.south) {$\Delta/\delta_{th}$};
			\node[rotate=90,anchor=south,yshift=0,xshift=0pt] at (Grafik.west) {$\Xi$};
		\end{tikzpicture} 
	\end{minipage}
	\caption{Effective wrinkling factors $\Xi$ as function $\Delta/\delta_{th}$ for $\tau=4.5$ (left) and $\tau=3$ (right); color code as in fig.(\ref{fig:wrinklingfactorXi}); solid/dashed lines represent modified Fureby/Keppeler models, respectively}
	\label{fig:wrinklingfactorXimodel}
\end{figure}
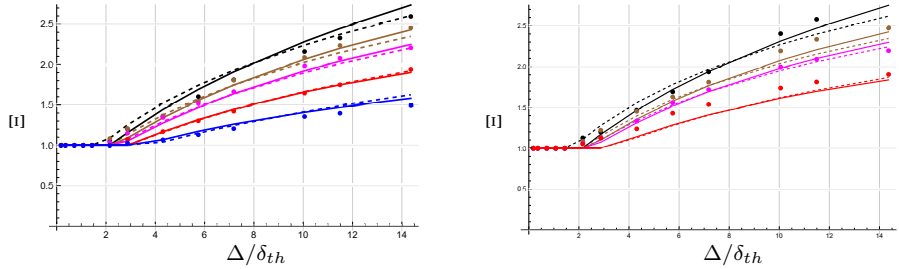 \\
Fig. (\ref{fig:wrinklingfactorXimodel}) shows that both models appear to represent the wrinkling factors $\Xi$ evaluated from the DNS data equally well. A saturation in $\Xi$ data seen for $u'/s_L=15$ at $\tau=4.5$ in the DNS data is not so apparent for $\tau=3$. The reason for this different behaviour is not clear. Further studies are necessary to investigate whether saturation of $\Xi$ vs. $u'/s_L$ will occur at very large $u'/s_L$. 

Since all the selected DNS datasets use the same chemistry and almost identical $\delta_{th}$, it can also not be decided whether the Fureby $u'_\Delta/s_L$ scaling or the Keppeler $Ka_\Delta$ scaling would be more physically correct. Analysis of other DNS data, particularly at different pressure, using other fuels and in different geometries might give some clue and such analyses are planned in the near future.

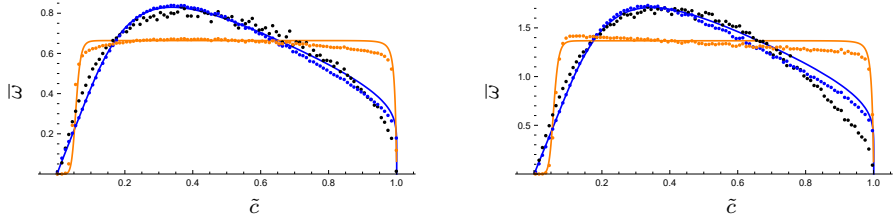
\begin{figure} [ht]
	\begin{minipage}[b]{.42\linewidth} 
		\begin{tikzpicture}
			\node[] (Grafik) at (0,0) {\includegraphics[width=1\textwidth]{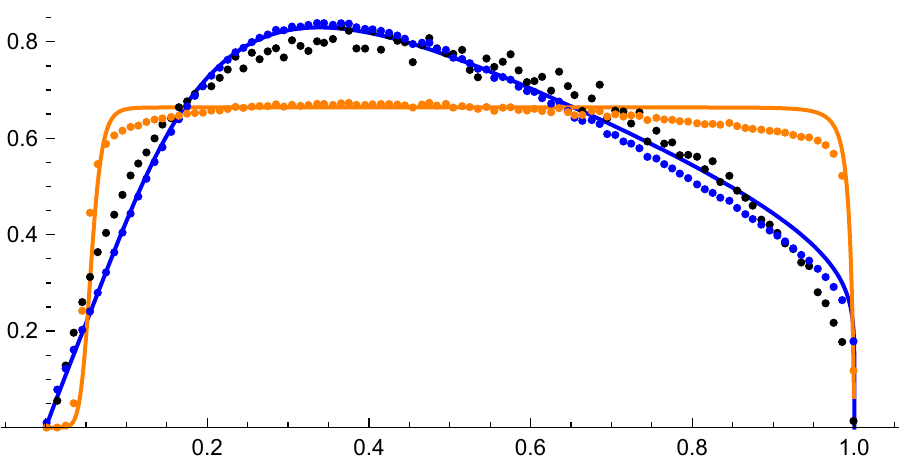}};
			\node[anchor=south,yshift=-10pt,xshift=10] at (Grafik.south) {$\tilde{c}$};
			\node[rotate=90,anchor=south,yshift=0,xshift=0pt] at (Grafik.west) {$\overline{\omega}$};
		\end{tikzpicture}
	\end{minipage}
	\hspace{.1\linewidth}
	\begin{minipage}[b]{.42\linewidth} 
		\begin{tikzpicture}
			\node[] (Grafik) at (0,0) { \includegraphics[width=\linewidth]{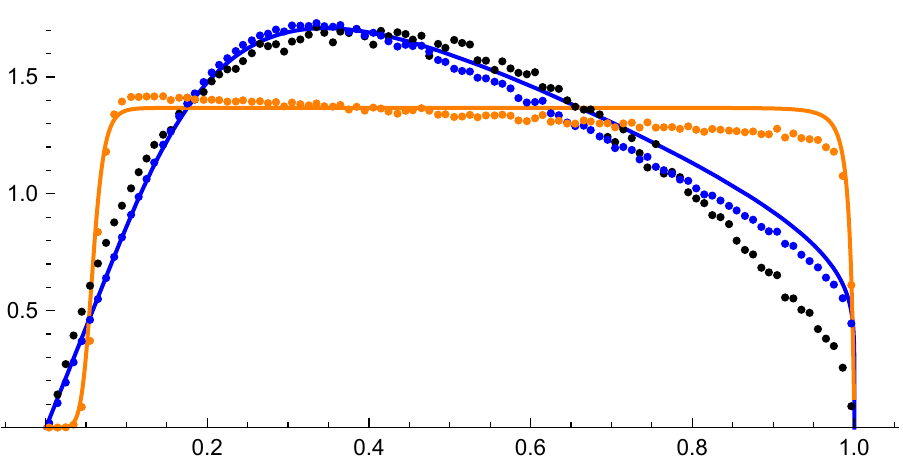}};
			\node[anchor=south,yshift=-10pt,xshift=10] at (Grafik.south) {$\tilde{c}$};
			\node[rotate=90,anchor=south,yshift=0,xshift=0pt] at (Grafik.west) {$\overline{\omega}$};
		\end{tikzpicture} 
	\end{minipage}
	\caption{$\overline{\omega}$ vs. $\tilde{c}$ for the $u'/sL=1, \Delta/\Delta_{DNS}=64$ (left) and $u'/s_L=15, \Delta/\Delta_{DNS}=128$ (right) cases; black: DNS, blue: Gaussian filter, orange: 1D filter; dots: $\Xi_F$ with $\tilde{c}$ binned $u'/s_L$, lines: constant $\Xi$ fitted to maximum of $\overline{\omega}$}.
		\label{fig:ctildeomXifu}
	\end{figure}

Fig.(\ref{fig:ctildeomXifu}) shows the effect of using the model wrinkling factor $\Xi_F$ with conditionally averaged  $u'_\Delta/s_L$ (coloured points) together with results using a constant $\Xi$ derived from mean $u'_\Delta/s_L$ (coloured lines); differences are marginal. Similar results are achieved when using the modified Keppeler wrinkling factor $\Xi_K$. 

\begin{figure} [ht]
	\begin{minipage}[b]{.42\linewidth} 
		\begin{tikzpicture}
			\node[] (Grafik) at (0,0) {\includegraphics[width=1\textwidth]{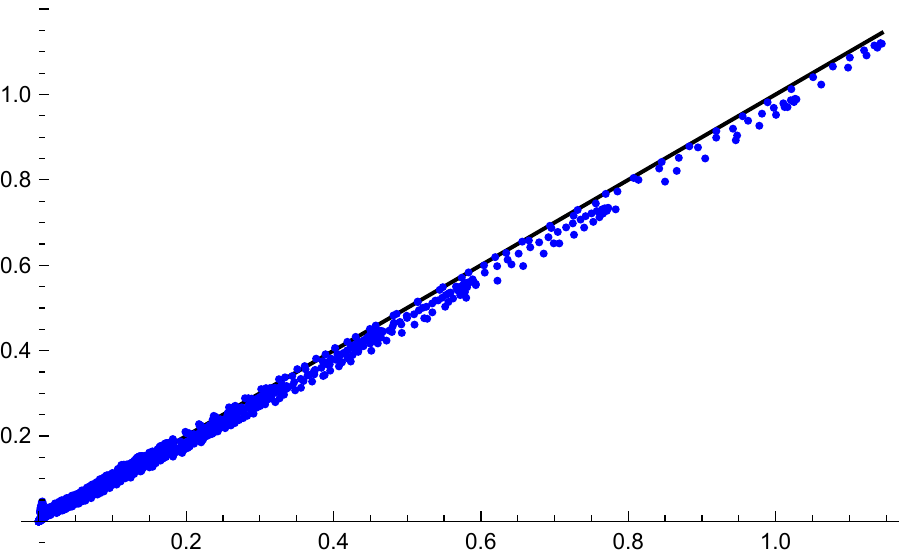}};
			\node[anchor=south,yshift=-10pt,xshift=10] at (Grafik.south) {$\overline{\omega}_{model}$};
			\node[rotate=90,anchor=south,yshift=0,xshift=0pt] at (Grafik.west) {$\overline{\omega}_{DNS}$};
		\end{tikzpicture}
	\end{minipage}
	\hspace{.1\linewidth}
	\begin{minipage}[b]{.42\linewidth} 
		\begin{tikzpicture}
			\node[] (Grafik) at (0,0) { \includegraphics[width=\linewidth]{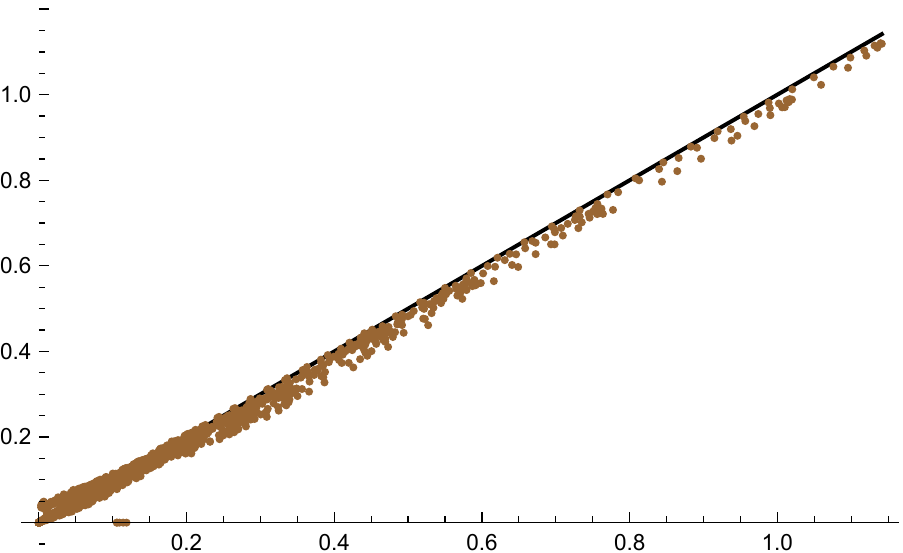}};
			\node[anchor=south,yshift=-10pt,xshift=10] at (Grafik.south) {$\overline{\omega}_{model}$};
			\node[rotate=90,anchor=south,yshift=0,xshift=0pt] at (Grafik.west) {$\overline{\omega}_{DNS}$};
		\end{tikzpicture} 
	\end{minipage}
	\caption{$\overline{\omega}_{DNS}$ vs. $\overline{\omega}_{model}$ for $\tau=4.5$, $u'/sL=1$ (left) and $u'/s_L=9$ (right)}.
	\label{fig:omDNSommodel}
\end{figure}
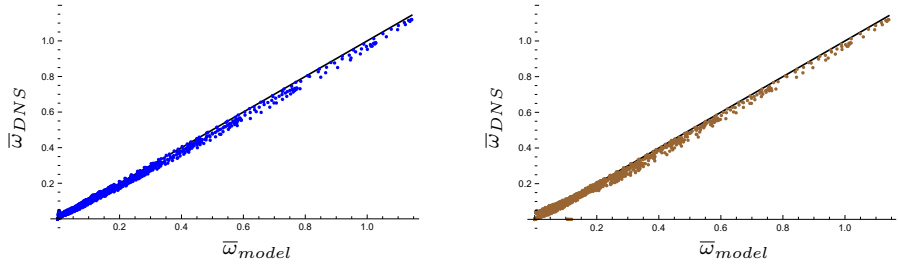

As a final and most salient validation test, fig.(\ref{fig:omDNSommodel}) presents plots of $\overline{\omega}_{DNS}$ vs. $\overline{\omega}_{model}$ for all evaluated filter widths $\Delta/\Delta_{DNS}$=(2, 4, 8, 12, 16, 24, 32, 48, 64, 80, 112, 128, 160), here using the $\Xi_k$ wrinkling factor model. A perfect model would put all data on the 45 degree diagonal line. It is seen that a very close agreement exists for all filter sizes and for both freestream turbulence levels $u'/s_L$ shown here. Similar agreement (not shown for brevity) is obtained for all other $u'/s_L$ cases, for $\tau=3$ and when using the $\Xi_F$ model. 

Compared to other results from the literature, this level of agreement is considered exceptionally good for a model without parameters adjustable to the particular $u'/s_L$ and $\Delta/\delta_{th}$. It is thus expected that the developed combustion model will perform well in a posteriori validations using similar assumptions as the present DNS.

Further work is necessary to evaluate the level of generality of the developed wrinkling factor models. Model constants in the original Fureby and Keppeler models were adapted based on LES of several experimental validation flames, the latter obviously featuring detailed chemistry, differential diffusion and non-constant $\lambda/c_p$ and where the flame is not folded by by decaying isotropic turbulence but by turbulence mostly generated in shear layers. The necessity to adapt model constant for a better fit to the present DNS results may be caused by any of the above reasons or a combination of them.

\section{Summary of premixed combustion model and its implementation}
This section presents a summary of the proposed combustion model. To facilitate its implementation into CFD codes, appendix 1 provides a step-by-step recipe. Although the preceding sections mostly used the canonical space variable $\xi$, also physical space coordinate $x$ can be used. 

In the preprocessing phase, a 1D premixed flame profile containing all species, temperature and chemical source terms (using e.g. CANTERA) is generated and a monotonous progress variable $c(x)$ with corresponding reaction source term $\omega(x)$ is chosen. The scaling of the spatial coordinate used in the flamlet calculations needs to be consistent with the scaling of $x,y,z$ coordinates in the LES simulation. Note that the relationships between $\rho$ and $\rho c$ in eq.'s (\ref{eq:rhocrho1},\ref{eq:rhocrho2},\ref{eq:ctilde1}) are only strictly valid for a progress variable derived from a normalized temperature. For other progress variables, $\tilde{c}$ has to be evaluated from independent calculations of $\overline{(\rho c)}$ and $\overline{\rho}$.

For efficient computation at runtime, the inverse of $\tilde{c}_\Delta(x_m)$, $x_m(\tilde{c},\Delta)$ should be provided during preprocessing as interpolating function or 2D table in the ranges of $0<\tilde{c}<1$ and $0<\Delta<\Delta_{max}$ where $\Delta_{max}$ is the largest filter size in the LES domain.

For small $\Delta/\delta_{th}$,  $x_m(\tilde{c},\Delta)$ reduces to $x(c)$, i.e. the inverse of $c(x)$. For large $\Delta$, $x_m(\tilde{c},\Delta)/\Delta$ approaches a limiting function of $\tilde{c}$ only. Inverting  eq.(\ref{eq:Deltalim}) for $\xi$ yields
\begin{equation}
	\xi_{m,\infty}(\tilde{c},\Delta)=\frac{\Delta  (\tilde{c} (\tau +2)-1)}{2 (\tilde{c} \tau +1)}
\end{equation}
\begin{figure}
	\begin{minipage}[b]{.42\linewidth} 
		\begin{tikzpicture}
			\node[] (Grafik) at (0,0) { \includegraphics[width=\linewidth]{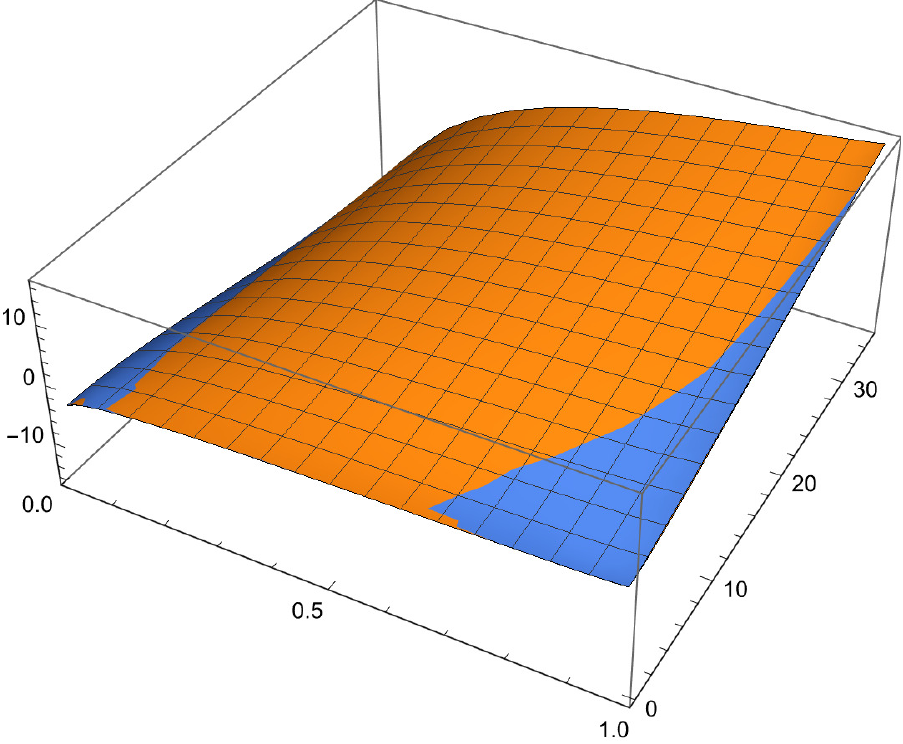}};
			\node[anchor=north,yshift=20pt,xshift=-30] at (Grafik.south) {$\tilde{c}$};
			\node[anchor=south,yshift=-30,xshift=140pt] at (Grafik.west) {$\Delta$};
			\node[anchor=west,yshift=60,xshift=-50pt] at (Grafik.east) {$\xi_m$};
		\end{tikzpicture} 
	\end{minipage}
	\begin{minipage}[b]{.42\linewidth} 
	\begin{tikzpicture}
		\node[] (Grafik) at (0,0) { \includegraphics[width=\linewidth]{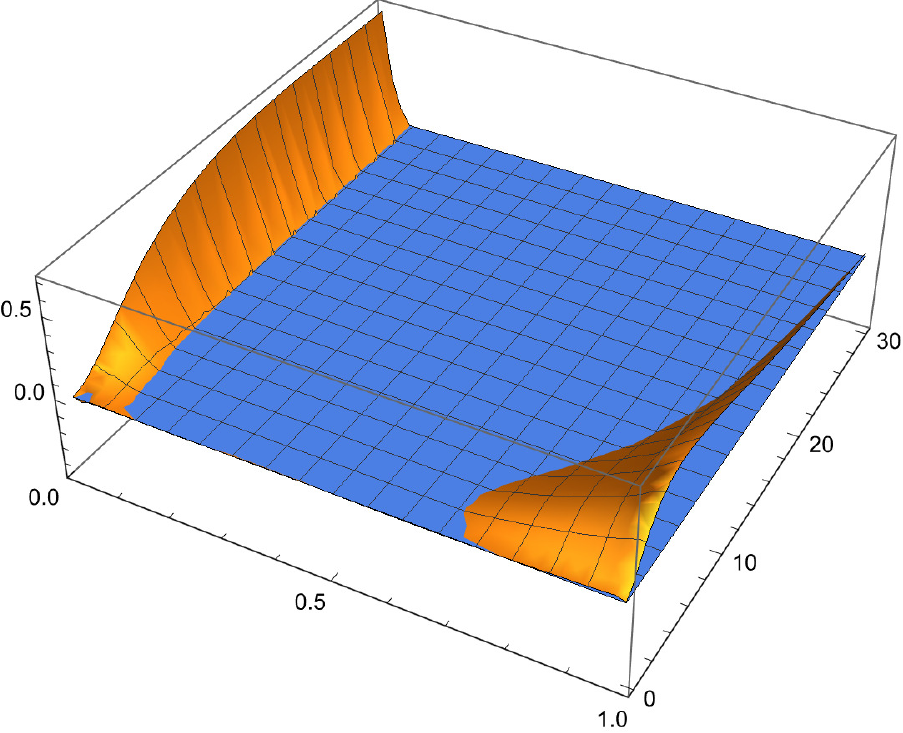}};
			\node[anchor=north,yshift=20pt,xshift=-30] at (Grafik.south) {$\tilde{c}$};
\node[anchor=south,yshift=-30,xshift=140pt] at (Grafik.west) {$\Delta$};
\node[anchor=west,yshift=60,xshift=-50pt] at (Grafik.east) {$\delta$};
	\end{tikzpicture} 
\end{minipage}
	\caption{left: $\xi_m(\tilde{c},\Delta)$ (orange) and approximation (blue) as function $\tilde{c},\Delta$; right: difference between $\xi_m(\tilde{c},\Delta)$ and eq.(\ref{eq:xmapprox}) (orange) and zero (blue)}
	\label{fig:xmplot}
\end{figure}
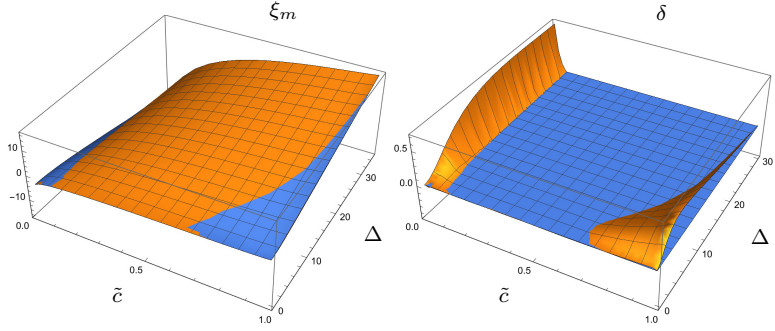
One can combine these two limiting forms to form an analytical approximation to $\xi_m(\tilde{c},\Delta)$ as:
\begin{equation}
	\xi_{m,a}(\tilde{c},\Delta) \approx \xi_n(\tilde{c})+\xi_{m,\infty}(\tilde{c},\Delta)\frac{atan(\frac{2}{5} \frac{\Delta}{\delta_{th}})}{\pi/2 }
	\label{eq:xmapprox}
\end{equation}
Fig.(\ref{fig:xmplot}) shows the numerically evaluated $\xi_m(\tilde{c},\Delta)$ for the present case together with the approximation eq.(\ref{eq:xmapprox}) and the difference between them. It can be seen that the largest differences occur  near $c=1$ and small $\Delta$ and near $c=0$ for all $\Delta$. The maximum error in $\xi_m$ is 0.71, the mean quadratic error is 0.0055. Considering that the range of $\xi_m$ is proportional to $\Delta$, a relative error can be calculated by dividing the difference by $\Delta$. In this metric, the mean quadratic error drops to 0.00043.

If using physical space coordinates $x$, $\xi_n(\tilde{c})$ in eq.(\ref{eq:xmapprox}) needs to be replaced by $x(\tilde{c})$ (i.e. the numerical inversion of the flame profile $c(x)$) and also $\Delta$ needs to be evaluated in physical space units.

During LES runtime, $\Delta$ is determined by the local cell size, $\tilde{c}_{LES}$ is provided by the $\tilde{c}$ transport equation. $u'_\Delta$ is calculated from the velocity subgrid turbulence model. The effective wrinkling factor $\Xi(\Delta/\delta_{th},u'_\Delta/s_L)$ can then be calculated from one of the two proposed models. With these ingredients, the model chemical source term is calculated as $\overline{\omega}=\overline{\omega}_{(\Delta/\Xi)}(x_m(\tilde{c}_{LES},\Delta/\Xi))$ using eq. (\ref{eq:zmean3DDelta}) with  filter kernel $r_g(x)$.

Accepting a slightly lower accuracy, the numerical zero search for $x_m(\tilde{c},\Delta)$ or tabulation / interpolation can be replaced by the correlation eq.(\ref{eq:xmapprox}) in the calculation of $\overline{\omega}=\overline{\omega}_{\Delta/\Xi}(x_m(\tilde{c}_{LES},\Delta/\Xi))$. In this case the source term model is fully explicit at runtime.

\section{Conclusions}
We have derived filter kernels to be used in the evaluation of variables filtered from 1D laminar flame profiles, which can represent the effect of the varying flame surface in 3D filter volumes. It is shown that filter kernels for planar flames moving obliquely through cubical and spherical filter volumes are qualitatively different from the kernel corresponding to the usually used 1D filtering operation.

While different filter kernels have little effect on $\overline{c}$ and $\tilde{c}$, $\overline{\omega}$ is sensitive to their form and actually mimics their shape at very large filter size. Representation of the spatial profiles of progress variable and source term through a series of suitable mathematical functions can yield analytic results for $\overline{c}$, $\overline{\rho c}$ and $\overline{\omega}$, thereby preventing numerical integrations during the runtime evaluation of the model. 

We discuss the variation of $\overline{\omega}_\Delta$ with $\tilde{c}_\Delta$ at different $\Delta$ and  the limiting cases of very small and very big filter size. Also, we show that the multidimensional effects introduce complicated modifications to the pdf $p(c)$ for moderate $\Delta$ values, while the 1D pdf is recovered for small and large $\Delta$.

Comparison of the model source term $\overline{\omega}_\Delta(\tilde{c})$ with filtered DNS data shows the expected good agreement for small filters $\Delta$ and at low turbulence levels, where there is a minimal level of subgrid flame folding. Evaluations of filtered DNS data show no effects from subgrid flame wrinkling for $\Delta/\delta_{th} < 1 .. 2$ while  at larger filter sizes, the wrinkling factor $\Xi$ increases roughly linearly with a slope increasing with freestream turbulence level at larger $\Delta$. The effect of subgrid flame folding can be mimicked in the 1D filtering operation by using a smaller effective filter $\Delta'=\Delta/\Xi$. 

We present two new wrinkling factor models as functions of subgrid turbulence intensity $u'_\Delta/s_L$ and $\Delta/\delta_{th}$, respectively, with fractal dimensions patterned on FSD models from the literature. 
Validation of the complete model yields excellent agreement with $\tilde{c}$ conditionally averaged $\overline{\omega}$ DNS data for all turbulence levels, filter widths and both investigated values of heat release parameter. Implementation of the new combustion model into CFD codes is straightforward. The single numerical zero search can be avoided at runtime by building a lookup table / interpolating function in a preprocessing step or by use of presented correlation $\xi_m(\tilde{c},\Delta/\delta_{th})$. \\ \\
The new model contains three elements of novelty: \\ \\
a) multidimensional effects caused by propagation of planar flame fronts through 3D filter volumes are represented through suitable filter kernels. \\ \\
b) The filtered chemical source term $\overline{\omega}$ of the model is derived as function of $\tilde{c}$ based on 1D filtering of $\rho(x) c(x)$ and $\rho(x)$ of the laminar flame profile.\\ \\
c) New wrinkling factor models depending on $u'_\Delta/s_L$ / $Ka_\Delta$ and $\Delta/\delta_{th}$ represent the effect of subfilter flame folding are derived and validated. \\ \\
In addition, alternative approximations to the $c$, $\rho c$ and $\omega$ spatial profiles are suggested, which yield analytical results upon 1D filtering. It is emphasized that the presented model will reproduce the DNS limit for filter sizes sufficiently smaller than the laminar flame thickness in contrast to models of the FSD type.

Future work will be devoted to a posteriori validations of the model for other operating conditions, flame geometries and more realistic transport assumptions. Generalizations of the method to stratified flames with spatially varying mixture fraction and to flames with non-unity Lewis number will also be investigated. In addition, we plan to look into the possibility to generalize the model to capture some of the remaining scatter of $\overline{\omega}$ seen in the filtered DNS data in the LES through use of additional variables.

\section{Appendix 1: step-by-step recipe for CFD implementation}
- Choose a chemical reaction mechanism. \\ \\
- Generate a 1D premixed flame profile. \\ \\
- Choose a suitable normalized progress variable $c$. Normalisation of $c$ is not strictly necessary, but conventional in premixed combustion.\\ \\
- Calculate $c(x)$, $\rho(x)$ and $\omega(x)$ profiles with $c_r(x)=\rho(x) c(x)$. \\ \\
- Optional: calculate $x(c)$, numerically inverting c(x). \\ \\
- Optional: approximate $c_r(x)$ by splines or function series yielding an analytical integral to avoid numerical integration during preprocessing.\\ \\
- Optional: approximate $\omega(x)$ by splines or function series yielding an analytical results upon integration with $r(x)$ (e.g. by a series of weighted, shifted, stretched Gaussians) to avoid numerical integration during preprocessing.\\ \\
- Calculate $\overline{(\rho c)}_\Delta(x_m)$ as $\int_{x_m-\Delta/2}^{x_m-\Delta/2}c_r(x)dx$. \\ \\
- Calculate $\overline{(\rho)}_\Delta(x_m)$ as $\int_{x_m-\Delta/2}^{x_m-\Delta/2}\rho(x)dx$; if $c(x)$ is a normalized temperature then eq.(\ref{eq:ctilde1}) can be used instead. \\ \\
- Calculate the $\tilde{c}_\Delta(x_m)=\overline{(\rho c)}_\Delta(x_m)/\overline{\rho}_\Delta(x_m)$.\\ \\
- Calculate the $\overline{\omega}_\Delta(x_m)$ as $\int_{-\infty}^{\infty}r_g(\frac{x-x_m}{\Delta})\omega(x)dx$ with $r_g(x)$ from eq.(\ref{eq:Agauss2}). \\ \\
- Tabulate $x_m(\tilde{c},\Delta)$ as solution of $\tilde{c} \stackrel{!}{=} \tilde{c}_\Delta(x_m)$ in the ranges of $0 \leq \tilde{c} \leq 1$ and $0 \leq \Delta \leq \Delta_{max}$; alternatively, use correlation from eq.(\ref{eq:xmapprox}) with previously generated $x(c)$.
 \\ \\
During the LES simulation, at given $\Delta$ from the LES grid and with $\tilde{c}_{LES},u'_\Delta$ from the transport equations, the effective wrinkling factor $\Xi$ is first locally evaluated from eq.(\ref{eq:xifmod}) or eq.(\ref{eq:xikmod}). The model source term is then evaluated as  $\overline{\omega}=\overline{\omega}_{\Delta/\Xi}(x_m(\tilde{c}_{LES},\Delta/\Xi))$.

\section{Appendix 2: Useful analytical results}
This appendix gives some analytical formulas useful to implement the proposed approximations to $c(\xi)$, $\omega(\xi)$. \\ \\
The solution to eq.(\ref{eq:cstrans}) with  together with Gaussian $\omega(\xi)$ from eq.(\ref{eq:omgauss}) and boundary conditions $c(\xi)=0,1$ for $\xi \rightarrow \mp \infty$ is given by
\begin{equation}
	c_a(\xi)=\frac{1}{2} \left(\text{erf}\left(\sqrt{a} \xi\right)-e^{\frac{1}{4 a}+\xi} \text{erf}\left(\frac{2 a \xi+1}{2 \sqrt{a}}\right)+e^{\frac{1}{4 a}+\xi}+1\right)
\end{equation}
$c_a(\xi)$ can be integrated analytically yielding exponentials and error functions. The integral can be more robustly evaluated numerically than hypergeometric functions resulting from integration of $c_n(\xi)$ in \cite{pfitzner2020pdf}.
\begin{equation*}
	i_1=-e^{-\xi} \text{erf}\left(\sqrt{a} \xi\right)+e^{\left.\frac{1}{4}\right/a} \xi-\frac{e^{-\xi (a \xi+1)}}{\sqrt{\pi } \sqrt{a}}-e^{-\xi}
\end{equation*}	
\begin{equation*}
	i_2=\frac{e^{\left.\frac{1}{4}\right/a} (2 a (\xi-1)+1) \text{erf}\left(\frac{2 a \xi+1}{2 \sqrt{a}}\right)}{2 a}
\end{equation*}	
\begin{equation}
	I_a(\xi)=\int c_a(\xi)d\xi=\frac{1}{2}\left(i_1-i_2\right)
\end{equation}
 \\ \\
If $\rho(x)c(x)$ (and possibly $\rho(x)$, $c(x)$) were represented by sum of scaled / shifted error/tanh/sigmoid functions and  $\omega(x)$ through a sum of scaled / shifted Gaussians, the following integrals can be useful to evaluate their 1D filtered counterparts: \\ \\
$\int \text{erf}(x)dx=x \text{erf}(x)+\frac{e^{-x^2}}{\sqrt{\pi }}$ \\
$\int tanh(x)dx=log[cosh(x)]$ \\
$\int \frac{dx}{1+exp(-x)}=log[1+exp(x)]$ \\
$\int exp(-c (x-d)^2)dx=-\frac{\sqrt{\pi } \text{erf}\left(\sqrt{c} (d-x)\right)}{2 \sqrt{c}}$ \\
$\int exp(-a (x-b)^2) exp(-c (x-d)^2)dx=\frac{\sqrt{\pi } e^{-\frac{a c (b-d)^2}{a+c}} \text{erf}\left(\frac{a (x-b)+c (x-d)}{\sqrt{a+c}}\right)}{2 \sqrt{a+c}}$ \\

\section{Acknowledgements}
The authors gratefully acknowledge funding of part of this work through Deutsche Forschungsgemeinschaft in projects PF443/9-1 and KL1456/5-1.

\section{Compliance with Ethical standards} The authors confirm that Ethical standards have been obeyed. 
%
\section{Conflict of interest} The authors declare that they have no conflict of interest.

\bibliographystyle{elsarticle-num}       
\bibliography{CNF_oblique_V3}

\begin{thebibliography}{10}
\expandafter\ifx\csname url\endcsname\relax
  \def\url#1{\texttt{#1}}\fi
\expandafter\ifx\csname urlprefix\endcsname\relax\def\urlprefix{URL }\fi
\expandafter\ifx\csname href\endcsname\relax
  \def\href#1#2{#2} \def\path#1{#1}\fi

\bibitem{nilsson2018structures}
T.~Nilsson, H.~Carlsson, R.~Yu, X.-S. Bai, Structures of turbulent premixed
  flames in the high karlovitz number regime--dns analysis, Fuel 216 (2018)
  627--638.

\bibitem{driscoll2008turbulent}
J.~F. Driscoll, Turbulent premixed combustion: Flamelet structure and its
  effect on turbulent burning velocities, Progress in Energy and Combustion
  Science 34~(1) (2008) 91--134.

\bibitem{luca2019statistics}
S.~Luca, A.~Attili, E.~L. Schiavo, F.~Creta, F.~Bisetti, On the statistics of
  flame stretch in turbulent premixed jet flames in the thin reaction zone
  regime at varying reynolds number, Proceedings of the Combustion Institute
  37~(2) (2019) 2451--2459.

\bibitem{colin:2000}
O.~Colin, F.~Ducros, D.~Veynante, T.~Poinsot, A thickened flame model for large
  eddy simulations of turbulent premixed combustion, Phys. Fluids 12~(7) (2000)
  1843--1863.
\newblock \href {https://doi.org/http://dx.doi.org/10.1063/1.870436}
  {\path{doi:http://dx.doi.org/10.1063/1.870436}}.

\bibitem{pitsch2002large}
H.~Pitsch, L.~D. De~Lageneste, Large-eddy simulation of premixed turbulent
  combustion using a level-set approach, Proceedings of the Combustion
  Institute 29~(2) (2002) 2001--2008.

\bibitem{poi05}
T.~Poinsot, D.~Veynante, Theoretical and numerical combustion, 2nd Edition,
  Edwards, 2005.

\bibitem{ma2013posteriori}
T.~Ma, O.~Stein, N.~Chakraborty, A.~Kempf, A posteriori testing of algebraic
  flame surface density models for les, Combustion Theory and Modelling 17~(3)
  (2013) 431--482.

\bibitem{moureau2011large}
V.~Moureau, P.~Domingo, L.~Vervisch, From large-eddy simulation to direct
  numerical simulation of a lean premixed swirl flame: Filtered laminar
  flame-pdf modeling, Combustion and Flame 158~(7) (2011) 1340--1357.

\bibitem{lapointe2017priori}
S.~Lapointe, G.~Blanquart, A priori filtered chemical source term modeling for
  les of high karlovitz number premixed flames, Combustion and Flame 176 (2017)
  500--510.

\bibitem{pfitzner2020pdf}
M.~Pfitzner, A new analytic pdf for simulations of premixed turbulent
  combustion, Flow, Turbulence and Combustion (2020).
\newblock \href {https://doi.org/https://doi.org/10.1007/s10494-020-00137-x}
  {\path{doi:https://doi.org/10.1007/s10494-020-00137-x}}.

\bibitem{klein2018flame}
M.~Klein, H.~Nachtigal, M.~Hansinger, M.~Pfitzner, N.~Chakraborty, Flame
  curvature distribution in high pressure turbulent bunsen premixed flames,
  Flow, Turbulence and Combustion 101~(4) (2018) 1173--1187.

\bibitem{klein2019priori}
M.~Klein, N.~Chakraborty, A-priori analysis of an alternative wrinkling factor
  definition for flame surface density based large eddy simulation modelling of
  turbulent premixed combustion, Combustion Science and Technology 191~(1)
  (2019) 95--108.

\bibitem{proch2017flame2}
F.~Proch, P.~Domingo, L.~Vervisch, A.~M. Kempf, Flame resolved simulation of a
  turbulent premixed bluff-body burner experiment. part t ii: A-priori and
  a-posteriori investigation of sub-grid scale wrinkling closures in the
  context of artificially thickened flame modeling, Combustion and Flame 180
  (2017) 340--350.

\bibitem{pfitzner2021near}
M.~Pfitzner, M.~Klein, A near-exact analytic solution of progress variable and
  pdf for single-step arrhenius chemistry, Combustion and Flame 226 (2021)
  380--395.

\bibitem{fiorina2010filtered}
B.~Fiorina, R.~Vicquelin, P.~Auzillon, N.~Darabiha, O.~Gicquel, D.~Veynante, A
  filtered tabulated chemistry model for les of premixed combustion, Combustion
  and Flame 157~(3) (2010) 465--475.

\bibitem{ferziger1993simplified}
J.~Ferziger, T.~Echekki, A simplified reaction rate model and its application
  to the analysis of premixed flames, Combustion science and technology
  89~(5-6) (1993) 293--315.

\bibitem{hansinger2020statistical}
M.~Hansinger, M.~Pfitzner, M.~Klein, Statistical analysis and verification of a
  new premixed combustion model with dns data, Combustion Science and
  Technology (2020).

\bibitem{bray2006finite}
K.~Bray, M.~Champion, P.~Libby, N.~Swaminathan, Finite rate chemistry and
  presumed pdf models for premixed turbulent combustion, Combustion and flame
  146~(4) (2006) 665--673.

\bibitem{pfitzner2021analytic}
M.~Pfitzner, P.~Breda, An analytic probability density function for partially
  premixed flames with detailed chemistry, Physics of Fluids 33~(3) (2021)
  035117.

\bibitem{keil2021comparison}
F.~B. Keil, M.~Amzehnhoff, U.~Ahmed, N.~Chakraborty, M.~Klein, Comparison of
  flame propagation statistics based on direct numerical simulation of simple
  and detailed chemistry. part 2: Influence of choice of reaction progress
  variable, Energies 14~(18) (2021) 5695.

\bibitem{KeKl2019}
F.~Keil, M.~Klein, N.~Chakraborty, Subgrid reaction progress variable variance
  closure in turbulent premixed flames, Flow Turbulence and Combustion (2020).
\newblock \href {https://doi.org/https://doi.org/10.1007/s10494-020-00121-5}
  {\path{doi:https://doi.org/10.1007/s10494-020-00121-5}}.

\bibitem{Jenkins:1999}
K.~Jenkins, R.~Cant, {DNS} of turbulent flame kernels, in: C.~Liu, L.~Sakell,
  T.~Beautner (Eds.), Proceedings of 2nd AFOSR Conference on {DNS} and {LES},
  Kluwer Academic Publishers, 1999, pp. 192--202.

\bibitem{Gao:2015a}
Y.~Gao, N.~Chakraborty, M.~Klein, Assessment of sub-grid scalar flux modelling
  in premixed flames for large eddy simulations, European Journal of Mechanics
  - B/Fluids 52 (2015) 97--108.

\bibitem{Klein:2016}
M.~Klein, N.~Chakraborty, Y.~Gao, Scale similarity based models and their
  application to subgrid scale scalar flux modelling in the context of
  turbulent premixed flames, International Journal of Heat and Fluid Flow 57
  (2016) 91--108.

\bibitem{chakraborty2011effects}
N.~Chakraborty, R.~Cant, Effects of lewis number on flame surface density
  transport in turbulent premixed combustion, Combustion and Flame 158~(9)
  (2011) 1768--1787.

\bibitem{Klein:2017a}
M.~Klein, N.~Chakraborty, S.~Ketterl, A comparison of strategies for direct
  numerical simulation of turbulence chemistry interaction in generic planar
  turbulent premixed flames, Flow Turbulence and Combustion 99 (2017) 955--971.

\bibitem{shin2021apriori}
J.~Shin, M.~Hansinger, M.~Pfitzner, M.~Klein, A priori analysis on deep
  learning of filtered reaction rate, Combustion and Flame (submitted).

\bibitem{keppeler2014low}
R.~Keppeler, E.~Tangermann, U.~Allaudin, M.~Pfitzner, Les of low to high
  turbulent combustion in an elevated pressure environment, Flow, turbulence
  and combustion 92~(3) (2014) 767--802.

\bibitem{bambauer2021vortex}
M.~Bambauer, N.~Chakraborty, M.~Klein, J.~Hasslberger, Vortex dynamics and
  fractal structures in reactive and nonreactive richtmyer--meshkov
  instability, Physics of Fluids 33~(4) (2021) 044114.

\bibitem{fureby2005fractal}
C.~Fureby, A fractal flame-wrinkling large eddy simulation model for premixed
  turbulent combustion, Proceedings of the Combustion Institute 30~(1) (2005)
  593--601.

\end{thebibliography}

\end{document}